\documentclass[11pt]{article}

\usepackage{amsfonts,amsmath,amssymb,graphics,bm}

\newtheorem{prop}{Proposition}
\newtheorem{lemma}{Lemma}

\newcommand{\R}{\mathbb{R}}

\newcommand{\fx}{f_X}
\newcommand{\gx}{g_X}
\newcommand{\fy}{f_Y}
\newcommand{\gy}{g_Y}
\newcommand{\hy}{h_Y}
\newcommand{\myprod}[2]{{#1}\circ{#2}}
\newcommand{\ffield}{A}
\newcommand{\xzero}{X_0}
\newcommand{\yzero}{Y_0}
\newcommand{\sstar}{\Delta s^\star}
\newcommand{\eval}{\epsilon}
\newcommand{\qone}{q_1}
\newcommand{\qtwo}{q_2}
\newcommand{\pone}{\sqrt{\qone}}
\newcommand{\ptwo}{\sqrt{\qtwo}}
\newcommand{\deltahat}{\hat{\delta}}
\newcommand{\gammahat}{\hat{\gamma}}
\newcommand{\gm}{\bm{a}}
\newcommand{\mI}{\bm{I}}
\newcommand{\mtheta}{\bm{\theta}}
\newcommand{\mq}{\bm{q}}
\newcommand{\mqone}{\bm{q}_{1}}
\newcommand{\mqtwo}{\bm{q}_{2}}
\newcommand{\spc}{5pt}
\newcommand{\muinfty}{\mu_\infty}
\newcommand{\msigma}{\bm{\sigma}}
\newcommand{\siminf}{\bumpeq}
\newcommand{\trs}{^\dagger}

\newcommand{\I}{\mathrm{i}}
\newcommand{\pr}{\mathbf{P}}
\newcommand{\ex}{\mathbf{E}}

\newcommand{\V}{\mathbf{V}}
\newcommand{\indic}{\mathbf{1}}

\newcommand{\notthis}[1]{}
\newcommand{\Real}{\mathrm{Re}\,}
\newcommand{\gap}{\vspace{3mm}}
\newcommand{\inv}{^{-1}}

\newcommand{\half}{\frac{1}{2}}

\newcommand{\cdl}{\,|\,}

\newcommand{\shalf}{{\textstyle\frac{1}{2}}}

\newcommand{\pderiv}[2]{\frac{\partial{#1}}{\partial{#2}}}
\newcommand{\deriv}[2]{\frac{d{#1}}{d{#2}}}

\newcommand{\pdderiv}[2]{\frac{\partial^2{#1}}{\partial{#2}^2}}
\newcommand{\sech}{\mathrm{sech}}

\newcommand{\arccosh}{\mathrm{arccosh}}

\newcommand{\Beta}{\mathrm{B}}
\newcommand{\tr}{\mathrm{tr}}

\begin{document}

\title{\bf Stochastic entropy production in diffusive systems}

\author{R. J. Martin\footnote{Department of Mathematics, Imperial College London, Exhibition Road, London, SW7 2AZ, U.K.}{ } and I. J. Ford\footnote{Department of Physics and Astronomy and London Centre for Nanotechnology, University College London, Gower Street, London, WC1E 6BT, U.K.}}

\maketitle

\begin{abstract}
Computing the stochastic entropy production associated with the evolution of a stochastic dynamical system is a well-established problem.
In a small number of cases such as the Ornstein--Uhlenbeck process, of which we give a complete exposition, the distribution of entropy production can be obtained analytically, but in general it is much harder. A recent development in solving the Fokker--Planck equation, in which the solution is written as a product of positive functions, enables the distribution to be obtained approximately, with the assistance of simple numerical techniques.
Using examples in one and higher dimension, we demonstrate how such a framework is very convenient for the computation of stochastic entropy production in diffusion processes.
\end{abstract}
\maketitle

\section{Introduction}

\label{sec:Intro}

The notion of the production of entropy as physical systems evolve dates from the time of Boltzmann and Gibbs, and underpins basic ideas of thermodynamics and statistical mechanics, including the celebrated second law of thermodynamics describing the irreversibility of events
on the macroscale \cite{Ford13}. The modern understanding of the production of entropy, based on effective stochastic dynamical models of system evolution \cite{Seifert05,Seifert08}, has to
a great extent clarified some of the puzzles raised by the development of statistical mechanics \cite{Brown09,Lebowitz99}, and has broadened
the meaning of the second law, especially at the nanoscale. While the entropy of the world is \emph{expected} to increase with time, this is subject to fluctuations, and for certain realisations of the dynamics of a system, entropy may decrease. 

The \emph{stochastic} \emph{entropy production} associated with a thermodynamic process is fundamentally a measure of the \emph{reversibility}
of the dynamics. It is defined for a specific path, or trajectory, taken by a system, and represents the `arrow of time' by reflecting the difference in likelihood between the trajectory in question and its time-reversed counterpart in suitable circumstaances \cite{Ford15}. It is very naturally
defined for a system that is subject to environmental influence by way of energy or particle exchanges, and may be divided conveniently into contributions associated with the flow of these quantities between the system and environment together with an internal production of
entropy within the system. It may also be defined for an isolated system if some procedure of coarse graining is adopted.

The degree to which the traditional second law is `broken' in the modern framework is quantified, at least to an extent, by fluctuation relations, symmetry requirements satisfied by the probability distribution of entropy production \cite{Carberry04,Harris07,SpinneyFord12a}.
Beyond this, a more detailed quantification of the fluctuations, necessary if we are to demonstrate how entropy behaves on different temporal scales and for systems of different size, typically requires extensive numerical investigation. The central problem is obtaining the solution to a Fokker--Planck equation that describes the evolving probability density function (pdf) of a system over its phase space, from which the probability distribution of entropy generated in realisations of the stochastic process may be derived. 

The computation of entropy production relies on assigning the correct probabilities to trajectories between points in system phase space followed over arbitrary time intervals. 
For simple models such as the Ornstein--Uhlenbeck model (OU, \cite{Uhlenbeck30}) the transition probability between arbitrary points over arbitrary time intervals is known in closed form and, as we show in \S\ref{sec:ou1}, the distribution of entropy distribution can be calculated analytically as well. 
In cases where the transition probability density is not known, the problem is much harder.
The estimation of a transition density by Monte Carlo simulation requires a very large number of simulations, and if one instead uses a partial differential equation (PDE) solver then solutions starting for each initial point will be needed to obtain the transition density to all possible final points, which is a considerable task.

Analytical approximations appear to confer a computational advantage, but they may suffer from another problem, which stems from the fact that we are after the \emph{logarithm} of the probability of a path, and this is particularly difficult when the probability is small. Any kind of approximation that assigns even the slightest negative (or zero) probability to some part of the transition density is destined to fail. In fact, most analytical approximations for PDEs are based
on orthogonal \emph{sum} expansions, and as the functions in question are oscillatory, this kind of error is hard to eradicate.
A variety of analytical and numerical methods have been employed in studies of the distribution of work performed or heat delivered by a system undergoing a stochastic process, which are associated with the production of entropy \cite{Auconi19,Chatterjee10,Crisanti17,Deffner10,Deffner08,Ford12,Imparato05,Imparato07,Manikandan17,Nicolis17,Pal13,Pal14,Qian13,Saha09,Talkner08}.

In a recent paper \cite{Martin18b}, a new method of dealing with Fokker--Planck PDEs was developed, in which the general
thesis is that one does better to write the solution as a \emph{product} of terms, all of which are positive. This naturally models the logarithm of the phase space density, so it obviates the difficulties described above, and is potentially very suitable for dealing with problems that pertain to entropy production. As the initial condition is a delta-function, it is very hard to make a sum of terms add up to zero at all points except at the starting-point where an enormous spike is required: truncation of such a series necessarily produces oscillatory artefacts. However, a product does not suffer from this disadvantage. One of the terms can, for $t\to0$, be zero except at the location of the delta-function---a Gaussian of variance $\propto$~time captures this effect perfectly, and indeed is an immediate consequence of the Fokker--Planck equation---and the initial condition will be correctly captured \emph{regardless of the other terms}. Another term can capture the long-time asymptote, and further terms `patch up the middle' without corrupting the short- and long-time behaviour. Indeed, a simple approximation containing only two terms (i.e.\ without intermediate-time corrections) is often very satisfactory, which is (\ref{eq:fapprox}) later, and this is indeed a product formula. The method also extends to higher-dimensional cases via (\ref{eq:newf2}) which is again clearly a product formula.
Intuitively, the method consists in expanding around an OU model, in the sense of finding the characteristics of a mean-reverting solution as exemplified by the OU case and capturing these characteristics for the general case, while reproducing the OU case exactly. It thereby provides a stepping-stone from the tractable OU model to the intractable general case.
Potentially, it also gives insight into Fokker--Planck diffusions in other contexts, and a comment in some very recent work on the related area of mutual information struck us as highly pertinent: ``Indeed, there is an interesting dichotomy in which we understand the intricate properties of the Ornstein--Uhlenbeck process since we have an explicit solution, whereas we know very little about the general Fokker--Planck process.'' \cite[\S{V}]{Wibisono19}.

This paper is, then, the first application in physics of the new method for approximately solving Fokker--Planck equations, and we apply it to the problem of stochastic entropy production in a range of different diffusive systems, all corresponding to the spreading of probability density from an earlier point source, but with different force fields and hence different stationary states (see Figure~\ref{fig:various}).
It is organised as follows. After a brief introduction to the nomenclature and methods, Section~\ref{sec:1d} deals with the one-dimensional theory. It begins with a complete exposition of the OU model, and provides new insights into that case and also into the arithmetic Brownian motion (drift-diffusion) which is a special case of it. After this it develops the theory for a general potential in \S\ref{sec:gp}, giving a brief account of the leading-order approximation shown in \cite{Martin18b} before showing a variety of examples in \S\ref{sec:ex1}.
Section~\ref{sec:md} describes the multidimensional theory, sgain starting with the multivariate OU model which is dealt with in some depth, with comparisons drawn to the one-dimensional case. Then the theory for a general potential is developed, along the same lines as in \cite{Martin18b} and \S\ref{sec:1d}. Our final section (\S\ref{sec:conc}) gives our conclusions and suggests opportunities for further research.


\section{Preliminaries}

\label{sec:prelim}
		
\subsection{Definitions}

For a stochastic process $X_t$, the dimensionless stochastic entropy production, associated with a transition from an initial to a final system coordinate during the interval between times $t_1>0$ and $t_2>t_1$, is defined as
\begin{equation}
\Delta s=\ln\left(\frac{\fx(t_{1},X_{t_{1}}){\cal T}(X_{t_{1}}\to X_{t_{2}},t_{2}-t_{1})}{\fx(t_{2},X_{t_{2}}){\cal T}(X_{t_{2}}\to X_{t_{1}},t_{2}-t_{1})}\right),\label{eq:entdef}
\end{equation}
where $\fx(t,X)$ is the pdf over $X$ at time $t$ and $\mathcal{T}(x_1\to x_2,\Delta t)$ is the transition probability density from $x_1$ to $x_2$ in time interval $\Delta t$. The argument of the logarithm is the probability of transitioning from $X_{t_1}$ at time $t_1$ to $X_{t_2}$ at time $t_2$, divided by the probability of a subsequent reversal, i.e.\ starting from $X_{t_2}$ at time $t_2$ and ending up at $X_{t_1}$ at time $t_2-t_1$ later. If the system were in a stationary statistical state (thermal equilibrium) then this ratio would be equal to unity and the stochastic entropy production would vanish for all transitions. In general, $\Delta s$ is nonzero and can take either sign, though it satisfies a second law in possessing a nonnegative expectation when averaged over all possible paths in $[t_1,t_2]$.

A condition of detailed balance holds in thermal equilibrium, defined by
\[
\fx(\infty,X_{t_{1}}){\cal T}(X_{t_{1}}\to X_{t_{2}},t_{2}-t_{1}) = \fx(\infty,X_{t_{2}}){\cal T}(X_{t_{2}}\to X_{t_{1}},t_{2}-t_{1}),
\]
such that we can write
\begin{equation}
\Delta s \big[(t_{1},X_{t_1})\to (t_2,X_{t_2})\big]=\ln\left(\frac{\gx(t_1,X_{t_1}\cdl \xzero)}{\gx(t_2,X_{t_2}\cdl \xzero)}\right),\label{eq:sdef}
\end{equation}
with 
\begin{equation}
\gx(t,x\cdl \xzero)=\frac{\fx(t,x\cdl \xzero)}{\fx(\infty,x)},\end{equation}
and where we explicitly note that we are considering a situation corresponding to evolution from a definite initial coordinate $X_0$ at $t=0$. More complicated scenarios can be constructed from this base case.
Our task is to obtain the probability distribution of $\Delta s$ for the interval between $t_{1}$ and $t_{2}$. Given these definitions, it makes sense to focus on the evolution of $\gx$ as a means of computing the entropy production. 
Indeed, when we come to approximate the density for analytically intractable models, it is $\gx$ that we choose to approximate.

The entropy production is transformation-invariant in the following sense.
If $Y_t=\psi(X_t)$ where $\psi$ has an inverse $\psi\inv$ and both $\psi,\psi\inv$ are appropriately smooth, then whenever $y=\psi(x)$ and $\yzero=\psi(\xzero)$ we must have
\[
\fy(t,y \cdl \yzero) = \fx(t,x \cdl \xzero) / |\psi'(x)|, \qquad
\gy(t,y \cdl \yzero) = \gx(t,x \cdl \xzero)
\]
and so over a particular time period the entropy production of $X$ and $Y$ is the same. Accordingly, we can take the general form of a time-independent diffusion, 
\begin{equation}
dX_t = \mu_X(X_t)\, dt + \sigma_X(X_t) \, dW_t
\label{eq:Xt}
\end{equation}
with $W_t$ a standard Wiener process,
and by applying the transformation $\psi$ given by\footnote{The so-called Lamperti transformation. This works provided $\sigma_X$ is bounded away from zero.}
\[
\psi'(x) = \deriv{y}{x} = \frac{\sqrt{2\kappa}}{\sigma_X(x)},
\]
it becomes
\begin{equation}
dY_t = \kappa \ffield(Y_{t})\,dt + \sqrt{2\kappa}\,dW_t.
\label{eq:Yt}
\end{equation}
By It\=o's lemma\footnote{It can also be derived purely algebraically by substituting $\gy\big(t,\psi(x)\big)=\gx(t,x)$ into the backward equation.} we have
\[
\kappa \ffield(y) = \psi'(x) \mu_X(x) + \shalf \psi''(x) \sigma^2_X(x) 
\]
and we call $\ffield$ the force field.
This transformation simplifies the model by making the volatility constant and does not affect the entropy production, so henceforth we work with (\ref{eq:Yt}); the parameter $\kappa$ has units time$\inv$ and is understood as a rate constant.

The density $\fy$ obeys the forward or Fokker--Planck equation, and the function $\gy$ obeys its adjoint, the backward or Feynman-Kac equation: 
\begin{eqnarray}
\pderiv{\fy}{\tau} & = & -\pderiv{}{y}\big[A(y)\fy\big]+\pdderiv{\fy}{y},\label{eq:pde_f}\\
\pderiv{\gy}{\tau} & = & A(y)\pderiv{\gy}{y}+\pdderiv{\gy}{y}\label{eq:pde_g}
\end{eqnarray}
with $\fy(0,y)=\delta(y-\yzero)$ and dimensionless time $\tau=\kappa t$. When we later present graphs showing the distribution of entropy production over a given time period, we are talking about $\kappa t$ or, equivalently, $t$ if we standardise on $\kappa=1$ throughout.
The invariant density (stationary state) is related to $\ffield$ by $\ffield(y)=\frac{d}{dy} \ln \fy(\infty,y)$.
Note also the
\emph{reciprocity condition} 
\begin{equation}
\gy(t,Y_{2}\cdl Y_{1})=\gy(t,Y_{1}\cdl Y_{2}).
\label{eq:revers}
\end{equation}
(Proof. Viewed as a function of $y$ (and $t$), the LHS obeys the adjoint forward equation, whereas the RHS obeys the backward equation. However, those are the same PDE, with the same initial condition. Alternatively, we can invoke the Kolmogorov criterion, which pertains to the transition probability around any closed loop being independent of the direction of travel: see e.g.~\cite[\S1.5]{Kelly79}.)

One device that we can use for studying the distribution of $\Delta s$ is its moment-generating function (mgf), 
\begin{equation}
M_{\Delta s}(\lambda) = 
\ex\left[\exp(\lambda\Delta s)\right] =
\ex\left[\Biggr(\frac{\gy(t_1,Y_{t_1}\cdl \yzero)}{\gy(t_2,Y_{t_2}\cdl \yzero)}\Biggr)^{\lambda}\right] ,
\end{equation}
where the expectation is over all transitions during the interval $t_{1}\le t\le t_{2}$.
It is easily seen  that $M_{\Delta s}(-1)=1$, known as the integral fluctuation relation \cite{Seifert05}.
The mgf relates to the density $p$ by
\begin{equation}
M_{\Delta s}(\lambda) = \int_{-\infty}^\infty p(\Delta s) e^{\lambda \Delta s} \, d\Delta s .
\label{eq:mgfint}
\end{equation}

\subsection{The $K$-, or Variance-Gamma, distribution}

When the dynamics are Gaussian, as in the OU process, the entropy production is a quadratic function and in the particular case where the starting-point is the equilibrium level (where the force field vanishes) its density can be found via the mgf using the following result, known variously as the $K$-distribution (in the physics literature) and the Variance-Gamma distribution (in the quantitative finance literature). We shall use it more than once:
\begin{lemma}
The pdf 
\begin{equation}
\frac{(b^2-a^2)^{\nu+1/2} e^{ax} |x/2b|^{2\nu} K_\nu(b|x|)}{\sqrt{\pi} \, \Gamma(\nu+\half)}
\label{eq:kdist}
\end{equation}
with $K_\nu$ denoting the modified Bessel function of the second kind \cite{Abramowitz64}, corresponds to the mgf
\begin{equation}
\bigg(\frac{b^2-a^2}{b^2-(\lambda+a)^2}\bigg)^{\nu+1/2},\qquad|a+\Real\lambda|<b,
\label{eq:kdist.mgf}
\end{equation}
and from this the mean and variance are
\[
\frac{(2\nu+1)a}{b^2-a^2}, \qquad 
\frac{(2\nu+1)(b^2+a^2)}{(b^2-a^2)^2}.
\]
\end{lemma}
\noindent Proof.
Immediate from the Fourier representation of $K_\nu$, e.g.\ \cite[\S8.432.5]{Gradshteyn94}.
$\Box$

\gap \noindent
When the order $\nu$ is a half-integer the function $K_\nu$ admits a representation using elementary functions, and when $\nu=\half$ the distribution is simply a double-expeonential.

\subsection{Inversion integrals}
\label{sec:fft}

When the mgf of the entropy production is known in closed form---the OU model is a case in point---it is possible to obtain the pdf of the entropy production by inverting (\ref{eq:mgfint}) via a Fourier integral: 
\begin{equation}
p(\Delta s) = 
\frac{1}{2\pi} \int_{-\infty}^\infty  M_{\Delta s}(\I\omega) e^{-\I\omega  \Delta s} \, d\omega =
\frac{1}{2\pi\I} \int_{\mathcal{C}}  M_{\Delta s}(\lambda) e^{-\lambda  \Delta s} \, d\lambda 
\end{equation}
where the contour $\mathcal{C}$ runs up the imaginary axis.

The most straightforward approach is to replace the integral with a finite sum, i.e.\ use the discrete Fourier transform, implemented via the Fast Fourier Transform (FFT) algorithm. There are a couple of points to be made about this. In going from an inversion integral to a finite, discrete approximation two undesirable effects are introduced. The first, \emph{aliasing}, causes the inverse transform to be periodised, with a period equal to $2\pi/\delta \omega$, where $\delta\omega$ is the spacing in $\omega$-space. If $d\omega$ is too large, features will appear in the wrong place. The second, \emph{leakage}, occurs as a result of truncating the range of integration (summation) to $[-\Omega/2,\Omega/2]$ say, and features are smeared out, losing resolution. It can be attenuated by multiplying $M(\I\omega)$ by a \emph{window function}, that is zero or nearly zero when $\omega$ is close to the ends of the interval $[-\Omega/2,\Omega/2]$, and  unity at the middle. A simple choice is the Hann window, given by $\cos^2(\pi\omega/\Omega)$; further details are in \cite[\S13]{NRC}. As an example of this in action: suppose that in the problem at hand we think that the entropy production can safely be ignored outside the range $[-5,5]$, and that we want a resolution of around 0.01 in entropy space. Then we shall need a 1024-point FFT and the spacing of the samples in $\omega$-space will need to be $\delta\omega=2\pi/10$. As an example of what aliasing means, suppose we were wrong about the entropy production being essentially oonfined to the interval $[-5,5]$, and that there were a pronounced spike at $\Delta s=6.3$, for the sake of argument: then this feature would incorrectly appear at $\Delta s=-3.7$.
Other techniques that numerically invert the Fourier integral are orthogonal expansion using Laguerre functions \cite{Weeks66}, provided that this is extended to the two-sided case, and numerical integration techniques discussed by Abate \& Whitt \cite{Abate95}.

Another, more specialised, approach requires analyticity of $M$, and deforms $\mathcal{C}$ so that it lies along a more convenient path. In many cases, including here, the mgf has branch points at $\lambda=\lambda_\pm$ say and is analytic in the complex plane with  $(-\infty,\lambda_-]$ and $[\lambda_+,\infty)$ deleted.
If the integrand is integrably singular at $\lambda_\pm$, we can collapse the integral around either branch cut, depending on the sign of $\Delta s$, and do a real integral (see e.g.\ \cite[\S3.1]{Martin19a}). However, this cannot be done if the integrand also possesses an essential singularity at $\lambda=\lambda_\pm$. 
In fact both cases can be seen in the OU model, depending on whether $\yzero$ is zero or not (see (\ref{eq:mgf2}) later).
This brings us to our final point, which is that it may be useful to deform $\mathcal{C}$ so that it lies along the path of steepest descent \cite{Bender78}, or at least fairly close to it. That way, the integrand is not too oscillatory, and is easily approximated by analytical methods or by numerical integration. This is what gives rise to \emph{saddlepoint} methods \cite{Daniels87}.

Interestingly, the problem of finding the pdf of a quadratically-transformed multivariate Normal variable---which is fundamental to stochastic entropy generation in the OU model---finds application in unrelated areas, notably mathematical finance. In \cite{Feuerverger00}, the authors use the moment-generating function as we do, and show by matrix algebra that the variable in question can be written as a weighted sum of independent Gamma-distributed variables. The authors discuss the use of saddlepoint methods as a way of approximating the pdf, though do not indicate why it is effective. In fact the Gamma distribution is very well approximated by its saddlepoint approximation, and this is why the technique works well \cite{Martin11b}.

\subsection{Entropy production from transition density by stratified sampling}

\label{sec:ss}

Much attention has been given in the literature to the derivation of the pdf of entropy production through efficient (semi-)analytical approaches, but in fact it is in principle straightforward to proceed numerically if the transition density is known exactly or approximately. For given $\yzero,t_{1},t_{2}$ we proceed as follows. 

Bound the phase space by $\mathcal{V}$ so that we disregard the possibility that the process goes outside this space.
Select a set of points $\{y_j\}_{j=1}^N$ (in our calculations we have used $N=10000$) that are distributed reasonably uniformly over $\mathcal{V}$. We enlarge on this point presently.
Next, divide entropy space into bins---throughout, we take the interval $[-10,10]$ and partition it into 2000 bins each of width 0.01---and write 0 in each bin.
For each pair $(j,k)$, calculate the entropy production using $\gy$, find which entropy bin this corresponds to, and add to that bin the quantity
\[
p_{jk} = \fy(t_1,y_j \cdl \yzero) \, \mathcal{T}(y_j \to y_k, t_2-t_1) .
\]
Note that an alternative notation for $\mathcal{T}(y_j\to  y_k, t_2-t_1)$ is $\fy(t_2-t_1, y_k \cdl y_j)$.
Write also
\[
\Delta s_{jk} = \ln \left(\frac{\gy (t_1,y_j \cdl \yzero)}{\gy(t_2,y_k \cdl \yzero)}\right).
\]
It is understood that the use the exact transition density and $\gy$ if these functions are known, or otherwise approximations to them.
Once this has been done for all pairs $(j,k)$, the probability of the entropy production lying in some interval $J$ is approximated by
\[
\pr(\Delta s \in J) \approx \frac{ \sum_{j,k} \indic [ \Delta s_{jk} \in J ] p_{jk}
}{
\sum_{j,k} p_{jk}
}.
\]
The denominator is just the sum over all bins, and this construction obviously ensures that the total probability mass for the entropy production is exactly unity, i.e.\ $M_{\Delta s}(0)=1$.

As promised we give a brief exposition of sampling techniques.
For a one-dimensional phase space this is easy as we dissect $\mathcal{V}$ into $N$ equal intervals. However, if $N$ is not known upfront, in the sense that one might want to run a certain number of samples, and then some more, and then some more after that, the use of pseudo- or quasi-random  procedures is preferred; this is also the case when the dimension is $>1$. We take these in turn.
\emph{Pseudo-random} sequences mimic the generation of truly random ones, and are typically obtained from whatever inbuilt linear congruential generator the user has to hand; for a discussion of these see e.g.\ \cite{NRC} and \cite[\S7]{Jackel02}. However, the random nature will in any one finite realisation of the random number generation give rise to nonuniformities, with some areas receiving more samples than others; the standard error of a Monte Carlo integral decays as $O(N^{-1/2})$ as $N\to\infty$.
\emph{Quasi-random} or \emph{low-discrepancy} sequences, which are not at all random, are designed to cover the space more uniformly by `filling in the gaps as they go'; the error from these decays as $S(N)/N$, where $S(N)/N^\varepsilon\to0$ for all $\varepsilon>0$ (often $S(N)$ is roughly a power of $\ln N$, but a deeper discussion of this would take us too far off track).
Two particular classes of low-discrepancy sequence are worth mentioning. The first are called Sobol sequences \cite[\S7.7]{NRC}, \cite[\S8.3]{Jackel02}. The second uses ideas from Diophantine approximation theory \cite{Niederreiter92}: the construction is 
\[
u_n = \big( \{n\alpha_1\}, \ldots, \{n\alpha_d\} \big)
\in [0,1)^d
\]
where $\{\cdot\}$ denotes the fractional part and $\alpha_1,\ldots,\alpha_d$ are appropriately-chosen irrational numbers.
Our preference is for this method on account of its simplicity.

We now return to the matter of entropy generation.
A final, optional, step is to ensure that the integral fluctuation relation holds exactly in spite of the various approximations (truncation of the phase space to $\mathcal{V}$, use of approximated transition density, use of finite sumber of samples). Defining
\[
\varepsilon = \ln \biggr( \sum_{j,k} p_{jk} e^{-\Delta s_{jk}} \biggr) ,
\]
which can be computed while the above calculation is being done, and shifting all the entropy bins by $\varepsilon$, will achieve this.

\notthis{
We shall employ this approach later on using an approximate transition density for a variety of cases using a high granularity ($N_1=N_2=5000$, $\delta y=0.002$ *** check ***), with entropy bins of width 0.01. Our problems are nondimensionalised, so they all have roughly the same length scale, but $\delta y$ still varies a little from problem to problem and we state with each result what value we used.
}

If we only need the \emph{expected} entropy production then we do not need to evaluate a double (nested) integral: instead we just need to compute a pair of single integrals as 
\begin{align}
\ex[\Delta s] & =\overline{s}(t_{2})-\overline{s}(t_{1}),\label{eq:mean}\\
\overline{s}(t) & =-\int_{-\infty}^{\infty}\ln \gy(t,y\cdl\yzero)\cdot \fy(t,y\cdl\yzero)\,dy.\nonumber 
\end{align}
The second law $\overline{s}(t_{2})-\overline{s}(t_{1})\ge0$ follows immediately from the Fokker--Planck equation or from the integral fluctuation relation\footnote{The log of the mgf, $K(\lambda)$ say, is necessarily convex; also $K(0)=0$ and $K'(0)$ is the mean. If, as here, $K(-1)=0$, then convexity requires that $K'(0)\ge0$.}.

\subsection{Note on density functions}

As will become apparent, the pdf of the entropy production is often singular at one or points, or more informally it may contain a large spike somewhere. 
More important than the pdf is the probability mass function or cumulative distribution function (cdf): one wants to know where the probability mass lies.
It is hard to visually estimate the probability mass under a narrow but infinitely high spike: such a feature may be visually distracting but yet represent only a very small probability mass, and of course if the spike is caused by a delta-function of strength $c$ then it is impossible to assess the value of $c$ from a density plot.
Furthermore two distributions can have very different pdf but quite similar cdf\footnote{On the principle that two functions may be close but their derivatives may not be.}. Indeed, for comparing two distributions, common methods such as Kolmogorov--Smirnov use cdf and not the pdf.
Finally, in higher-dimensional cases the granularity of the method described in \S\ref{sec:ss} gives rise to a `hairiness' in the pdf which is difficult to eradicate, as seen for example in Figures~\ref{fig:student2d} and \ref{fig:student3d}.
We have therefore chosen to additionally plot the \emph{logit function}, commonly encountered in statistics and defined as $L(x)=\ln \frac{F(x)}{1-F(x)}$ where $F$ is the cdf. Obviously this is one order of differentiability smoother than the pdf.
The median is immediately apparent as $L\inv(0)$, the interquartile range is roughly $[L\inv(-1),L\inv(1)]$, and a two-sided 99.5\% confidence interval is roughly $[L\inv(-6),L\inv(6)]$. This method of plotting is also convenient for examining the shapes of the tails.


\section{Theory in one dimension}

\label{sec:1d}

\subsection{Ornstein--Uhlenbeck}

\label{sec:ou1}

The OU model with $A(y)=-\theta y$ is analytically tractable, and we discuss it in detail as a foundational step. We have 
\begin{eqnarray}
\fy(t,y) & = & \frac{1}{\sqrt{2\pi(1-q)/\theta}}\exp\left(-\frac{(y-\sqrt{q}\,\yzero)^{2}}{2(1-q)/\theta}\right), \\
\gy(t,y) & = & \frac{1}{\sqrt{1-q}}\exp\left(-\frac{qy^{2}-2\sqrt{q}\,y\yzero+q\yzero^{2}}{2(1-q)/\theta}\right),
\end{eqnarray}
with $q=e^{-2\theta\tau}$, $\tau=\kappa t$, and we drop the $\cdl \yzero$ notation for simplicity. The entropy production $\Delta s$ is simply a quadratic in the triplet $(\yzero,Y_{1},Y_{2})$ (where $Y_{1}$ is short for $Y_{t_{1}}$, etc.): it is 
\begin{equation}
\Delta s =
\half\ln\frac{1-\qtwo}{1-\qone} - \frac{\theta}{2}
\begin{bmatrix}\yzero \\ Y_{1} \\ Y_{2} \end{bmatrix} \trs
\begin{bmatrix}\frac{\qone}{1-\qone}-\frac{\qtwo}{1-\qtwo} & \frac{-\pone}{1-\qone} & \frac{\ptwo}{1-\qtwo} \\[\spc]
\frac{-\pone}{1-\qone} & \frac{\qone}{1-\qone} & 0 \\[\spc]
\frac{\ptwo}{1-\qtwo} & 0 & \frac{-\qtwo}{1-\qtwo}
\end{bmatrix}
\begin{bmatrix} \yzero \\ Y_1 \\ Y_2 \end{bmatrix} \end{equation}
with $q_i=e^{-2\theta \tau_i}$.
(Throughout this paper $\trs$ denotes the transpose.)
From this we can see that the mean entropy production is\footnote{Positivity, which is anticipated on fundamental grounds, can be seen from the fact that $\ell(x) = \ln x-x$ is an increasing function for $0<x\le 1$, and so $\ell(1-q_1) < \ell(1-q_2$).} 
\begin{equation}
\ex[\Delta s]= \half \ln\frac{1-\qtwo}{1-\qone}+\frac{\qone-\qtwo}{2}(\theta\yzero^2-1)\ge0.
\end{equation}
Analytical results can be derived by appeal to the moment generating function $M_{\Delta s}$: we are to find the double integral 
\begin{equation}
M_{\Delta s}(\lambda) = \left(\frac{1-\qtwo}{1-\qone}\right)^{\lambda/2}\frac{\theta}{2\pi(1-\qone)^{1/2}(1-\qtwo/\qone)^{1/2}} \int_{-\infty}^{\infty}\int_{-\infty}^{\infty} e^{-\mathcal{Q}/2} \,dy_{1}\,dy_{2},
\label{eq:mgfou}
\end{equation}
where $\mathcal{Q}$ is the quadratic form 
\[
\mathcal{Q}= 
\begin{bmatrix}\yzero\\y_1\\y_2\end{bmatrix}\trs 
\begin{bmatrix}\frac{\theta(\lambda+1)\qone}{1-\qone}-\frac{\theta\lambda\qtwo}{1-\qtwo} & -\frac{\theta(\lambda+1)\pone}{1-\qone} & \frac{\theta\lambda\ptwo}{1-\qtwo}\\[\spc]
-\frac{\theta(\lambda+1)\pone}{1-\qone} & \frac{\theta\lambda\qone+1}{1-\qone}+\frac{\theta\qtwo}{\qone-\qtwo} & -\frac{\theta\sqrt{\qone\qtwo}}{\qone-\qtwo}\\[\spc]
\frac{\theta\lambda\ptwo}{1-\qtwo} & -\frac{\theta\sqrt{\qone\qtwo}}{\qone-\qtwo} & -\frac{\theta\lambda\qtwo}{1-\qtwo}+\frac{\theta\qone}{\qone-\qtwo}
\end{bmatrix}
\begin{bmatrix}\yzero\\
y_{1}\\
y_{2}
\end{bmatrix} .
\]
Write the matrix in the above expression as 
\[
Q = \begin{bmatrix}Q_{00} & Q_{01} & Q_{02}\\
Q_{10} & Q_{11} & Q_{12}\\
Q_{20} & Q_{21} & Q_{22}
\end{bmatrix},
\]
and define the following two determinants: 
\[
\Delta_Q=\left|\begin{matrix}Q_{00} & Q_{01} & Q_{02}\\
Q_{10} & Q_{11} & Q_{12}\\
Q_{20} & Q_{21} & Q_{22}
\end{matrix}\right|, \qquad 
\delta_Q=\left|\begin{matrix}Q_{11} & Q_{12}\\
Q_{21} & Q_{22}
\end{matrix}\right|.
\]
The double-integral of $e^{-\mathcal{Q}/2}$ above evaluates to 
\[
\frac{2\pi}{\delta_Q^{1/2}}\exp\biggr(-\frac{\yzero^{2}\Delta_Q}{2\delta_Q}\biggr),
\]
and so the mgf of the entropy production is 
\begin{equation}
M_{\Delta s}(\lambda)=\left(\frac{1-\qtwo}{1-\qone}\right)^{\lambda/2}\frac{\exp(-\yzero^{2}\Delta_Q/2\delta_Q)}{[(1-\qone)(1-\qtwo/\qone)\hat{\delta}_Q]^{1/2}}
\label{eq:mgf2}
\end{equation}
with $\hat{\delta}_Q=\delta_Q/\theta^2$.
This is understood as follows. By the shifting theorem (of the Laplace transform) the term on the front represents a shift to the right by an amount 
\begin{equation}
\sstar=\half\ln\frac{1-\qtwo}{1-\qone}>0.
\end{equation}
We now examine the rest of the expression, starting with $\yzero=0$.

\subsubsection{Case $\yzero=0$}

\begin{figure}
\noindent
\begin{tabular}{rr}
\scalebox{0.625}{\includegraphics*{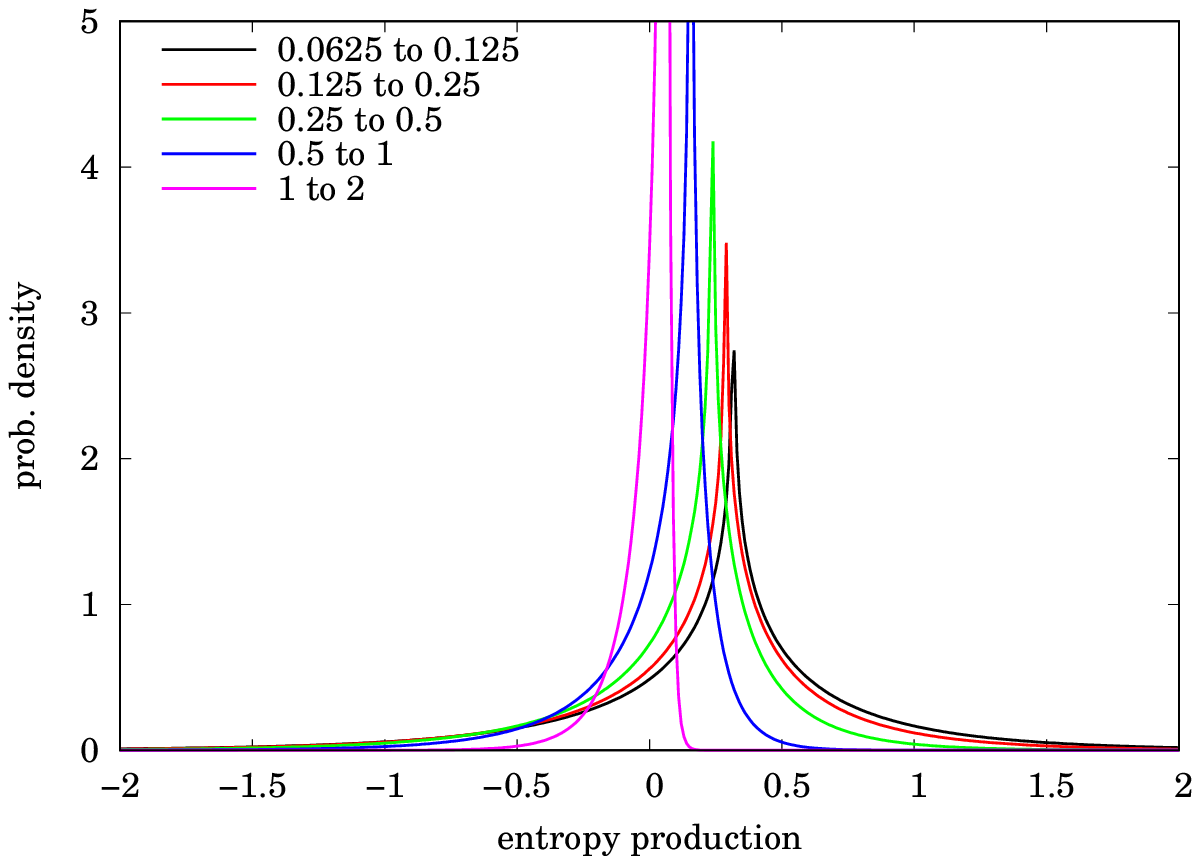}} &
\scalebox{0.625}{\includegraphics*{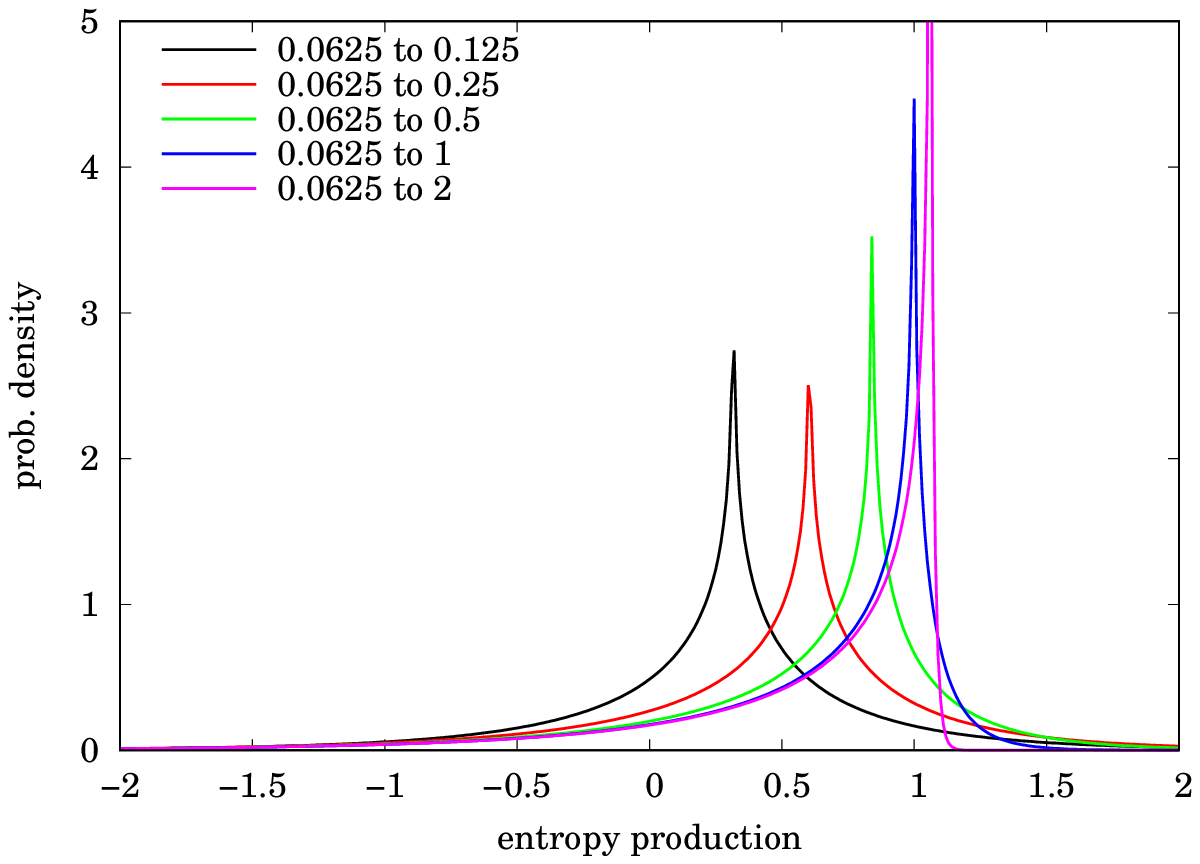}}
\end{tabular}
\caption{\small Probability density of entropy production for OU process ($\ffield(y)=-y$) starting from the origin. (Left) Incremental entropy in time intervals $[0.0625,0.125]$, $[0.125,0.25]$ etc., as labelled; (Right) entropy production from time 0.0625 to times $0.125,0.25,0.5,1,2$. 
The density has a very narrow spike up to $\infty$: on account of the finite plotting resolution this is not fully captured, but its location is clear in each case.
}
\label{fig:ou0}
\end{figure}

When $\yzero=0$ the numerator of (\ref{eq:mgf2}) is unity, and
we are left with the denominator, which is the square root of a quadratic in $\lambda$. By evaluating $\delta_Q$ we establish 
\begin{eqnarray}
M_{\Delta s}(\lambda) & = & \left(\frac{1-\qtwo}{1-\qone}\right)^{\lambda/2}\!\!\bigg(1+(\qone-\qtwo)\lambda-\frac{\qtwo(\qone-\qtwo)}{1-\qtwo}\lambda^{2}\bigg)^{-1/2}\nonumber \\
 & \equiv & \left(\frac{1-\qtwo}{1-\qone}\right)^{\lambda/2}(1-\lambda/\lambda_{+})^{-1/2}(1-\lambda/\lambda_{-})^{-1/2},\label{eq:lampm}
\end{eqnarray}
as the quadratic must have real roots $\lambda_{\pm}$ of opposite sign. Indeed, observing that the quadratic is positive (and in fact equal to $(1-\qone)/(1-\qtwo)$) at the two values $\lambda=-1$ and $\qtwo\inv$, we deduce something stronger: 
\[
\lambda_- < -1 < 1 < \qtwo\inv < \lambda_+.
\]
Further, as $\lambda_++\lambda_-= \qtwo\inv-1>0$, the right-hand tail decays more quickly than the left.
The case $t_2\to\infty$ is special, as the positive root goes to $\infty$ and the negative one to $-\qone\inv$, and the entropy production is given by
\[
\half \ln \frac{1}{1-\qone} - \frac{\qone\Xi}{2}, \qquad \Xi \sim \chi^2_1,
\]
i.e.\ $\Xi$ is distributed as $\chi^2$ with one degree of freedom. Consequently in this limit the production of entropy is bounded above but not below. This behaviour is suggested by the right-hand plot in Figure~\ref{fig:ou0}.

Comparison of $M_{\Delta s}$ with (\ref{eq:kdist.mgf}) yields:
\begin{prop}
\label{prop:ou1_0}
For the OU model with $\yzero=0$, the distribution of the entropy production $\Delta s$ for the interval $[t_1,t_2]$
is given by: 
\begin{equation}
p(\Delta s) =
\pi\inv(b^{2}-a^{2})^{1/2}e^{a(\Delta s-\sstar)}K_0(b| \Delta s-\sstar|), \qquad \Delta s\in\R, \quad0\le|a|<b
\label{eq:k0distsh}
\end{equation}
with $\lambda_\pm$ given by (\ref{eq:lampm}) and
\begin{equation}
\begin{array}{lclcl}
a & = & {\displaystyle -\frac{\lambda_+ +\lambda_-}{2}} & = & {\displaystyle -\frac{1-\qtwo}{2\qtwo},}\\
b & = & {\displaystyle \frac{\lambda_+ -\lambda_-}{2}} & = & {\displaystyle \left(\frac{1-\qtwo}{\qtwo(\qone-\qtwo)}+\bigg(\frac{1-\qtwo}{2\qtwo}\bigg)^{2}\right)^{1/2},}\\
\sstar & = & {\displaystyle \half\ln\frac{1-\qtwo}{1-\qone}},
\end{array}  
\end{equation}
and $q_1=e^{-2\kappa \theta t_1}$, $q_2=e^{-2\kappa\theta  t_2}$.
The mean is
\[
\ex[\Delta s] =  \half \left( \ln \frac{1-q_2}{1-q_1} + (q_2-q_1) \right) > 0
\]
and the variance is
\[
\V[\Delta s] = \frac{(q_1-q_2)(q_1+q_2-q_1q_2+q_2^2)}{2(1-q_2)} .
\]
The density has a logarithmic singularity (spike) at $\sstar$ the origin; the probability of being to the right (resp.\ left) of this is 
\[
\pr(\Delta s \gtrless \sstar) = \frac{\arccos(\mp a/b)}{\pi},
\]
and the mean conditional on being positive (resp.\ negative) is 
\[
\ex[\Delta s \cdl \Delta s \gtrless \sstar] = \frac{a}{b^{2}-a^{2}}\pm\frac{\sqrt{b^{2}-a^{2}}}{b^{2}\arccos(\mp a/b)}.
\]
By analysis of the singularities of the mgf the asymptotes of the density are\footnote{Recall $\lambda_-<0<\lambda_+$.}
\begin{equation}
p(\Delta s) \sim \mathrm{const} \times |\Delta s|^{-1/2} \exp(-\lambda_\pm \Delta s), \qquad \Delta s\to\pm\infty .
\label{eq:k0distasym}
\end{equation}
The distribution arises as the weighted difference of two independent central $\chi_{1}^{2}$ random variables (with positive weights).
$\Box$
\end{prop}
Figure~\ref{fig:ou0} shows some examples of this analytical computation (with $\theta=1$).

\subsubsection{Case $\yzero \ne0$}

The effect of starting at $\yzero\ne0$ is found by convolving the previous result (\ref{eq:k0distsh}) with the function whose mgf is $\exp(-\yzero^{2}\Delta_Q/2\delta_Q)$, so to make further deductions we must examine this expression in detail. Now $\Delta_Q$ is a cubic in $\lambda$ that vanishes at $\lambda=0,-1$, because $M_{\Delta s}(0)=M_{\Delta s}(-1)=1$ for all $\yzero$; also, the coefficient of $\lambda^{3}$ vanishes.
(Write the entropy production as a quadratic form in $(\yzero,Y_{1},Y_{2})$: as it can be written as a sum of only two squares, its determinant vanishes.) So $\Delta_Q$ must be of the form $\lambda(\lambda+1)$ multiplied by a constant, and in fact 
\begin{equation}
\Delta_Q=\frac{-\qone \theta^3}{1-\qone}\lambda(\lambda+1).\label{eq:DQ}
\end{equation}
This can be seen directly by from elementary row and column operations on the matrix:
\begin{equation}
\begin{bmatrix} 1 & \pone & \ptwo \\[\spc] 0 & 1 & 0 \\[\spc] 0 & 0 & 1 \end{bmatrix}
Q 
\begin{bmatrix} 1 & 0 & \ptwo \\[\spc] 0 & 1 & \sqrt{\qtwo/\qone} \\[\spc] 0 & 0 & 1 \end{bmatrix}
= \theta 
\begin{bmatrix} 0 & -\lambda\pone & 0 \\[\spc]
\frac{-(\lambda+1)\pone}{1-\qone} & \frac{\lambda\qone+1}{1-\qone}+\frac{\qtwo}{\qone-\qtwo} & 0 \\[\spc]
\frac{\lambda\ptwo}{1-\qtwo} & \frac{-\sqrt{\qone\qtwo}}{\qone-\qtwo} & 1
\end{bmatrix} 
 .
\label{eq:Qfact}
\end{equation}
Returning to (\ref{eq:mgf2}), we can now conclude that the term in the exponential is simply the ratio of two quadratics in $\lambda$:
\begin{equation}
-\frac{\yzero^2 \Delta_Q}{2 \delta_Q} =
\frac{(\qone-\qtwo)\lambda(\lambda+1) \theta\yzero^2/2}{1+(\qone-\qtwo)\lambda - \displaystyle\frac{\qtwo(\qone-\qtwo)}{1-\qtwo}\lambda^2} .
\label{eq:Dqdq}
\end{equation}

To find what pdf corresponds to the mgf that is the exponential of the ratio of two quadratics, we recall the compound Poisson distribution.
Let $P$ follow a Poisson distribution of mean $\mu$, and let $(Z_j)$ be iid (independent and identically distributed) random variables, independent also of $P$, with mgf $M_Z$.
Form the random variable\footnote{There is no standard notation for this. Our choice is motivated by the idea that if we take the sum of $P$ copies of $Z$ then we have a sort of product of $Z$ by $P$; note also that if $K$ denotes the log of the mgf then we have $K_{\myprod{P}{Z}} = K_P \circ K_Z$, i.e.\ the composite of the two functions.} 
\[
\myprod{P}{Z}=\sum_{j=1}^{P}Z_{j}
\]
and note that its mgf can be obtained by conditioning on $P$ and then integrating out:
\[
M_{\myprod{P}{Z}}(\lambda)=\exp\big(\mu M_{Z}(\lambda)-\mu\big).\]
Consider now the distribution formed of two exponentials as follows: with probability $\pi_+$ it is exponential with mean $\lambda_+$, and with probability $\pi_-$ it is ${-1}\times$ an exponential variable of mean ${-\lambda_-}$, where $\lambda_-<0<\lambda_+$.
Thus the density of $Z$ is 
\begin{equation}
f_{Z}(x) = \pi_+ \lambda_+ e^{-\lambda_+ x}\mathbf{1}_{x>0}- \pi_- \lambda_- e^{-\lambda_- x}\mathbf{1}_{x<0},
\label{eq:dblexp}
\end{equation}
and its mgf is 
\begin{equation}
M_{Z}(\lambda)=\frac{\pi_+}{1-\lambda/\lambda_+} + \frac{\pi_-}{1-\lambda/\lambda_-}, \qquad \lambda_- < \Real \lambda < \lambda_+, \qquad \lambda_- < \Real \lambda < \lambda_+.
\end{equation}
This will do what we want, because the ratio of two quadratics can be written as a constant plus an expression of the above form, by partial fractions. Accordingly, we have synthesised a function whose mgf is the exponential of the ratio of two quadratics, and have proven:
\begin{prop}
\label{prop:ou1_ne0}
For the OU model, if $\yzero\ne0$ the distribution of the entropy production is obtained by convolving the $\yzero=0$ result (\ref{eq:k0distsh}) with the compound Poisson distribution $\myprod{P}{Z}$, where $P$ has a Poisson distribution of mean $\mu$ and $Z$ has a double-exponential distribution (\ref{eq:dblexp}), parametrised thus: 
\begin{equation}
\pi_+ = \frac{\lambda_+ +1}{\lambda_+ - \lambda_-},\quad\pi_- = -\frac{\lambda_- +1}{\lambda_+ - \lambda_-}, \qquad 
\mu=\frac{1-\qtwo}{\qtwo}\frac{\theta\yzero^2}{2},
\end{equation}
and $\lambda_{\pm}$ as earlier.
$\Box$
\end{prop}
(Recall that $\lambda_{-}<-1$, so $\pi_{\pm}$ are both positive.)

While this result is unhelpful in writing down a closed-form expression for the density, it provides very clear intuition about what it looks like, as follows. If we start the OU process near its equilibrium point and/or observe entropy production over a short time, then $\mu$ is small, so the main contribution to $\myprod{P}{Z}$ is a delta-function at the origin of strength $e^{-\mu}$, with exponential wings on either side. The effect of convolving with such a density is to move probability mass to the right \emph{without displacing the spike}. On the other hand if we start a long way from the origin and/or observe the entropy production over a longer time, then $\mu$ is larger, and $P$ is more likely to be high: so $\myprod{P}{Z}$ is approximated as a multiple convolution of iid double-exponential distributions, which by the Central Limit Theorem must be somewhat Gaussian in shape. So the distribution becomes more bulbous in the middle.

The extra \emph{mean} entropy production that arises from starting at $\yzero\ne0$  is $(q_1-q_2)\theta\yzero^{2}/2$, which is positive. The form of this is unsurprising, because it increases with $(t_2-t_1)$ but only if one has not waited a long time since inception, as otherwise reversion will have occurred and the starting-place become irrelevant: hence the form $(q_1-q_2)$. It is a simple matter to verify this expression, as it pertains to the $O(\lambda)$ term in the Maclaurin expansion of the mgf.

A minor but nonetheless interesting point is that the effect of starting away from equilibrium is neither to simply shift the whole distribution to the right (because the spike in the pdf does not move\footnote{As can be seen from letting $\lambda\to\I\infty$: to shift the spike over by an amount $c$ would require the mgf to oscillate as $e^{\I\lambda c}$ in that limit, but this has been ruled out as the $\exp(\cdot)$ term simply tends to a constant.}); nor does it alter the exponential decay-rates in the tails which, referring to (\ref{eq:k0distsh}), remain as $\lambda_\pm$.

\begin{figure}
\noindent
\begin{tabular}{rr}
\scalebox{0.625}{\includegraphics*{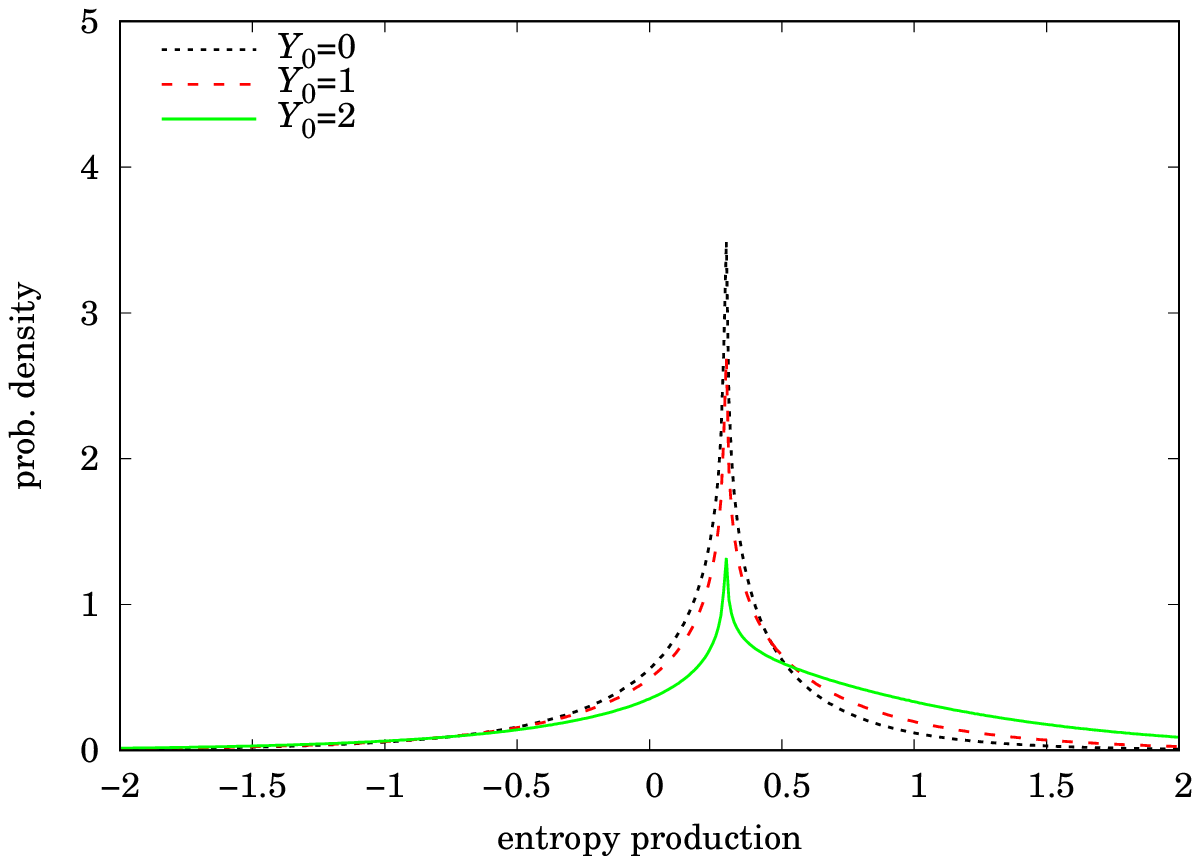}} &
\scalebox{0.625}{\includegraphics*{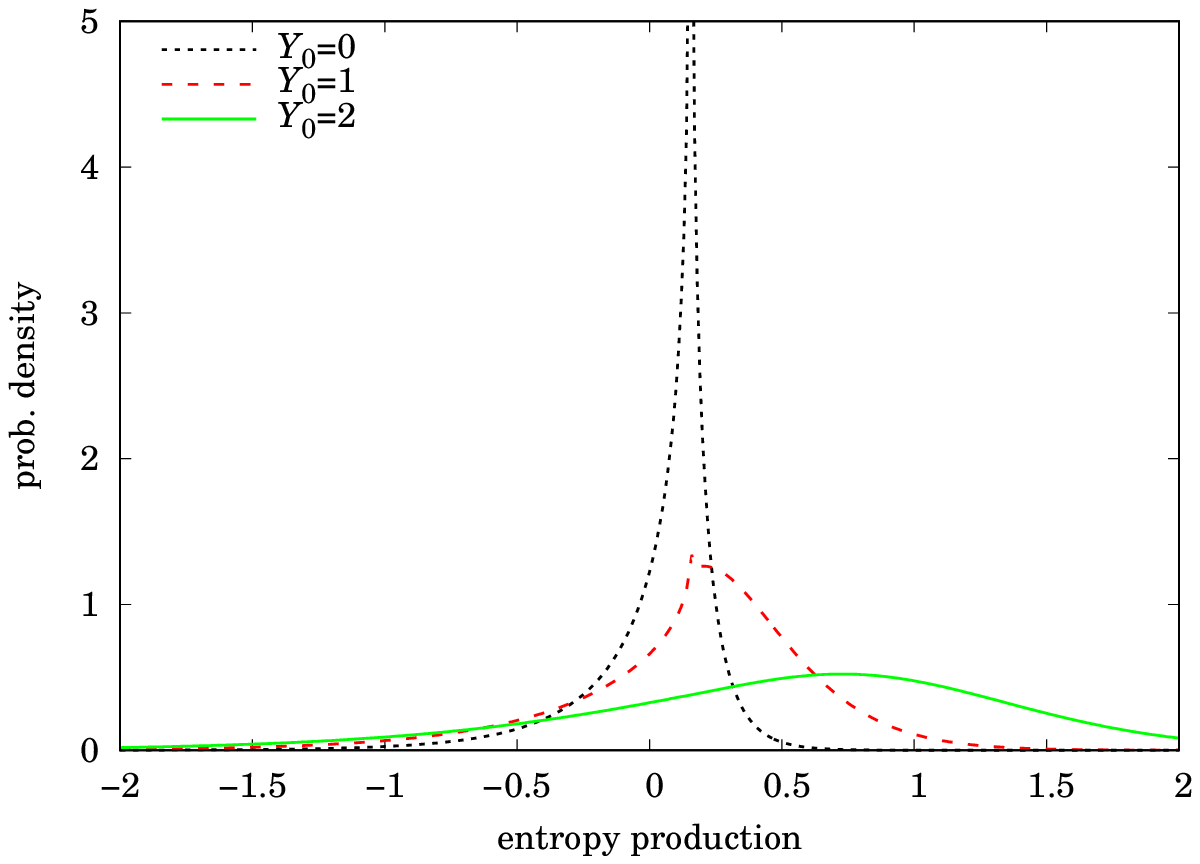}}
\end{tabular}
\caption{\small Probability density of entropy production for OU process ($\ffield(y)=-y$), starting at various points ($\yzero=0,1,2$). Time intervals: (left) $[0.125,0.25]$; (right) $[0.5,1]$.
}
\label{fig:ou1}
\end{figure}

The distribution of entropy production can be calculated in two ways. The first is to invert the mgf (Fourier integral) numerically using the FFT as discussed in \S\ref{sec:fft}.  
The second is to employ the stratified sampling method discussed earlier (\S\ref{sec:ss}), which needs only the transition density and not the mgf. As a check, both methods were used. 
Results are shown in Figure~\ref{fig:ou1} for a few cases.

\subsubsection{Limit of zero mean reversion}

When $\theta\ll 1$ we have
\[
M_{\Delta s}(\lambda) \sim \left(\frac{t_2}{t_1}\right)^{\lambda/2} \big( 1 - (1-t_1/t_2) \lambda^2 \big)^{-1/2}
\exp \Biggr( \frac{\lambda(\lambda+1)  \theta^2\yzero^2 \kappa(t_2-t_1)}{1 - (1-t_1/t_2) \lambda^2}\Biggr).
\]
To interpret this result, let us write $\sigma=\sqrt{2\kappa}$ for the volatility and also shift the process so that it starts from zero and has reversion level $y_\infty$. 
Then
\[
dY_t = -(\theta\sigma^2/2) (Y_t-y_\infty) \, dt + \sigma\, dW_t
\]
for which the mgf of the entropy production is
\[
M_{\Delta s}(\lambda) \sim \left(\frac{t_2}{t_1}\right)^{\lambda/2} \big( 1 - (1-t_1/t_2) \lambda^2 \big)^{-1/2}
\exp \Biggr( \frac{\lambda(\lambda+1) (\sigma^2  \theta^2 y_\infty^2/2) (t_2-t_1)}{1 - (1-t_1/t_2) \lambda^2}\Biggr).
\]
Now set $\theta y_\infty = 2\mu/\sigma^2$, with $\mu$ fixed (and representing the drift), and let $\theta \to 0$ with $y_\infty\to\pm\infty$ according as $\mu$ is positive or negative. 
Then we end up with the familiar arithmetic Brownian motion,
\begin{equation}
dY_t = \mu \, dt + \sigma\, dW_t,
\label{eq:abm}
\end{equation}
for which the mgf of the entropy production is  (cf.\ \cite[\S3.1]{Nicolis17})
\begin{equation}
M_{\Delta s}(\lambda) \sim \left(\frac{t_2}{t_1}\right)^{\lambda/2} \big( 1 - (1-t_1/t_2) \lambda^2 \big)^{-1/2}
\exp \Biggr( \frac{\lambda(\lambda+1) (2\mu^2/\sigma^2) (t_2-t_1)}{1 - (1-t_1/t_2) \lambda^2}\Biggr).
\end{equation}
The mean entropy production is
\[
\shalf \ln (t_2/t_1) + (2\mu^2/\sigma^2)(t_2-t_1).
\]
The first term represents, as usual, the broadening-out of the pdf. In the second, $2\mu^2/\sigma^2$ is identified as the rate of accretion of entropy resulting from the drift.
This is analogous to the dissipation of work as heat of a particle being moved through a viscous medium by the influence of a conservative field: for example, a charged particle, in an oil bath, by an electric field.

In the driftless case we have the simple result
\[
p(\Delta s) = \pi\inv b \, K_0(b|\Delta s-\sstar|)
\]
with
\[
\sstar = \shalf \ln (t_2/t_1); \qquad b= (1-t_1/t_2)^{-1/2}.
\]
The distribution is symmetrical about $\sstar$, the mean, and the variance is $1-t_1/t_2$.
As expected, for zero drift the result depends only on $t_1/t_2$.

All of the above can be derived directly from (\ref{eq:entdef},\ref{eq:abm}) using essentially the same techniques as used here. Indeed, writing $Z_t=X_t-\mu t$ we have
\[
\Delta s = \half \ln \frac{t_2}{t_1} + \frac{2\mu^2(t_2-t_1)}{\sigma^2} + \frac{Z_{t_2}^2}{2\sigma^2 t_2} - \frac{Z_{t_1}^2}{2\sigma^2 t_1} + \frac{2\mu(Z_{t_2}-Z_{t_1})}{\sigma^2}
\]
and the joint density of $(Z_{t_1},Z_{t_2})$, conditionally on starting from the origin at time zero (which we may assume without loss of generality), is
\[
\frac{\exp\big( -Z_{t_1}^2/2\sigma^2t_1 - (Z_{t_2}-Z_{t_1})^2/2\sigma^2(t_2-t_1)\big)}
{2\pi\sigma^2 \sqrt{t_1(t_2-t_1)}}.
\]
The mgf of the entropy production is then obtained by completing the square and doing a bivariate Gaussian integral.
It is worth noting, though, that the question of entropy production for the arithmetic Brownian motion is not materially simpler than for the OU model\footnote{But, as an aside (and a sort of `health warning'!), the same is \emph{not} true of matters pertaining to first-passage times. The introduction of mean reversion makes things very much more difficult \cite{Martin19a}.}.


\subsection{General potential}

\label{sec:gp}


When the transition density is not known---which is the general case---it has to be approximated, and the approach in \cite{Martin18b}, developed initially in \cite{Martin15b}, provides a framework for this.
In the interest of stating the main result upfront, the approximated transition density from $\yzero$ at time zero to $y$ at time $t$ is given by
\begin{equation}
\fy(t,y \cdl \yzero) \sim  \frac{(\theta/2\pi)^\frac{\scriptstyle \sqrt{q}}{\scriptstyle 1+\sqrt{q}}}{\sqrt{1-q}} 
\exp\left( 
\frac{-\half\theta \!\sqrt{q}(y-\yzero)^2}{1-q}    
\right)
\fy(\infty,y)^\frac{\scriptstyle 1}{\scriptstyle 1+\sqrt{q}} \fy(\infty,\yzero)^\frac{\scriptstyle -\sqrt{q}}{\scriptstyle 1+\sqrt{q}} 
\label{eq:fapprox}
\end{equation}
and also
\begin{equation}
\gy(t,y \cdl \yzero) \sim  \frac{1}{\sqrt{1-q}} 
\exp\left( 
\frac{-\half\theta \!\sqrt{q}(y-\yzero)^2}{1-q}    
\right)
\left(\frac{\theta/2\pi}{\fy(\infty,y) \fy(\infty,\yzero)} \right)^\frac{\scriptstyle \sqrt{q}}{\scriptstyle 1+\sqrt{q}}
\label{eq:gapprox}
\end{equation}
where as before
\[
q=e^{-2\theta\tau}
\]
and $\theta>0$ is now a constant that controls the average strength of mean reversion, as will be explained presently.
The first part of the expression (prefactor and Gaussian term) deal with the short-time behaviour, and the rest ensures that the long-time asymptote is correct. Furthermore, the result is exact for the OU model $\ffield(y)=\theta(y_\infty-y)$ regardless of the reversion level $y_\infty$.

We now give some further details. 
The reader who wishes only to use (\ref{eq:fapprox},\ref{eq:gapprox}) may do so without reading the rest of this section, except for noting the definition of $\theta$ in (\ref{eq:theta}).

As we are concerned with a product representation of the transition pdf, and as entropy relates directly to the logarithm of the probability density, it makes sense to deal not with $\fy$ directly but instead with its logarithmic derivative. Recall that $\tau=\kappa t$. Then, writing
\[
\hy=-(\partial/\partial y)\ln \gy,
\]
we have 
\begin{equation}
\pderiv{h_{Y}}{\tau}=\pderiv{}{y}\left\{ A(y)h_{Y}+\pderiv{h_{Y}}{y}-h_{Y}^{2}\right\} .
\label{eq:pde_h}
\end{equation}
Now this equation appears to be harder than (\ref{eq:pde_f}) because it is nonlinear and also has a singular initial condition, in the sense that $\hy \sim (y-\yzero)/2\tau$ as $\tau\to0$, which can be seen by dominant balance in (\ref{eq:pde_h}). However, it turns out that $\hy$ is easier to approximate than $\fy$.

Consider the class of OU models $\ffield(y)=\theta(y_\infty-y)$, with the proportionality constant $\theta$ controlling the strength of mean reversion and the constant $y_\infty$ denoting the long-term mean---alternatively the stationary state is Normal with mean $y_\infty$ and variance $1/\theta$. For this, we have exactly
\begin{equation}
\mbox{(General OU)}\qquad \hy(\tau,y) = \frac{\theta \!\sqrt{q} (y-\yzero)}{1-q} + \frac{\theta\!\sqrt{q}(y_\infty-y)}{1+\!\sqrt{q}} 
\label{eq:h_OU}
\end{equation}
as is easily verified by substituting it into (\ref{eq:pde_h}), or writing it down directly from the known OU solution. Clearly $\hy$ is just a linear function of $y$. The first term is singular as $\tau\to0$, and captures the initial Gaussian behaviour as the process spreads out from its point source. The second term refers to mean reversion, for it vanishes when the process is at its equilibrium level ($y=y_\infty$).
Note also that $\hy\to0$ as $\tau\to\infty$, as it must, because we require $\gy\to1$.

This inspires the approximation for the \emph{general} case:
\begin{equation}
\hy(\tau,y) = \frac{\theta \!\sqrt{q} (y-\yzero)}{1-q} + \frac{\sqrt{q}}{1+\!\sqrt{q}} \ffield(y) + \sqrt{q}\,o(1)_{q\to1} ,
\label{eq:hynew}
\end{equation}
where now $\theta$ is understood as an \emph{arbitrary} parameter. However, we must now explain what $\theta$ corresponds to in this general case.

When (\ref{eq:hynew}) is inserted into (\ref{eq:pde_h}), and a Laurent expansion performed around $\tau=0$, the LHS and RHS agree at $O(\tau^{-2})$ and $O(\tau^{-1})$, explaining why we are writing the error term in (\ref{eq:hynew}) as $o(1)$ in the short-time limit.
In so doing, we find
\[
\hy(\tau,y) = \frac{y-\yzero}{2\tau} + \frac{\ffield(y)}{2} + o(1), \qquad \tau\to0,
\]
and we observe that $\theta$ is absent from both the first two terms. So all $\theta$'s are equally good from this point of view and we cannot say anything about $\theta$ simply by looking at the first two terms in the short-time expansion. As $\theta$ does not affect the long-time asymptote, it must therefore control the intermediate-time behaviour. 

It turns out that the next-order term (i.e.\ expanding the $o(1)$ term in (\ref{eq:hynew}) in powers of $1-q$) is a rather complicated expression involving $\frac{d}{dy} \big(\ffield(y)+\theta y\big)$, which is unsurprising as if that quantity vanished identically then we would be back to the OU model, for which (\ref{eq:hynew}) is exact. Given that we are going to truncate the series before this term, it makes sense to minimise it, and so we want to choose $\theta$ so as to make $\ffield'(y)+\theta$ as close as possible to zero `on average'. This motivates the choice
\begin{equation}
\theta = \langle -\ffield' \rangle_\infty = \langle \ffield^2 \rangle_\infty 
\label{eq:theta}
\end{equation}
where $\langle \cdot\rangle_\infty$ denotes an average over the stationary distribution $f_Y(\infty,\cdot)$;  this relates to the force field via $\ffield(y) = \frac{d}{dy} \ln \fy(\infty,y)$.
As is apparent from the RHS of this expression (which follows from integration by parts), this choice of $\theta$ is positive, which is necessary for (\ref{eq:fapprox},\ref{eq:gapprox}) to be valid\footnote{Though as we have already seen we can take $\ffield(y)=\theta(y_\infty-y)$ and allow $\theta\to0$ with $\theta y_\infty=\mu$ fixed to obtain the arithmetic Brownian motion. The invariant density is formally $\ffield(y)\propto e^{\mu y}$ which is non-normalisable, but (\ref{eq:fapprox}) nevertheless gives the correct transition density. One cannot, however, permit $\theta<0$.}.  In \cite{Martin18b} it is pointed out that (\ref{eq:theta}) relates directly to the Fisher information for the estimation of the long-term mean of a mean-reverting process\footnote{
Although in principle we could average $-\ffield'$ over some other distribution---perhaps varying over time and/or space---this can present its own difficulties, as it is essential that $\theta$ be positive, and it also needs to be symmetric in $y$ and $\yzero$ so as to preserve the reciprocity condition (\ref{eq:revers}).}.

Now that we have approximated $\hy$, we simply integrate w.r.t.~$y$ and identify the implied constant of integration from the fact that the initial condition is a delta-function of unit strength.  The derivation can be simplified by recalling the reciprocity condition, namely that $\gy$ must be symmetric in $y$ and $\yzero$. This gives (\ref{eq:gapprox}) and thence (\ref{eq:fapprox}); full details can be found in \cite{Martin18b}.

\subsection{Examples}

\label{sec:ex1}

\begin{figure}
\noindent
\begin{tabular}{rr}
\scalebox{0.625}{\includegraphics*{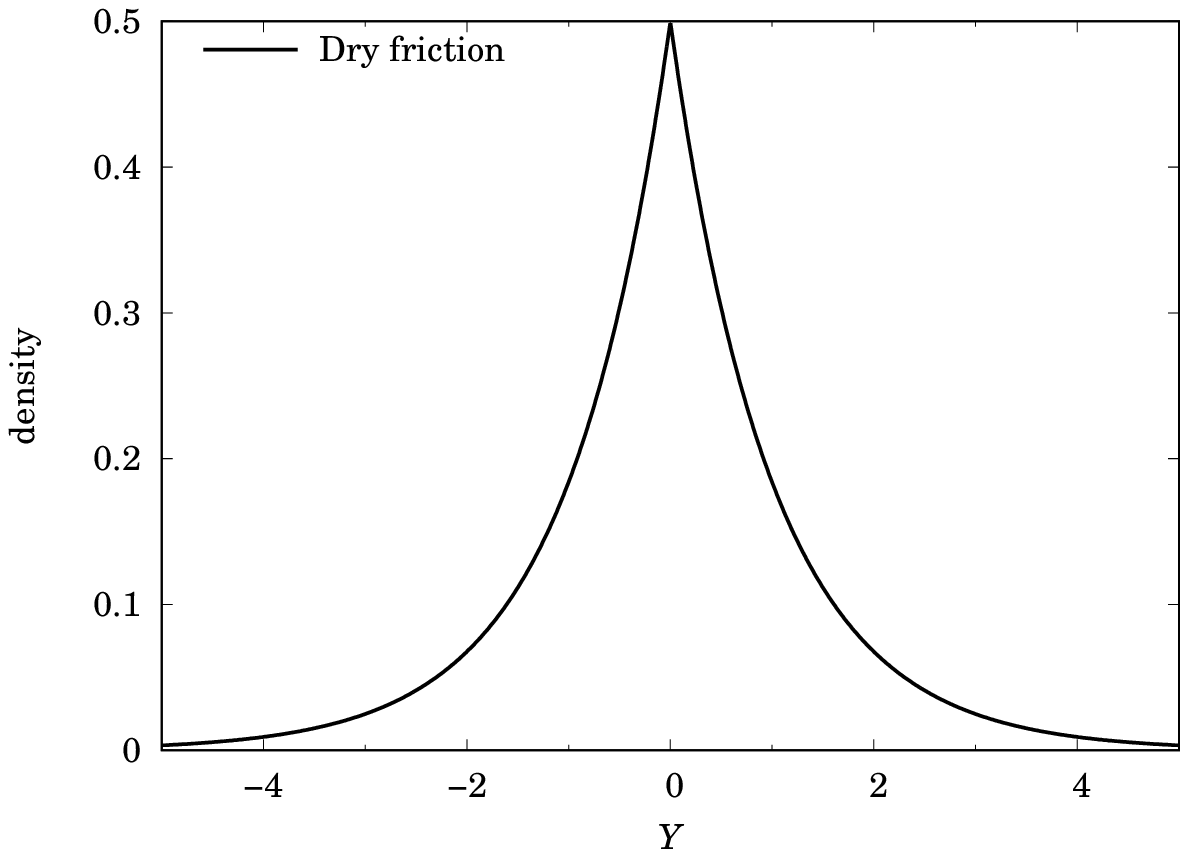}} 
&
\scalebox{0.625}{\includegraphics*{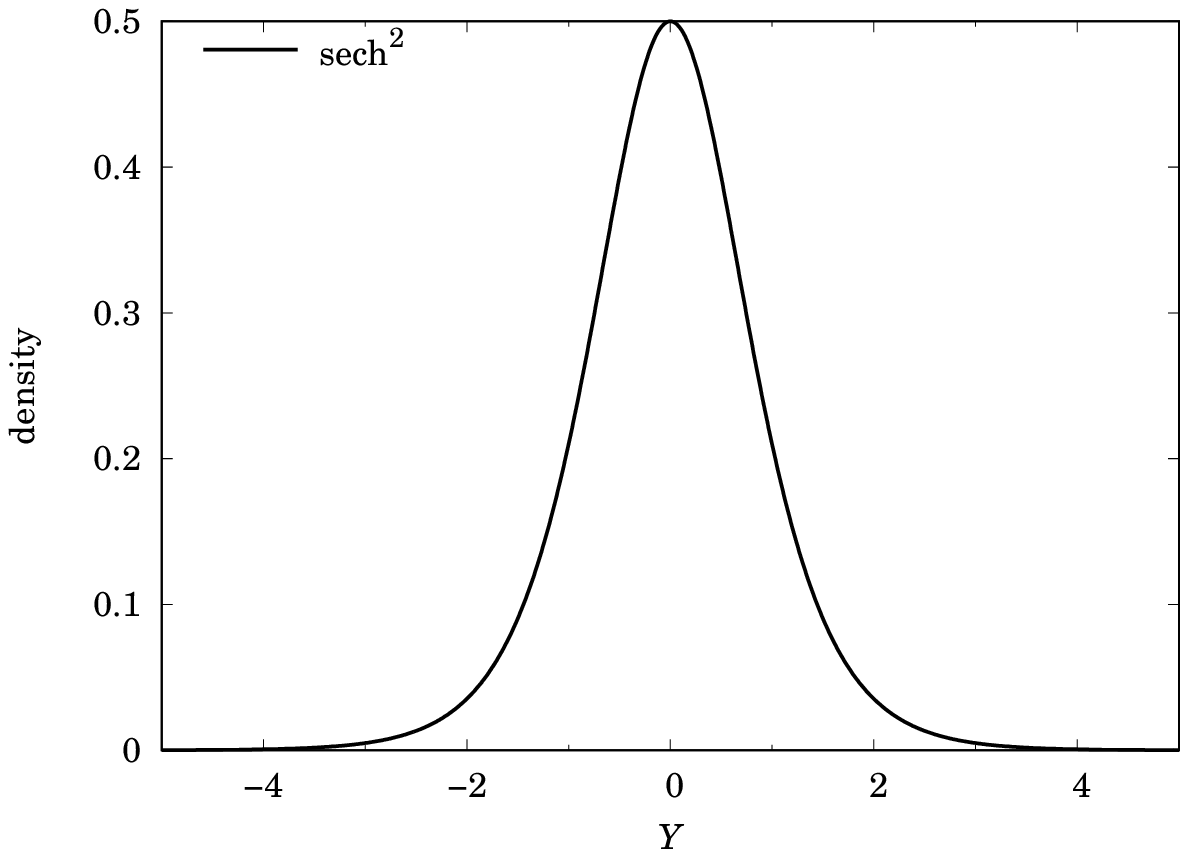}}
\\
\scalebox{0.625}{\includegraphics*{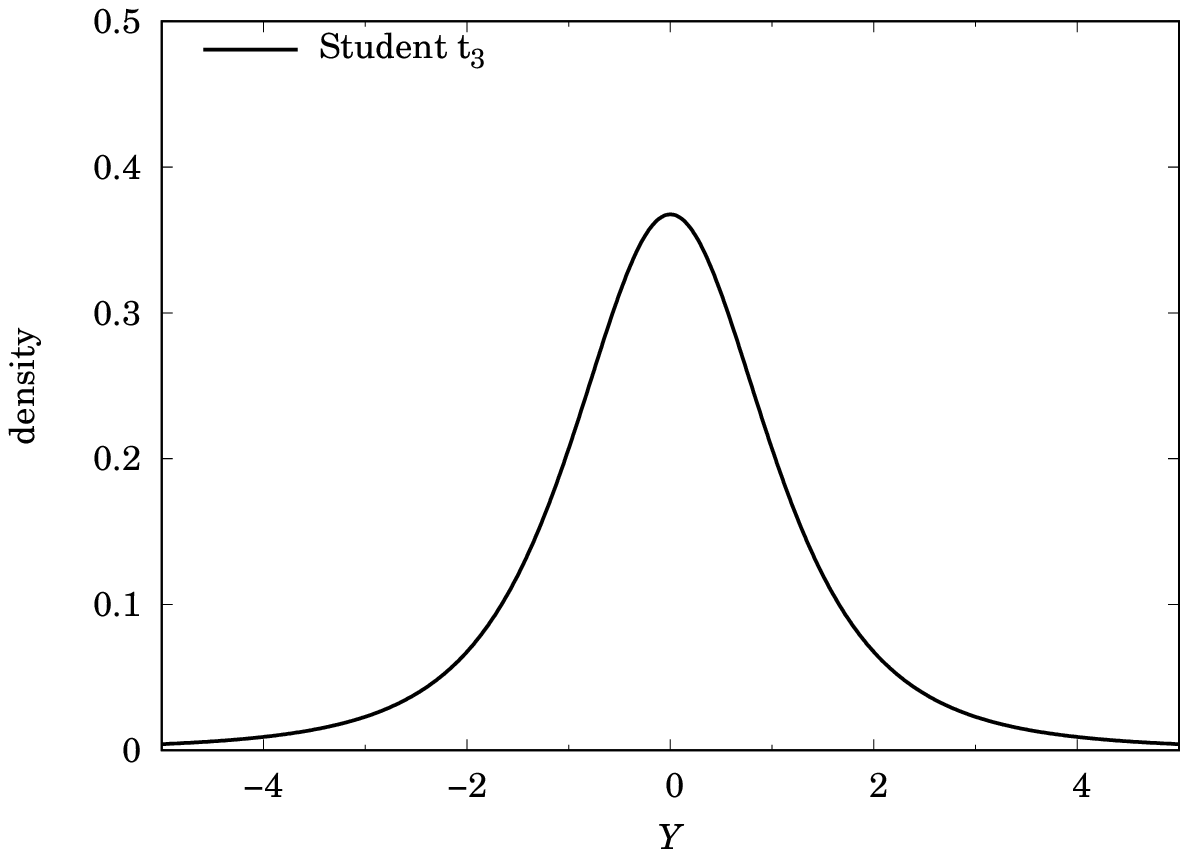}}
&
\scalebox{0.625}{\includegraphics*{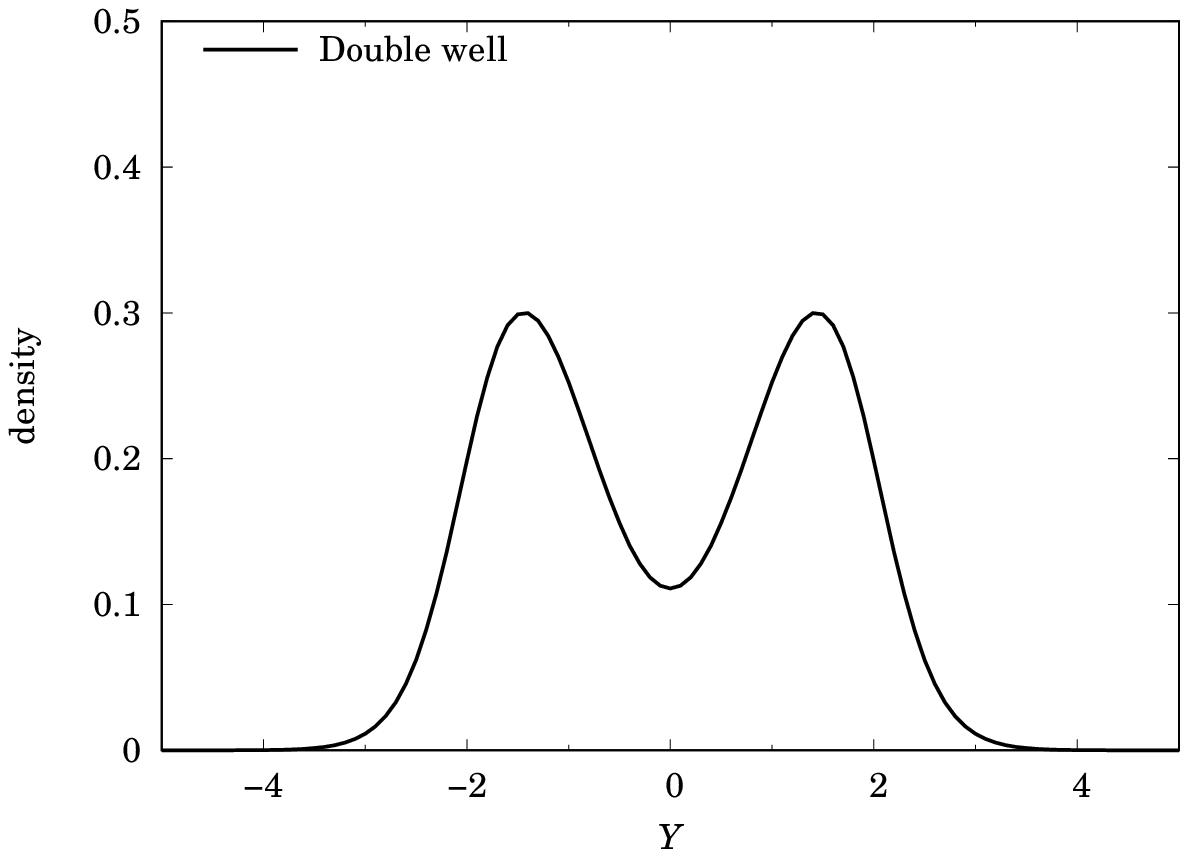}}
\end{tabular}
\caption{\small Stationary probability densities for the various models considered in this paper (see text for explanation): dry friction, $\sech^2$, Student $\mathrm{t}_3$, double well.
}
\label{fig:various}
\end{figure}

\label{sec:other}

In this section we present a variety of results for models differing from the OU process: see Figure~\ref{fig:various}.  These models were considered in \cite{Martin18b} and the approximate transition density was found to correspond well with the exact (as calculated numerically where necessary).
For the reader's convenience, the diagrams are assembled at the end of this section.

\subsubsection{Dry-friction}

\begin{figure}
\noindent
\begin{tabular}{rr}
(a)\scalebox{0.625}{\includegraphics*{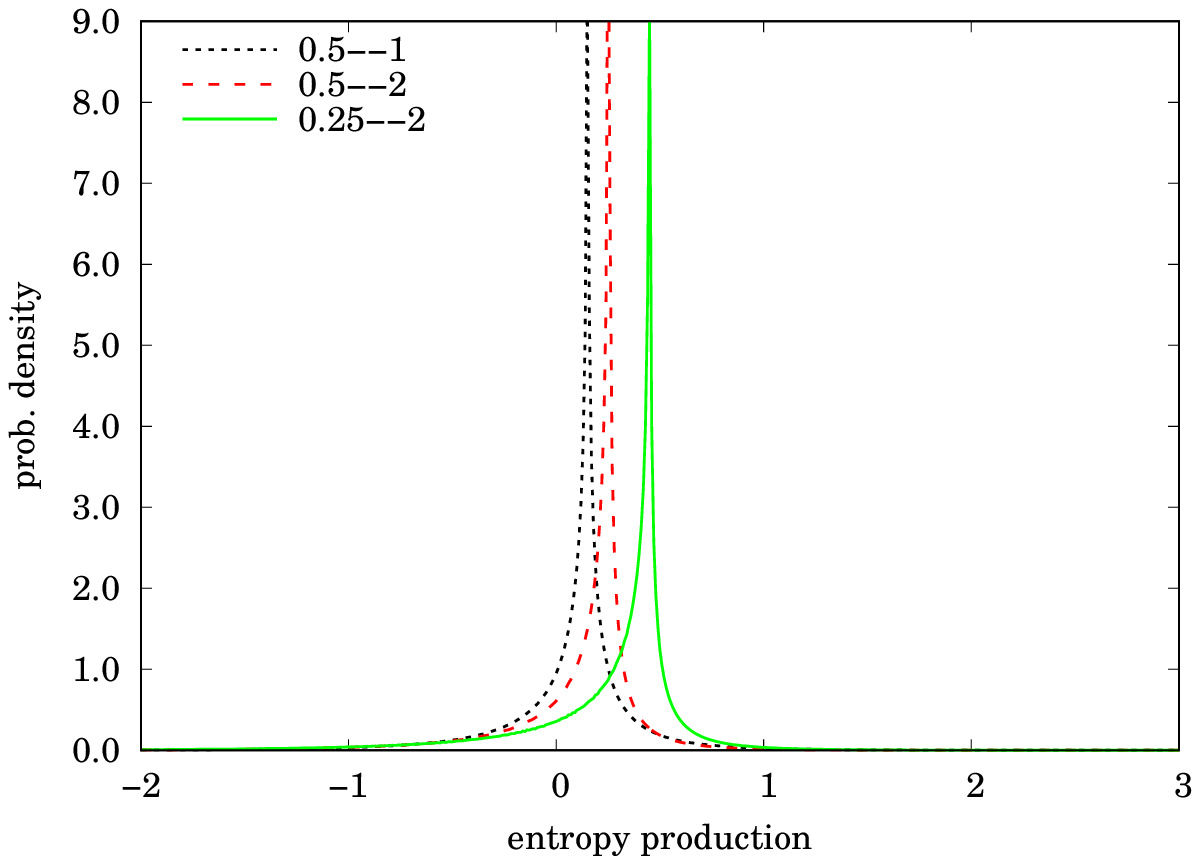}} &
(b)\scalebox{0.625}{\includegraphics*{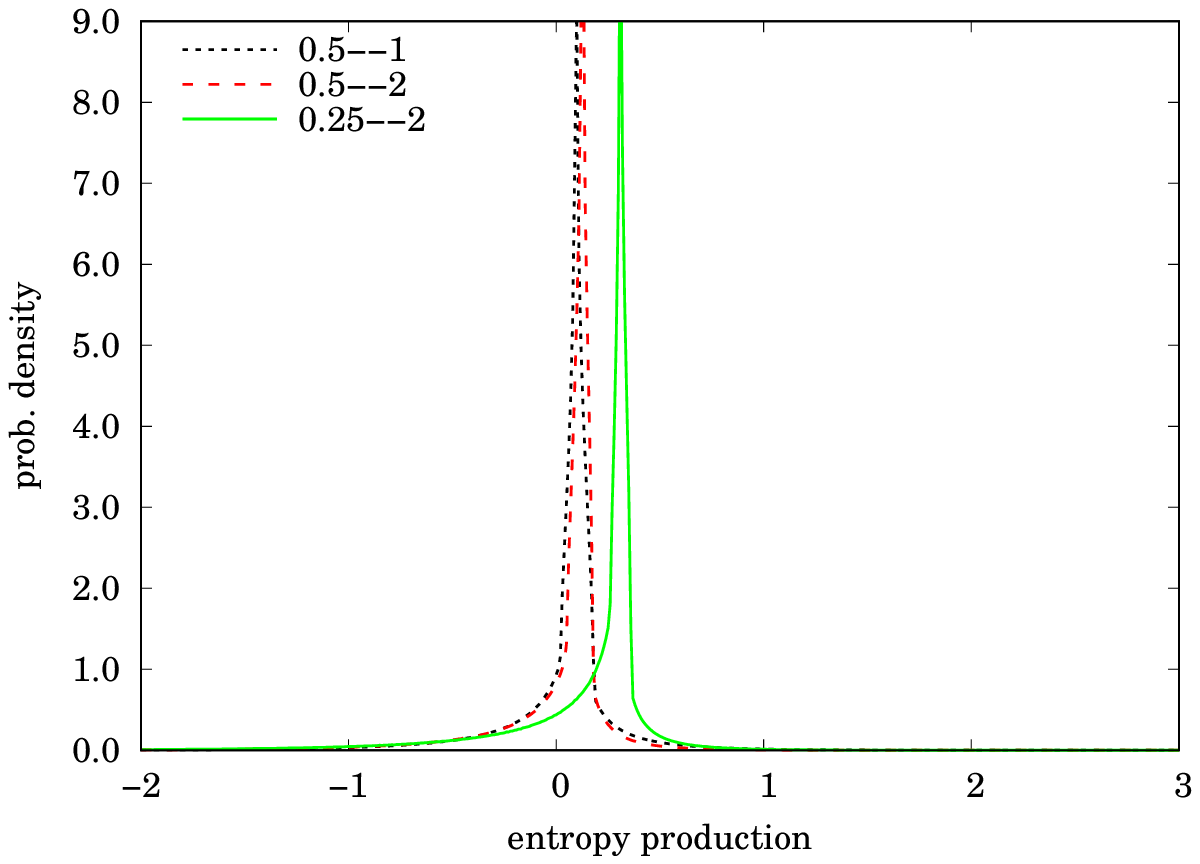}} \\
(c)\scalebox{0.625}{\includegraphics*{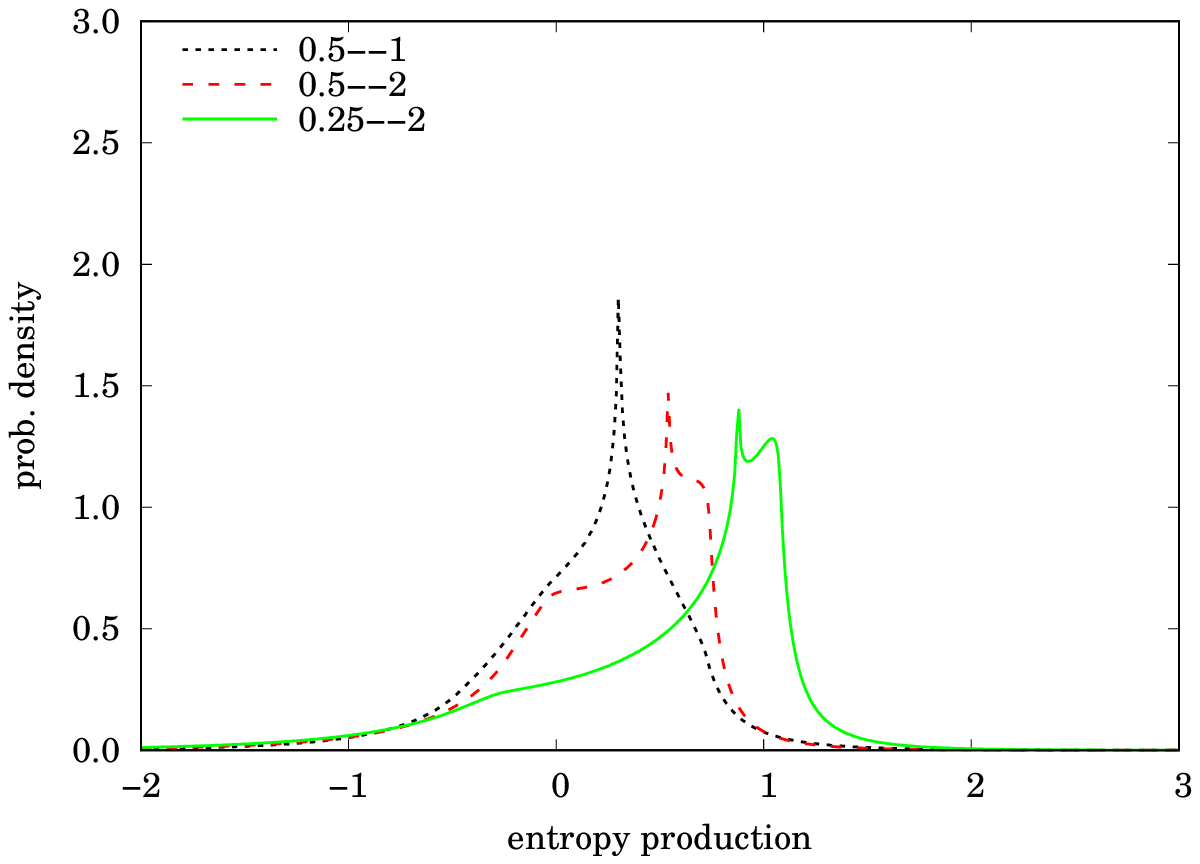}} &
(d)\scalebox{0.625}{\includegraphics*{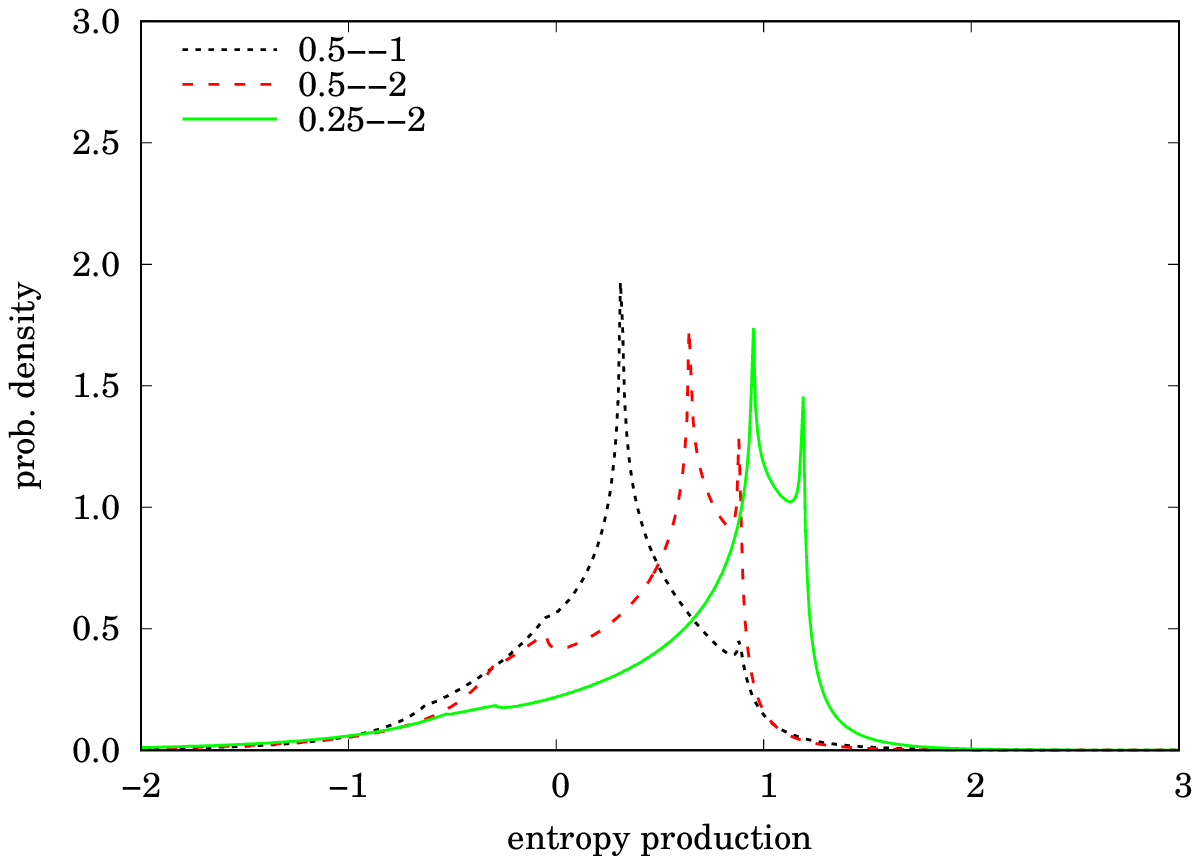}} \\
(e)\scalebox{0.625}{\includegraphics*{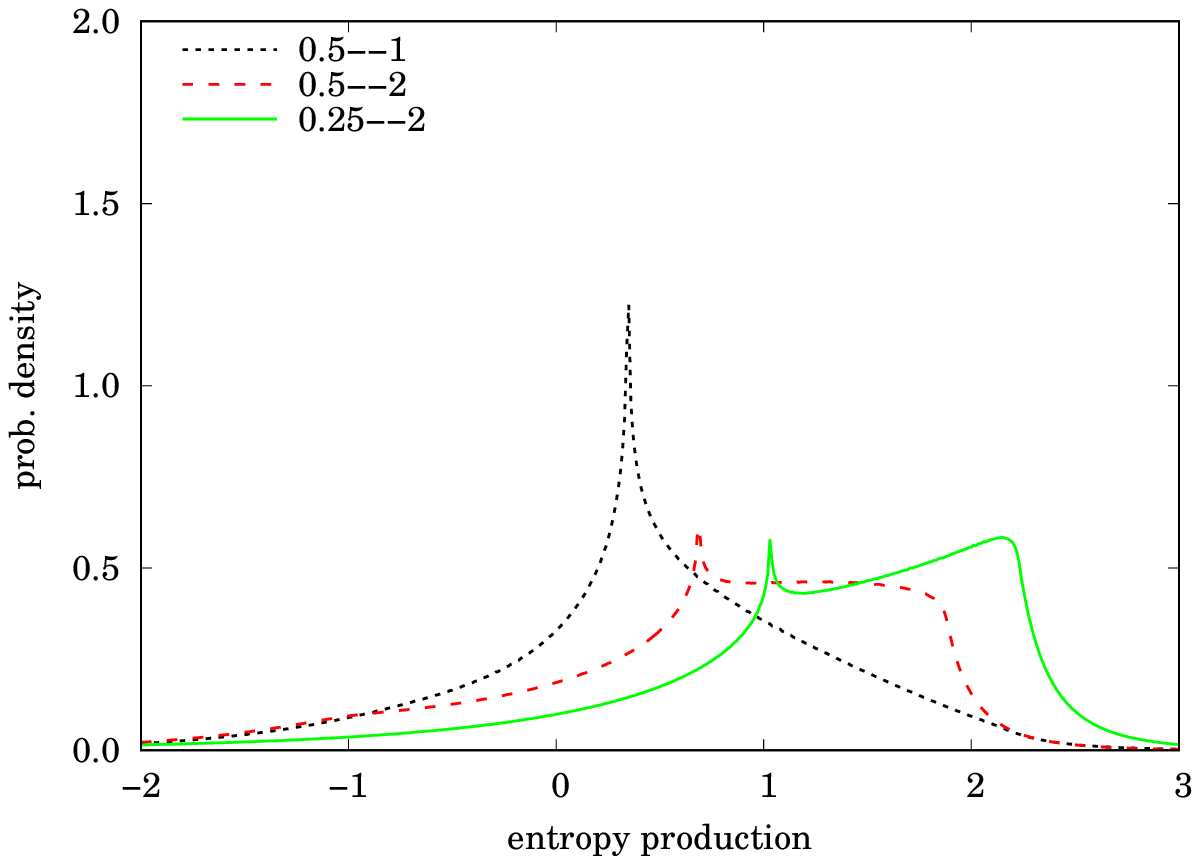}} & 
(f)\scalebox{0.625}{\includegraphics*{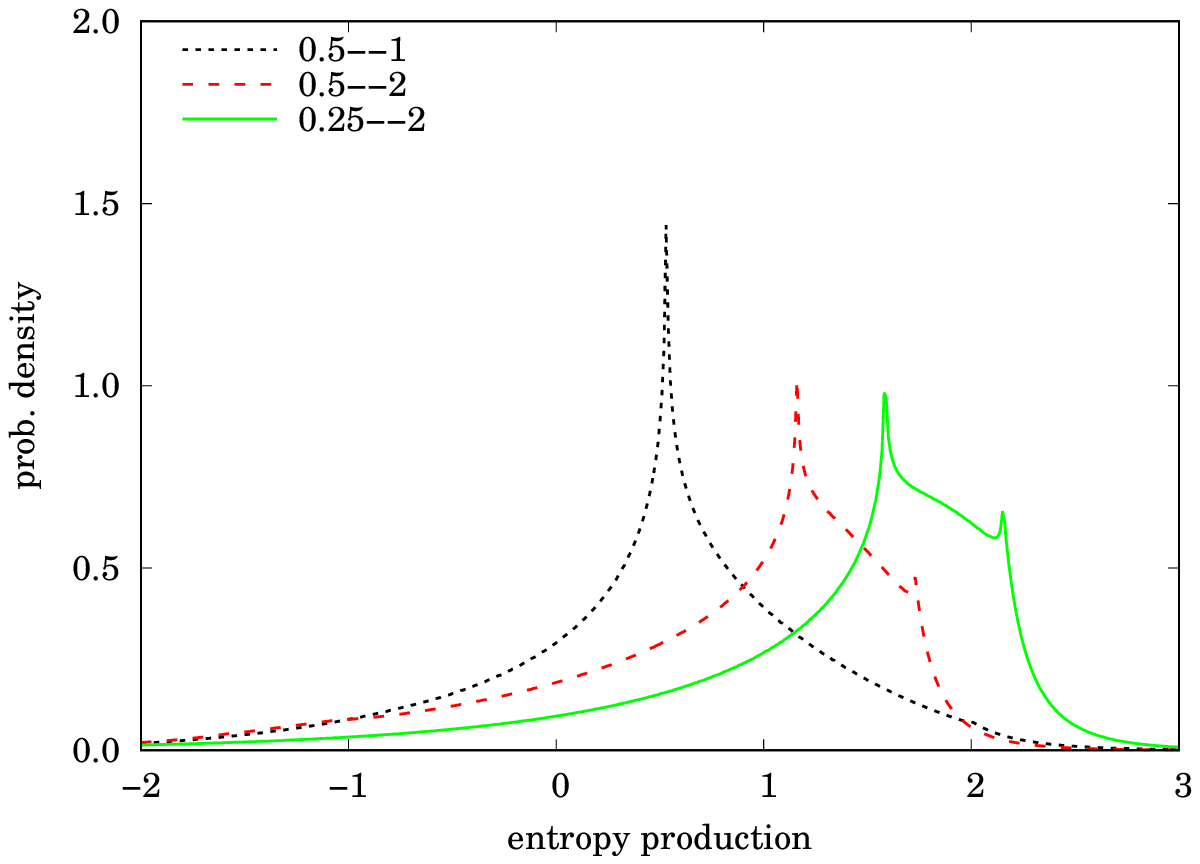}} 
\end{tabular}
\caption{\small Dry friction: exact and approximation compared, over different time intervals as indicated on the plots, and starting from different $\yzero$. Starting-points: (a,b) $\yzero=0$, (c,d) $\yzero=1$, (e,f) $\yzero=2$. In (a,c,e) the exact transition density is used and in (b,d,f) the approximations (\ref{eq:fapprox},\ref{eq:gapprox}).
}
\label{fig:dryfric}
\end{figure}

\begin{figure}
\noindent
\begin{tabular}{rr}
(a)\scalebox{0.625}{\includegraphics*{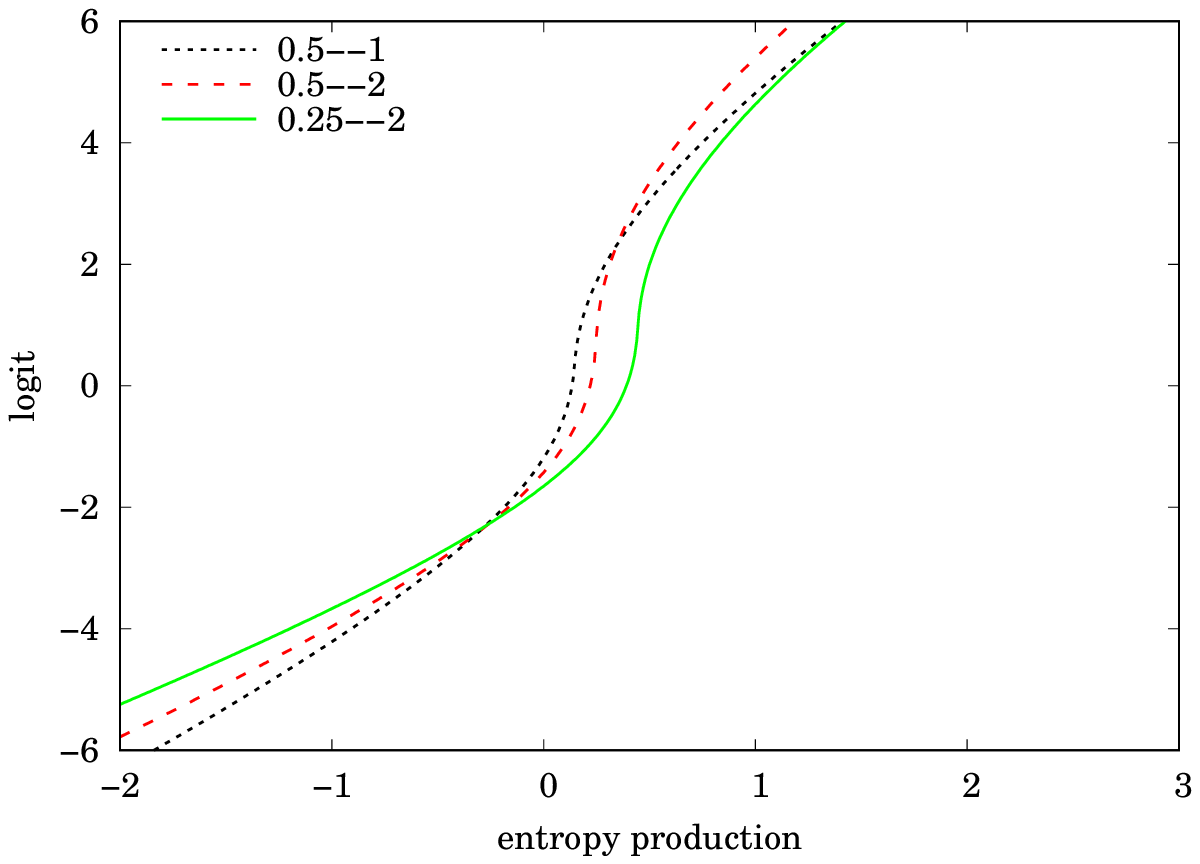}} &
(b)\scalebox{0.625}{\includegraphics*{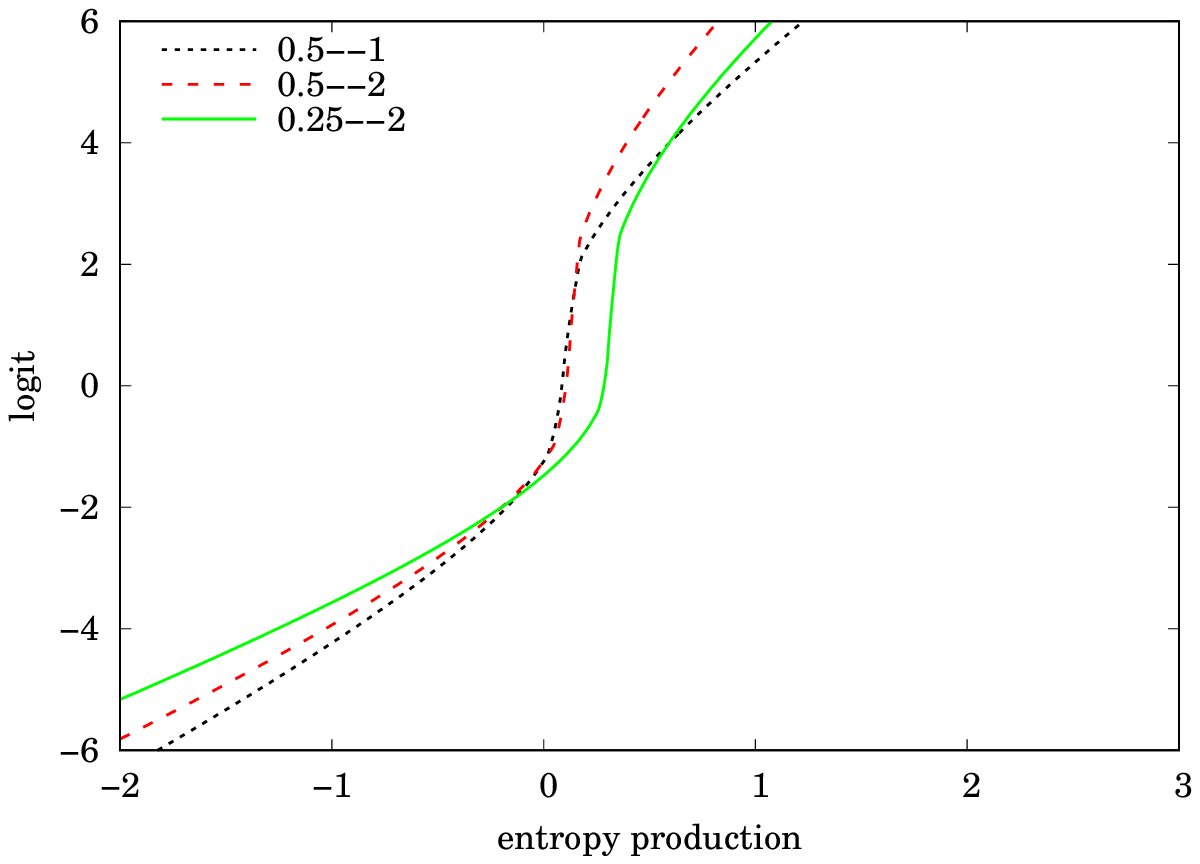}} \\
(c)\scalebox{0.625}{\includegraphics*{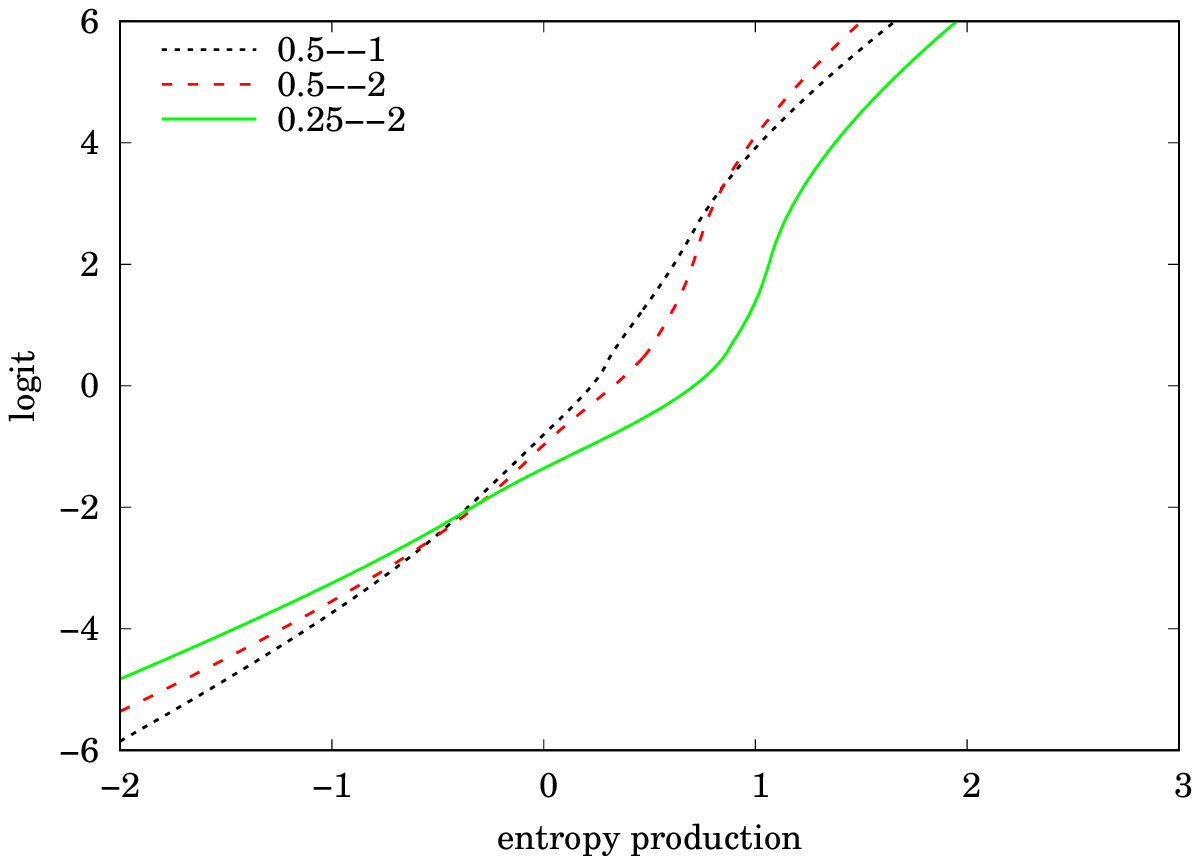}} &
(d)\scalebox{0.625}{\includegraphics*{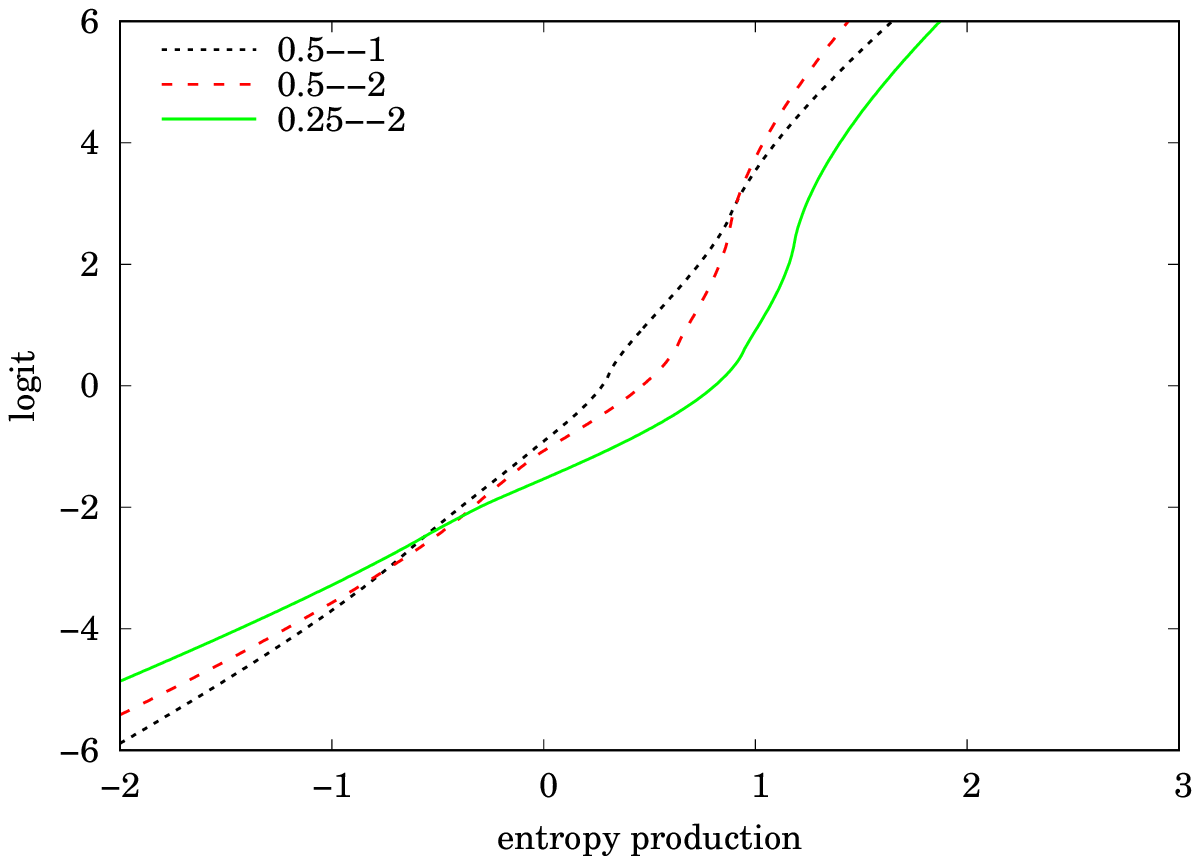}} \\
(e)\scalebox{0.625}{\includegraphics*{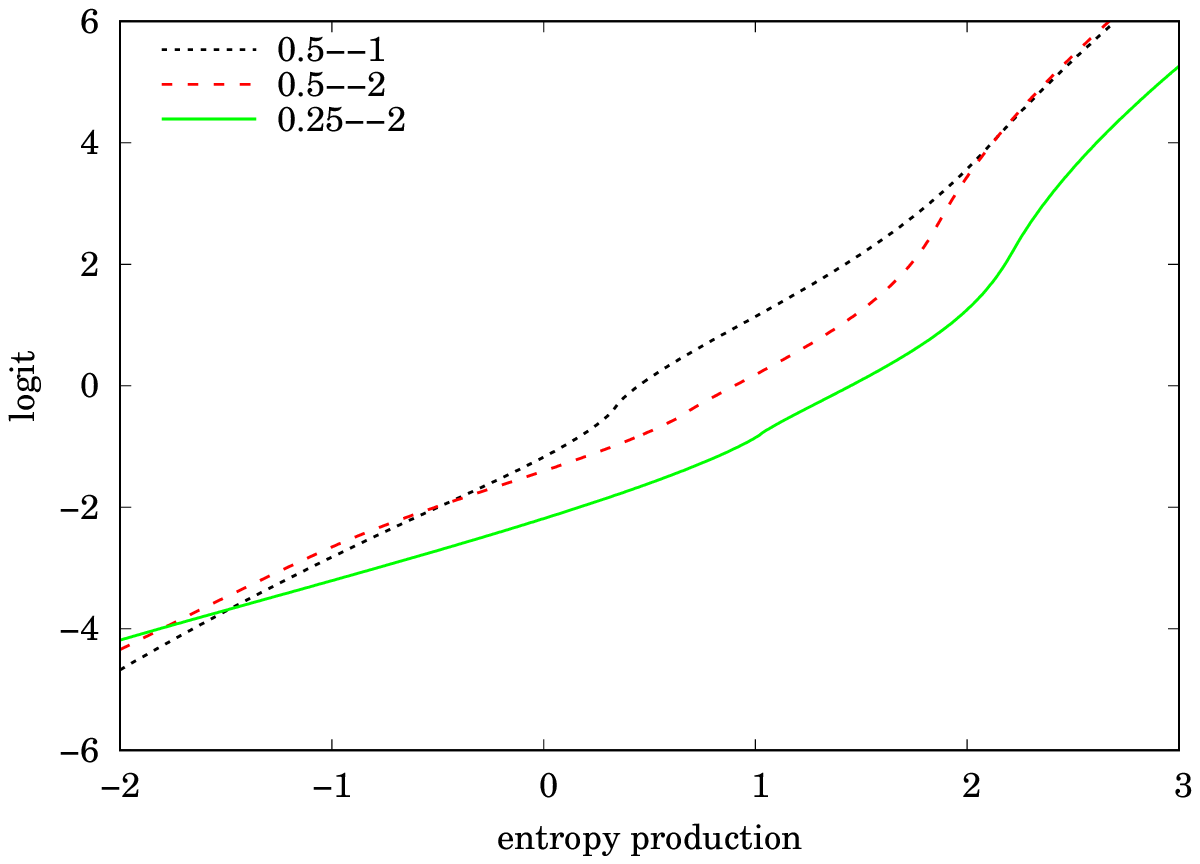}} & 
(f)\scalebox{0.625}{\includegraphics*{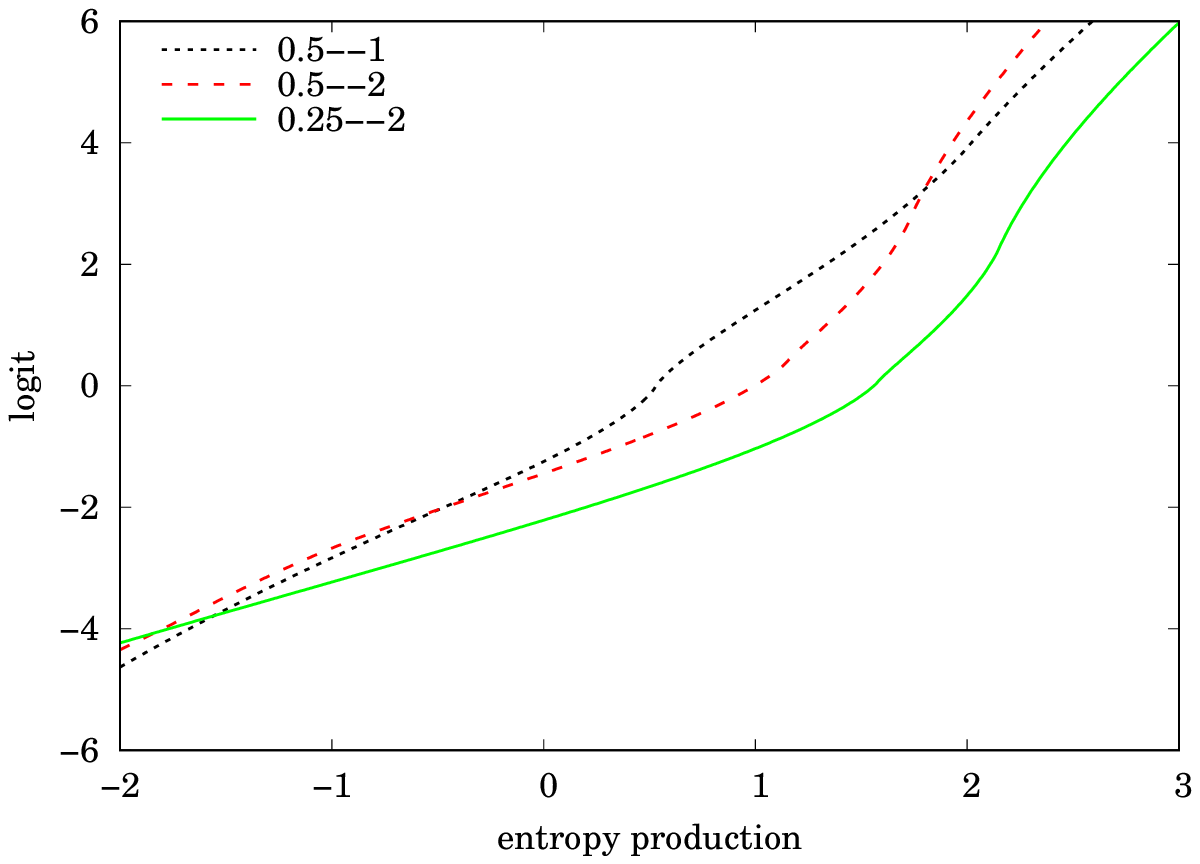}} 
\end{tabular}
\caption{\small As Figure~\ref{fig:dryfric} but on logit scale.
}
\label{fig:dryfric2}
\end{figure}

In the dry-friction model,
\[
A(y)=-\mathrm{sgn}\,y,\qquad \fy(\infty,y)=\frac{e^{-|y|}}{2},\qquad\langle-A'\rangle_{\infty}=1.
\]
Conveniently the transition density can be obtained in closed form \cite{Touchette10a}, e.g.~by the usual route of Laplace transforming the Fokker--Planck equation: 
\[
\fy(t,y\cdl\yzero)=\frac{e^{-(y-\yzero)^{2}/4\tau}}{\sqrt{4\pi\tau}}e^{-\tau/4}e^{(|\yzero|-|y|)/2}+\frac{e^{-|y|}}{2}\Phi\left(\frac{\tau-|y|-|\yzero|}{\sqrt{2\tau}}\right)
\]
with $\Phi$ denoting the cdf of the standard Normal distribution.
The entropy distribution still has to be obtained numerically, as per \S\ref{sec:ss}, but we can compare the results using the exact transition density with those using the approximated transition density (\ref{eq:fapprox}). These are shown in Figures~\ref{fig:dryfric},\ref{fig:dryfric2}, for different time periods and starting-points.
The agreement is particularly good when starting near the equilibrium point; it is less so when starting away from it, but the difference is smaller when viewed on the logit scale (Figure~\ref{fig:dryfric2}(e,f)) than when viewed as a pdf (Figure~\ref{fig:dryfric}(e,f)).

\subsubsection{Stationary state sech-power}

\begin{figure}
\noindent
\begin{tabular}{rr}
(a)\scalebox{0.625}{\includegraphics*{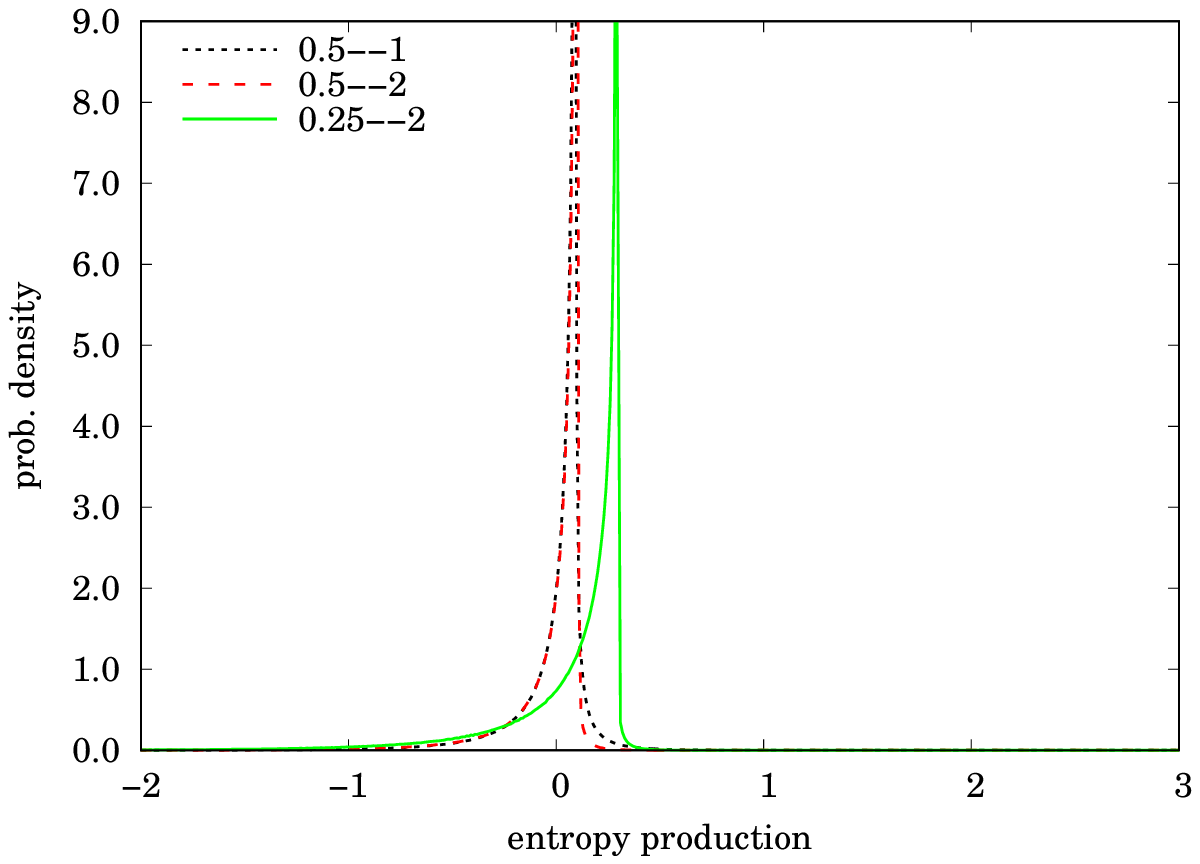}} &
(b)\scalebox{0.625}{\includegraphics*{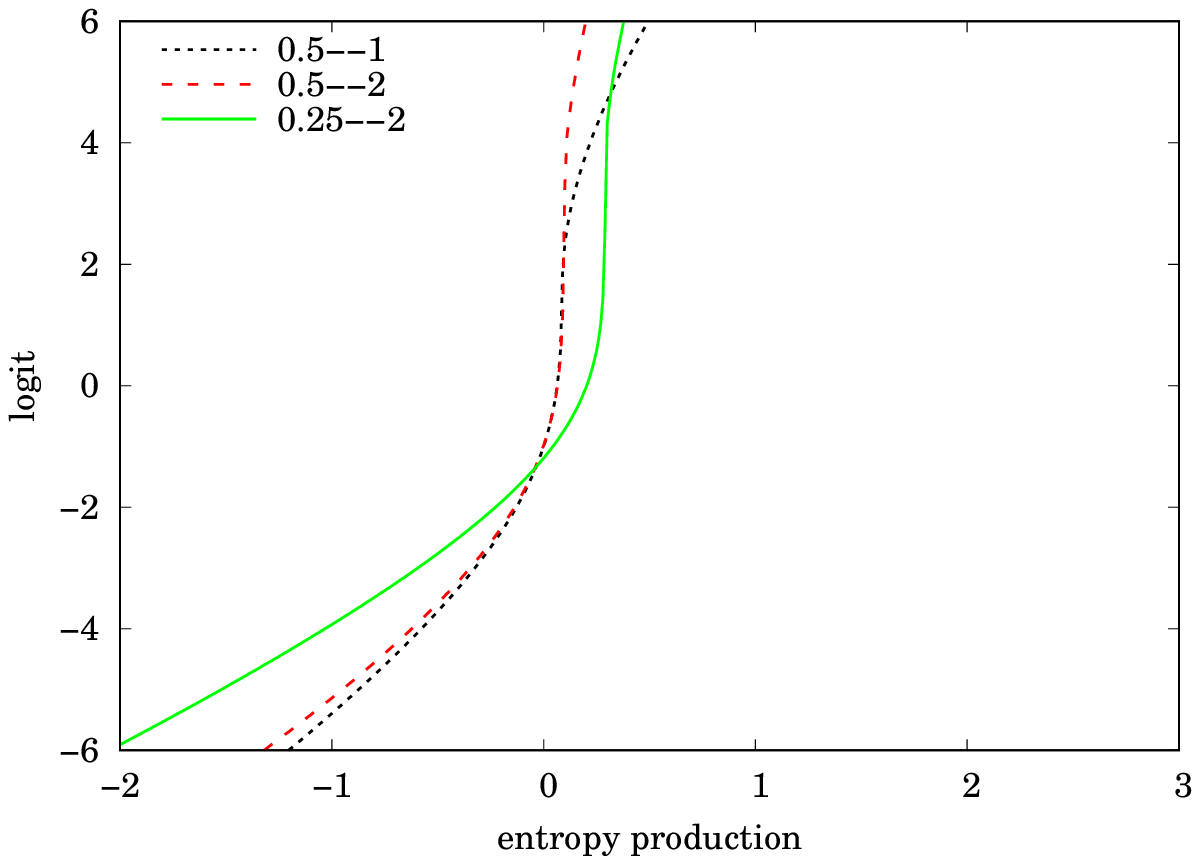}} \\
(c)\scalebox{0.625}{\includegraphics*{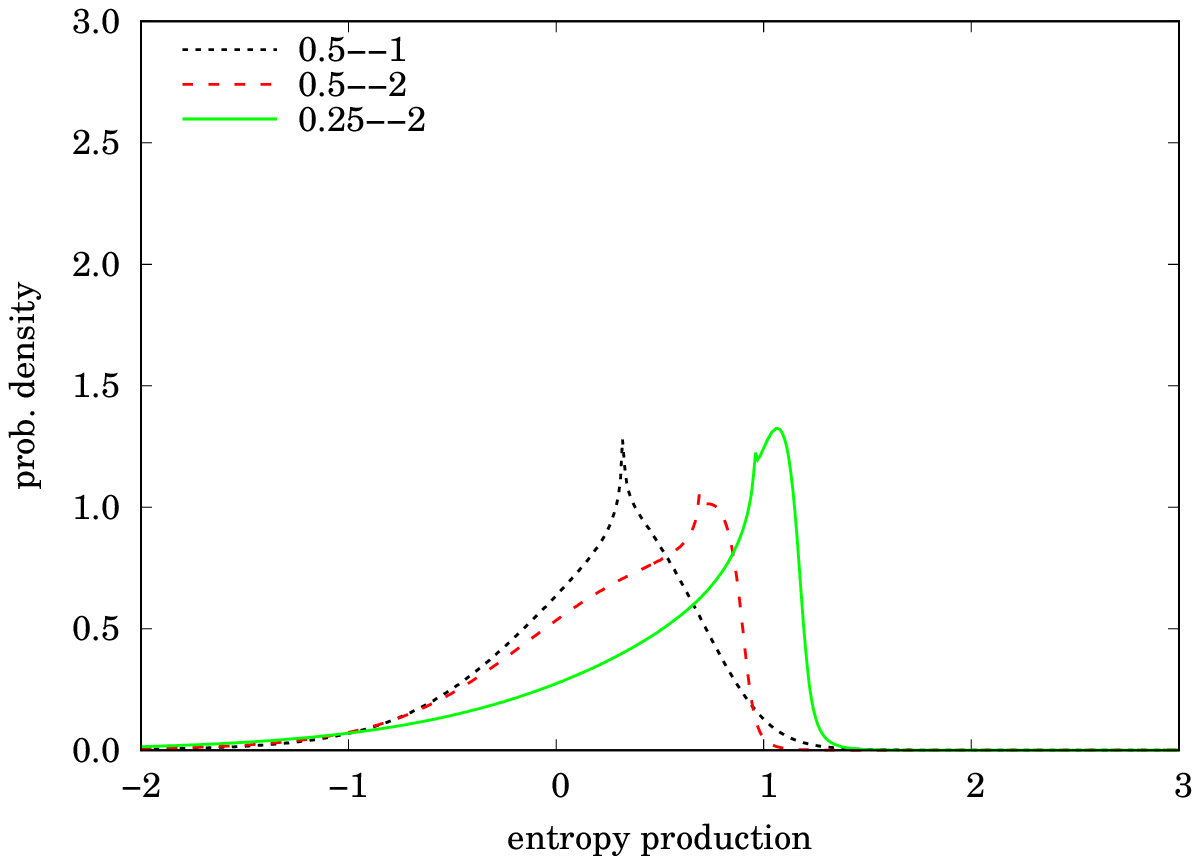}} &
(d)\scalebox{0.625}{\includegraphics*{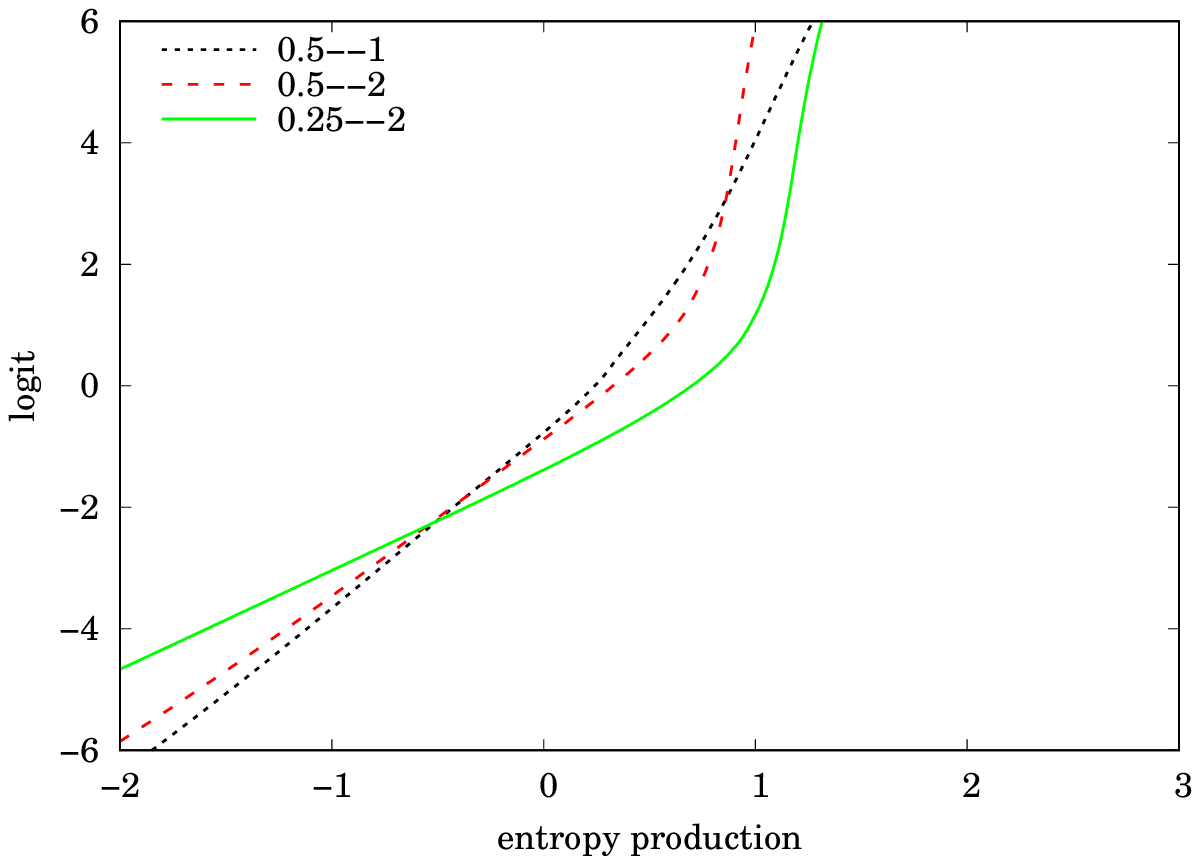}} \\
(e)\scalebox{0.625}{\includegraphics*{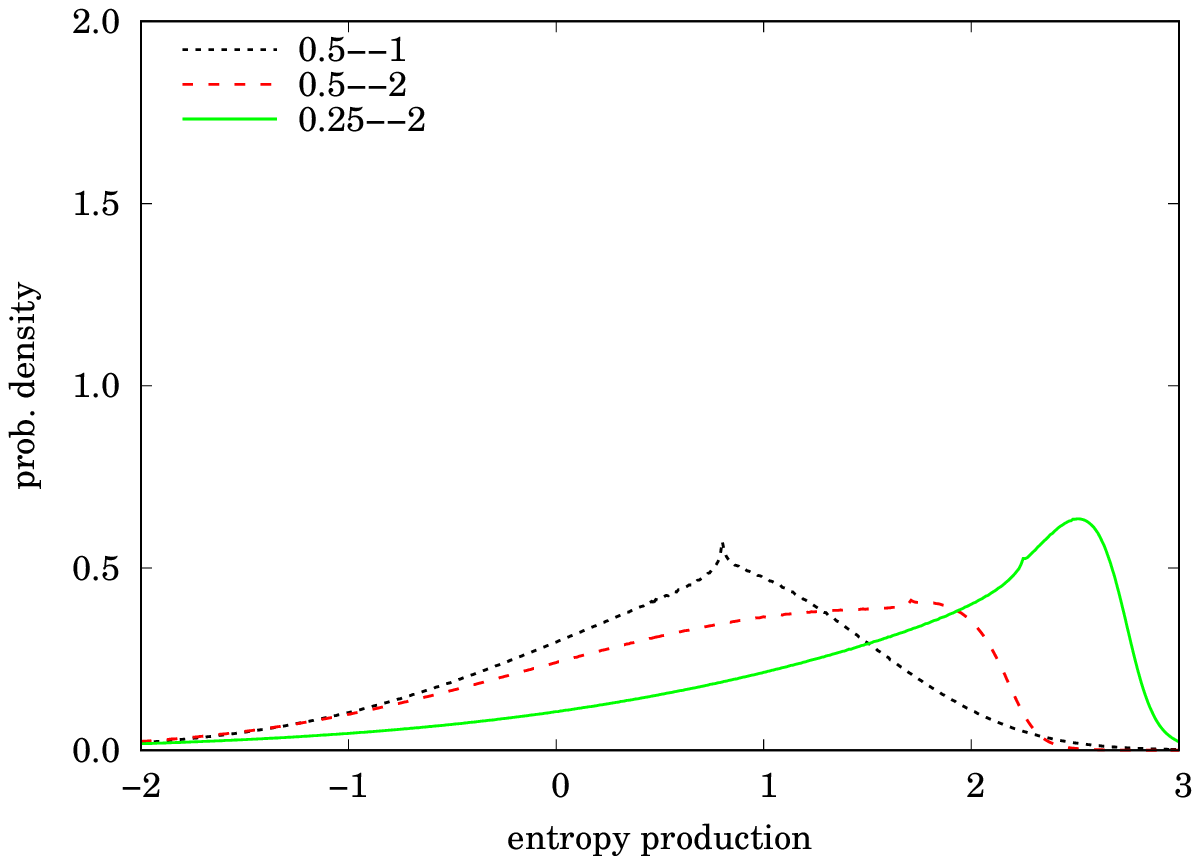}} &
(f)\scalebox{0.625}{\includegraphics*{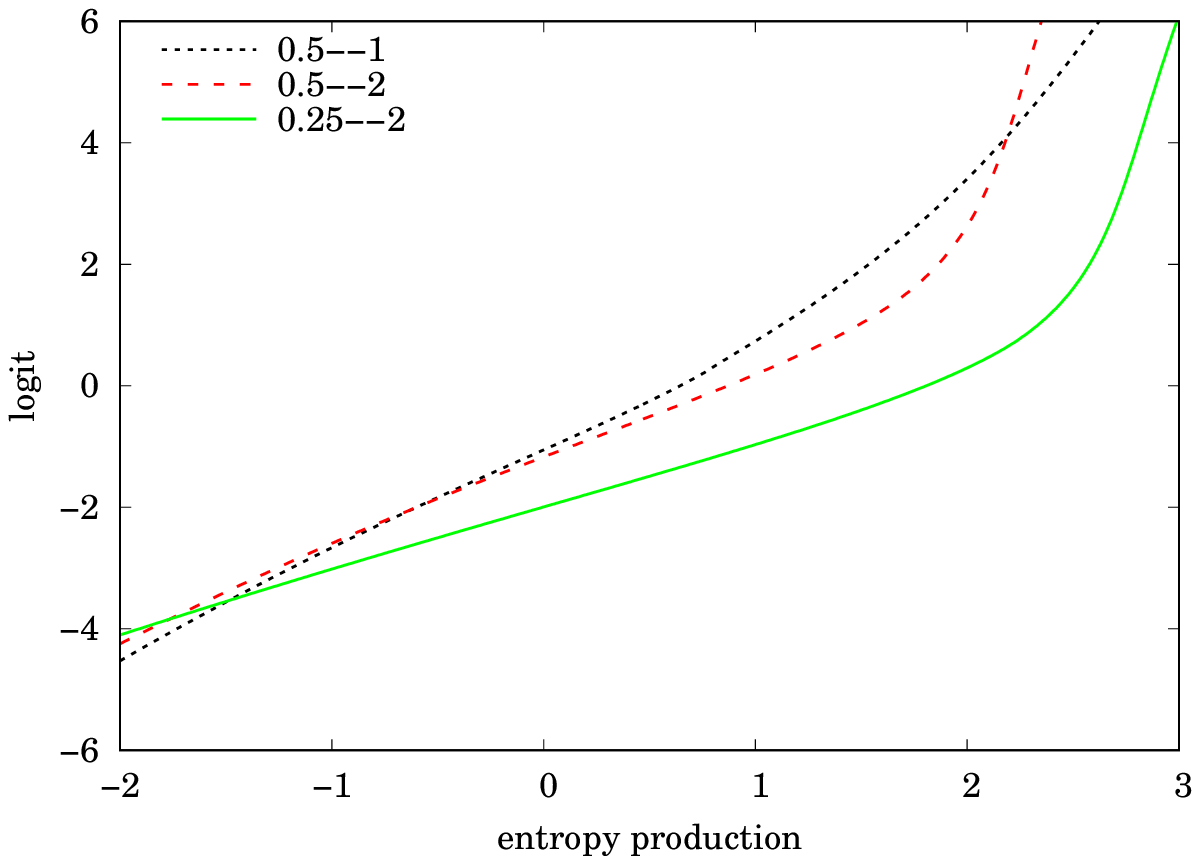}} 
\end{tabular}
\caption{\small $\sech^2$ example. (a,c,e) Approximate pdf of entropy production over various time periods, starting from $\yzero=0,1,2$. (b,d,f) As (a,c,e) but on logit scale.
}
\label{fig:sechsq}
\end{figure}

One way of moving away from the linear force field (quadratic potential well) of the OU model is to make the force field grow less rapidly away from equilibrium by setting 
\[
A(y)=-\frac{\deltahat}{\gammahat}\tanh\gammahat y.
\]
(Incidentally this can be obtained from the local volatility model 
\[
dX_{t}=-\kappa X_{t}\,dt+\sigma\sqrt{1+\gamma^{2}X_{t}^{2}}\,dW_{t}\label{eq:39}
\]
in which volatility increases away from equilibrium, and changing variable by $\gamma X=\sinh\gammahat Y$.)
The stationary state is a sech-power: more precisely, 
\[
\fy(\infty,y)=\frac{\gammahat(\cosh\gammahat y)^{-\deltahat/\gammahat^{2}}}{\Beta\big(\frac{\deltahat}{2\gammahat^{2}},\half\big)},\qquad\langle-A'\rangle_{\infty}=\frac{\deltahat^{2}}{\deltahat+\gammahat^{2}}
\]
with B denoting the Beta function. In the limit $\gammahat\to0$ we recover the OU model, so $\gammahat$ measures the deviation from OU-ness.
Also in the limit $\deltahat=\gammahat\to\infty$ we arrive at the dry-friction case considered above.
We have used  $\gammahat=1$, $\deltahat=2$, which gives $\theta=\frac{4}{3}$, making the stationary distribution $\half\,\sech^2 y$. See Figure~\ref{fig:sechsq}.

Qualitatively the results are similar to those of the OU model. However, there is a minor difference: whereas in the OU model, starting further from equilibrium does not affect the the position of the spike in the density, in this model it is shifted to the right.

\subsubsection{Stationary state Student~t}

\begin{figure}
\noindent
\begin{tabular}{rr}
(a)\scalebox{0.625}{\includegraphics*{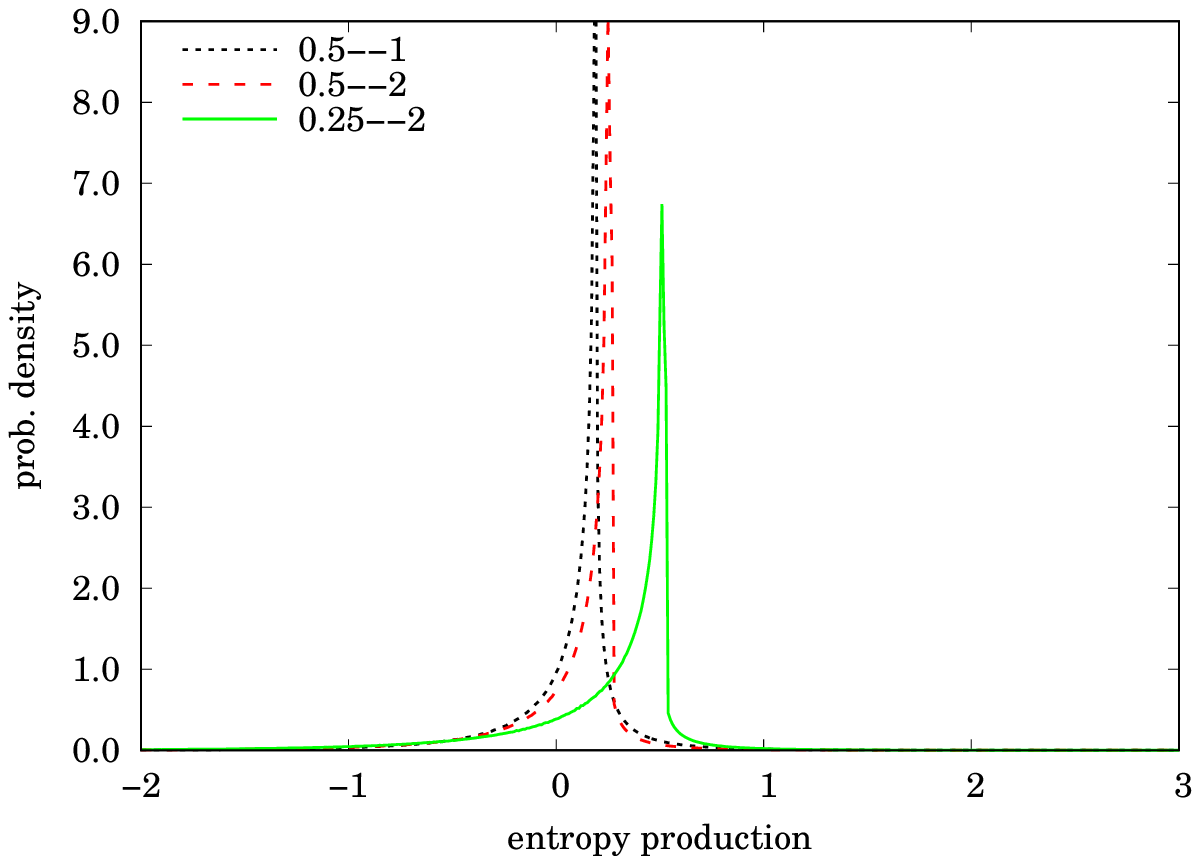}} &
(b)\scalebox{0.625}{\includegraphics*{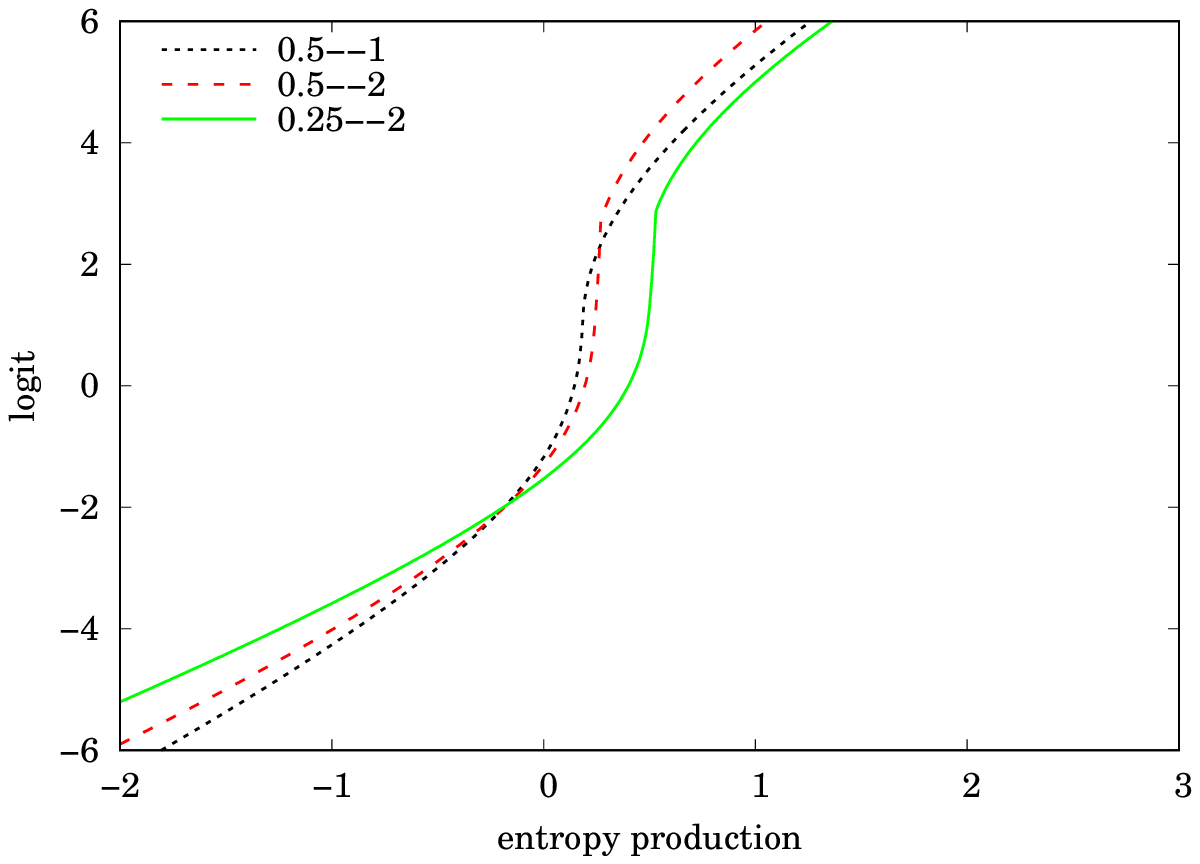}} \\
(c)\scalebox{0.625}{\includegraphics*{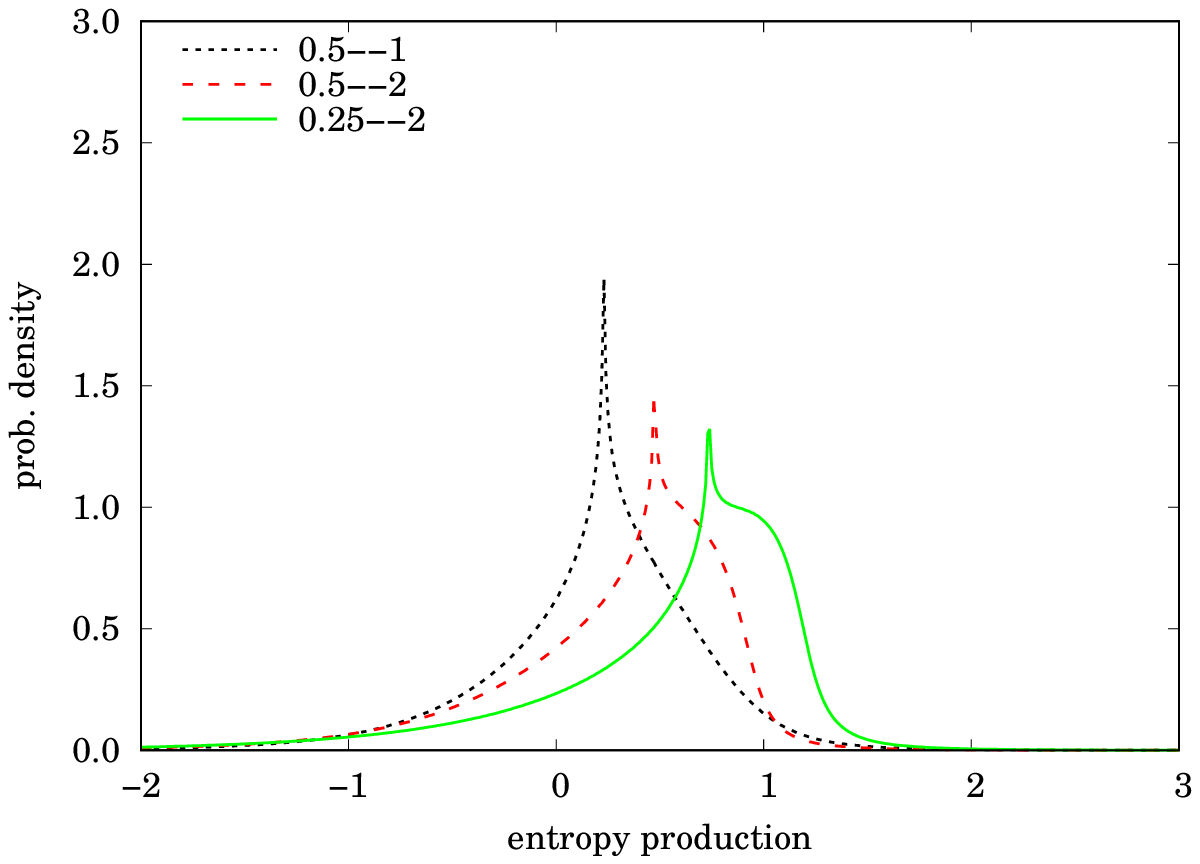}} &
(d)\scalebox{0.625}{\includegraphics*{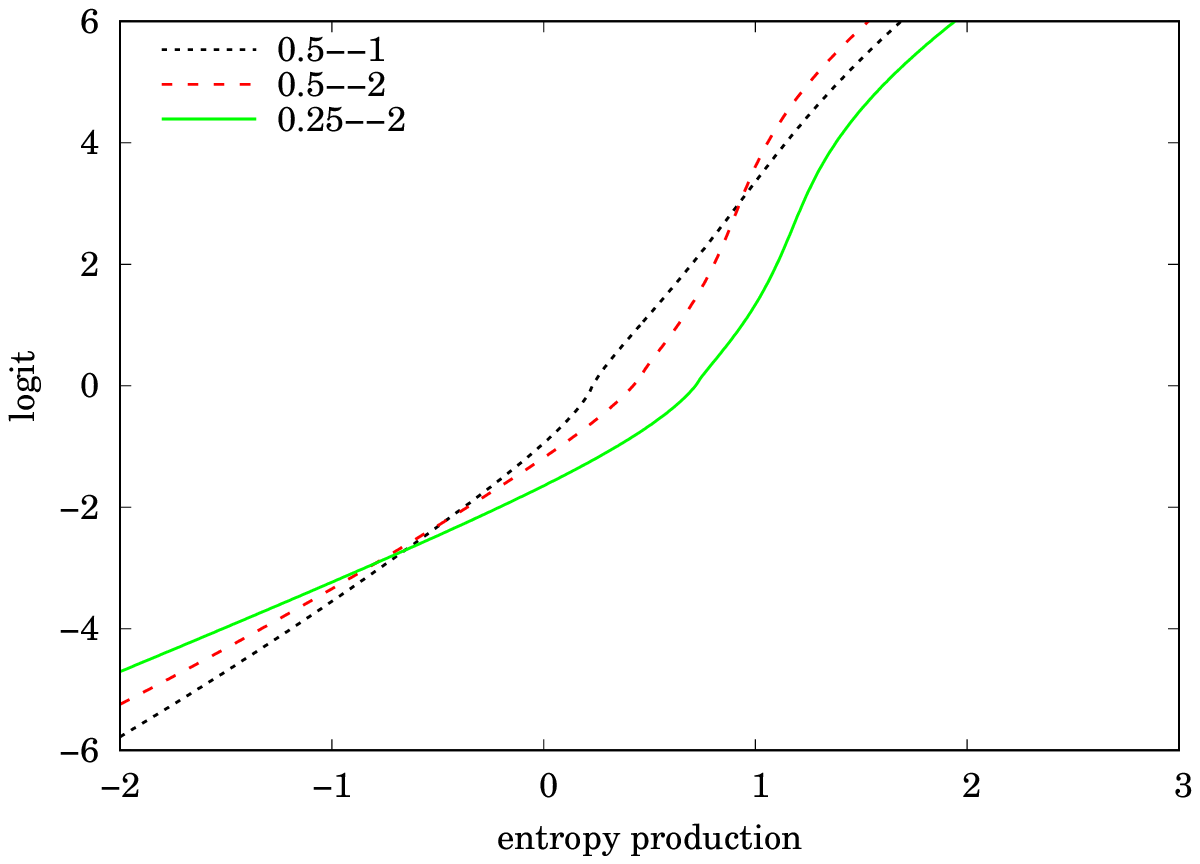}} \\
(e)\scalebox{0.625}{\includegraphics*{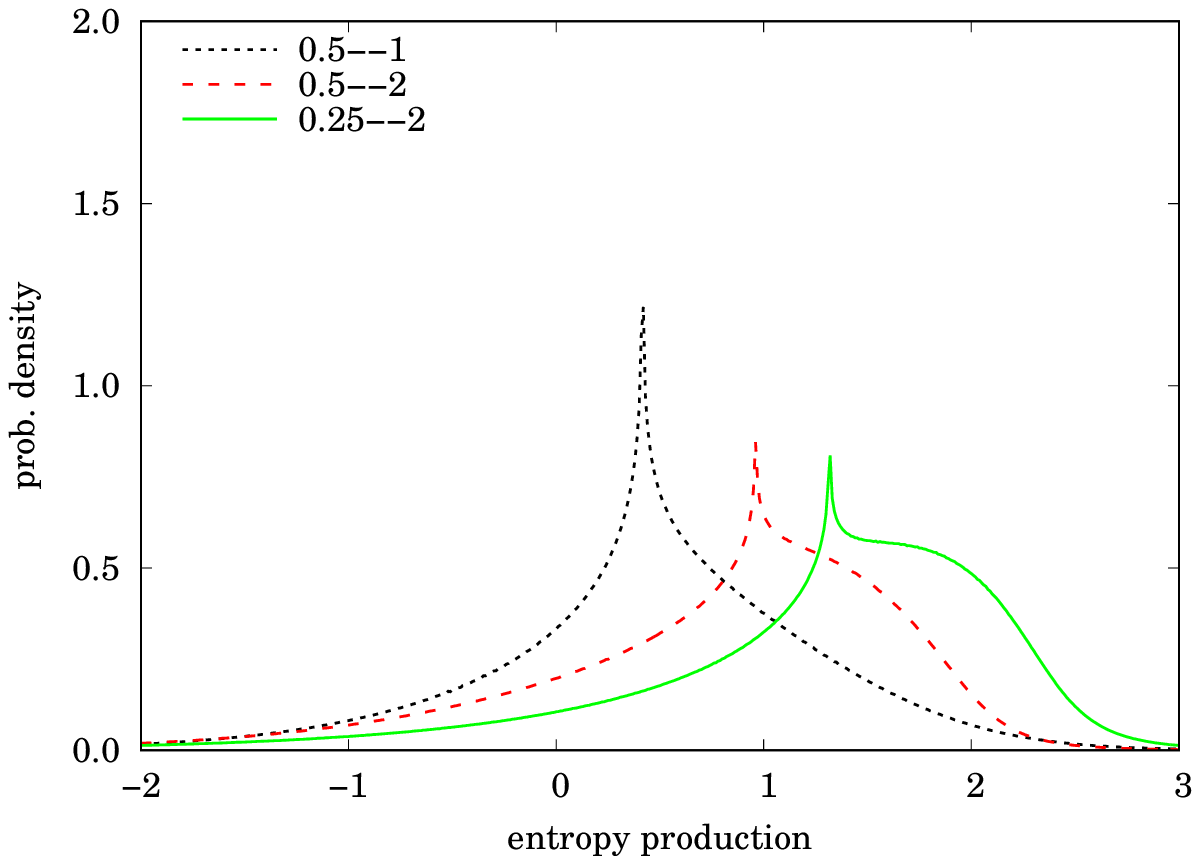}} &
(f)\scalebox{0.625}{\includegraphics*{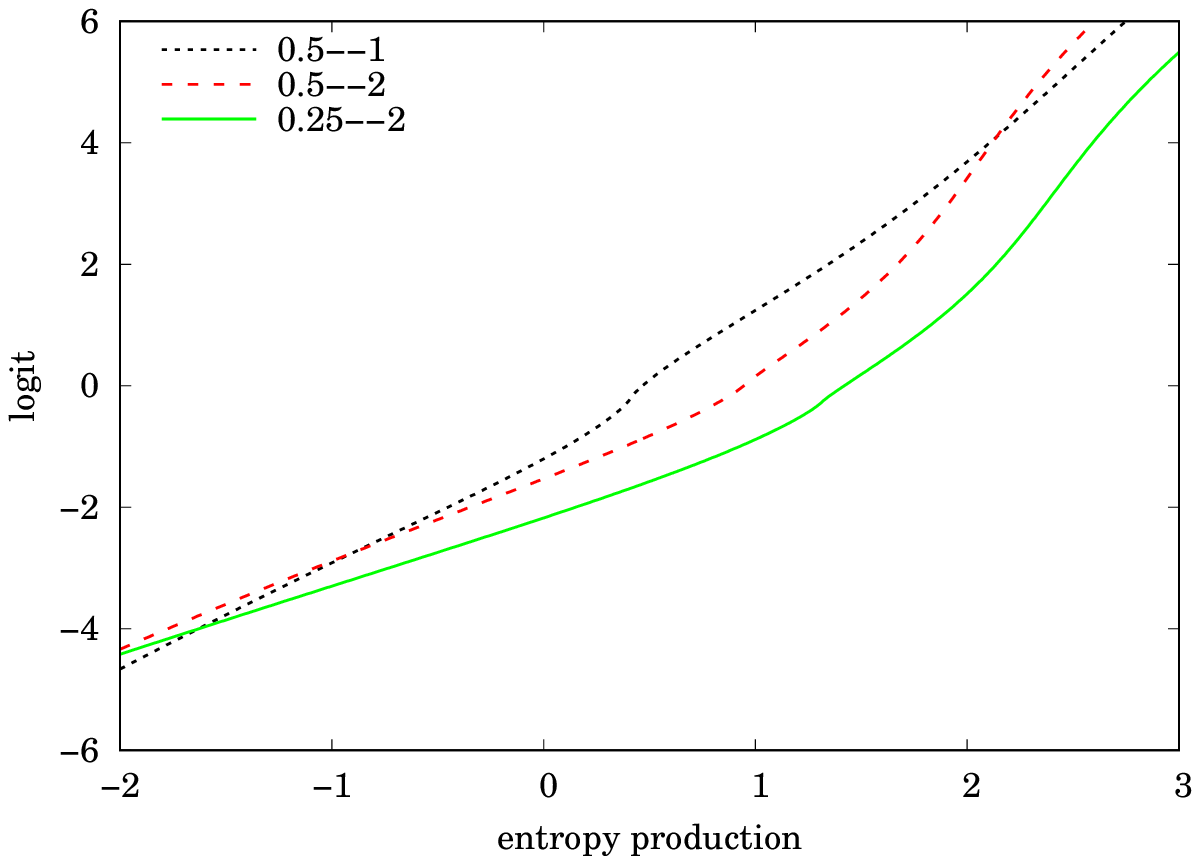}} 
\end{tabular}
\caption{\small Student $\mathrm{t}_3$ example. (a,c,e) Approximate pdf of entropy production over various time periods, starting from $\yzero=0,1,2$. (b,d,f) As (a,c,e) but on logit scale.
}
\label{fig:student}
\end{figure}

We can also write down a model that has Student~$\mathrm{t}_\nu$ as its steady state: 
\[
\ffield(y) = -\frac{\frac{\nu+1}{\nu} \, y}{1+ y^2 /\nu} , \qquad
\fy(\infty,y) = \frac{(1+y^2/\nu)^{-(\nu+1)/2}}{\sqrt{\nu}\, \Beta\big(\frac{\nu}{2},\half\big)}, 
\qquad 
\langle-\ffield'\rangle_{\infty}=\frac{\nu+1}{\nu+3}.
\]
This model gives rise to fatter tails than the sech-power example, because the force field decays to zero as $|y|\to\infty$. 
We have used\footnote{The definition of $\nu$ was different in \cite{Martin15b} and \cite[Fig.3]{Martin18b}, but this should not cause confusion.} $\nu=3$.
See Figure~\ref{fig:student}.

\subsubsection{Double-well potential}

\begin{figure}
\noindent
\begin{tabular}{rr}
(a)\scalebox{0.625}{\includegraphics*{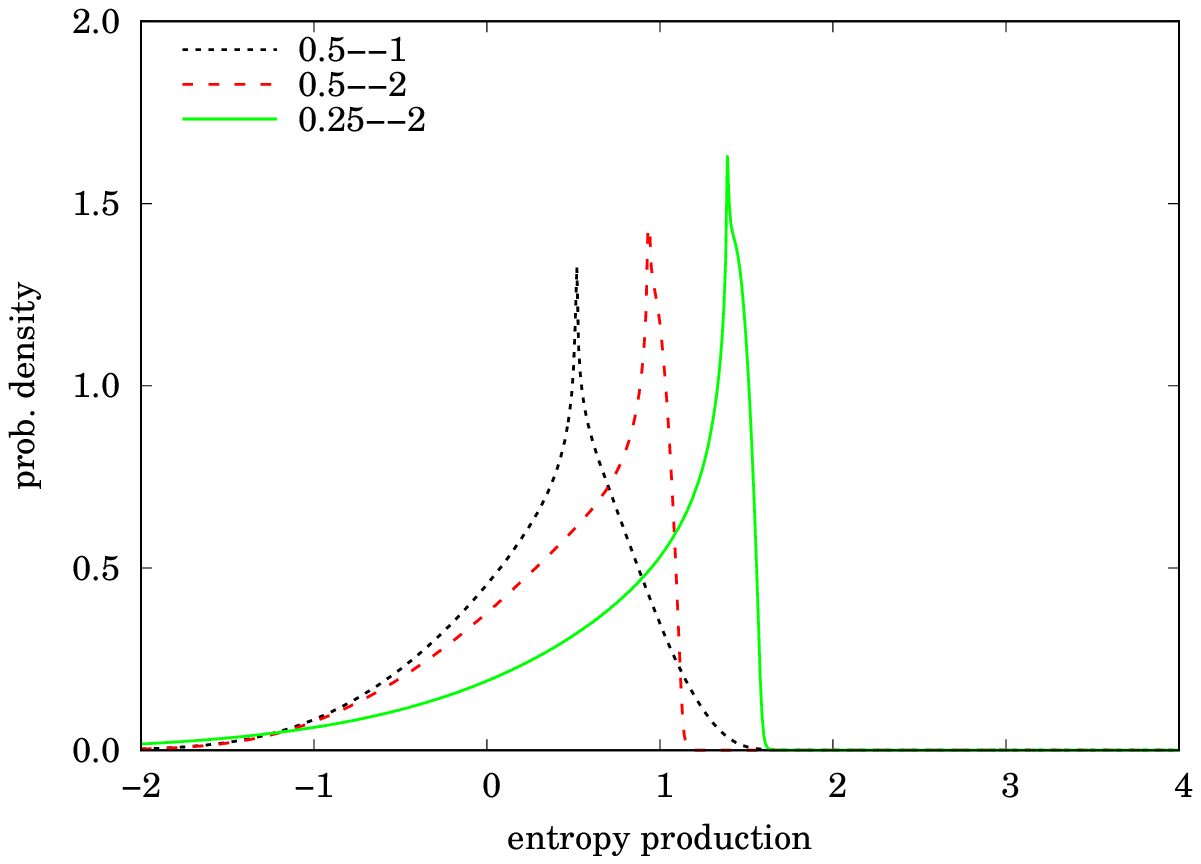}} &
(b)\scalebox{0.625}{\includegraphics*{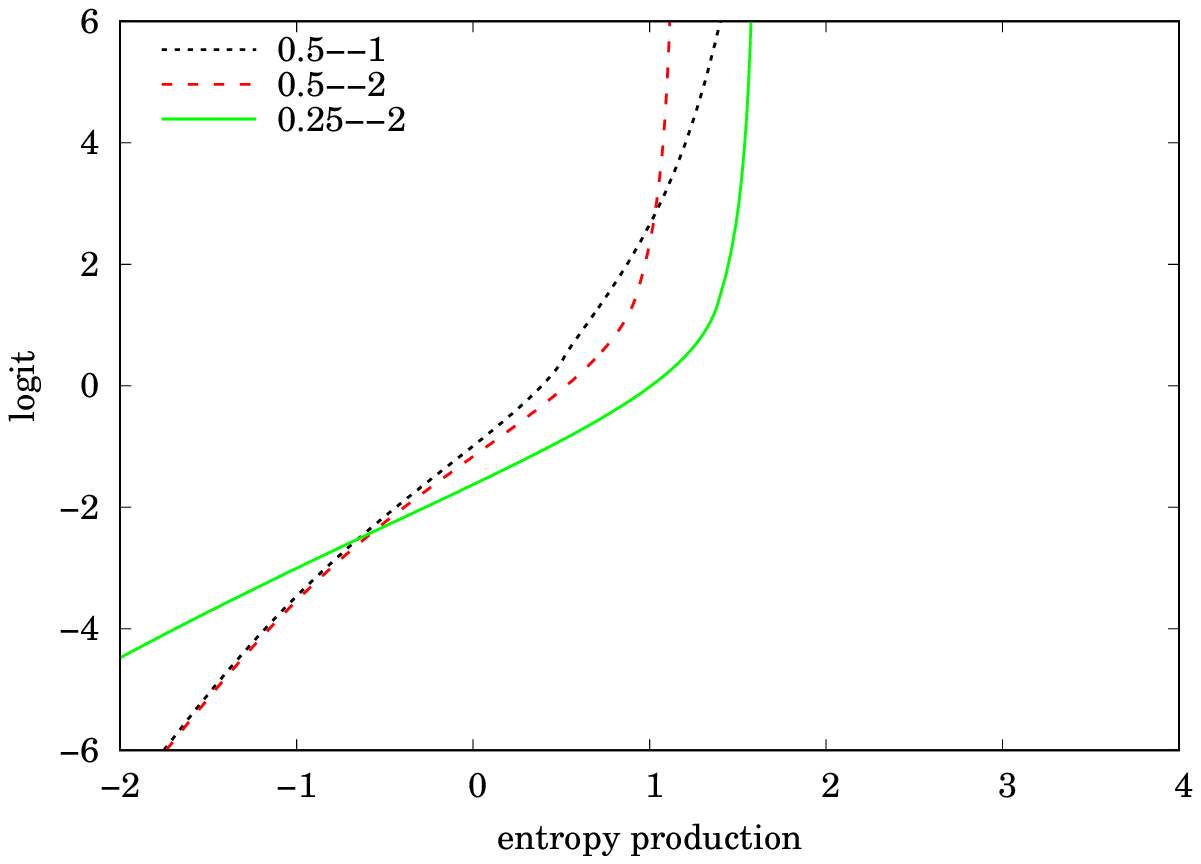}} \\
(c)\scalebox{0.625}{\includegraphics*{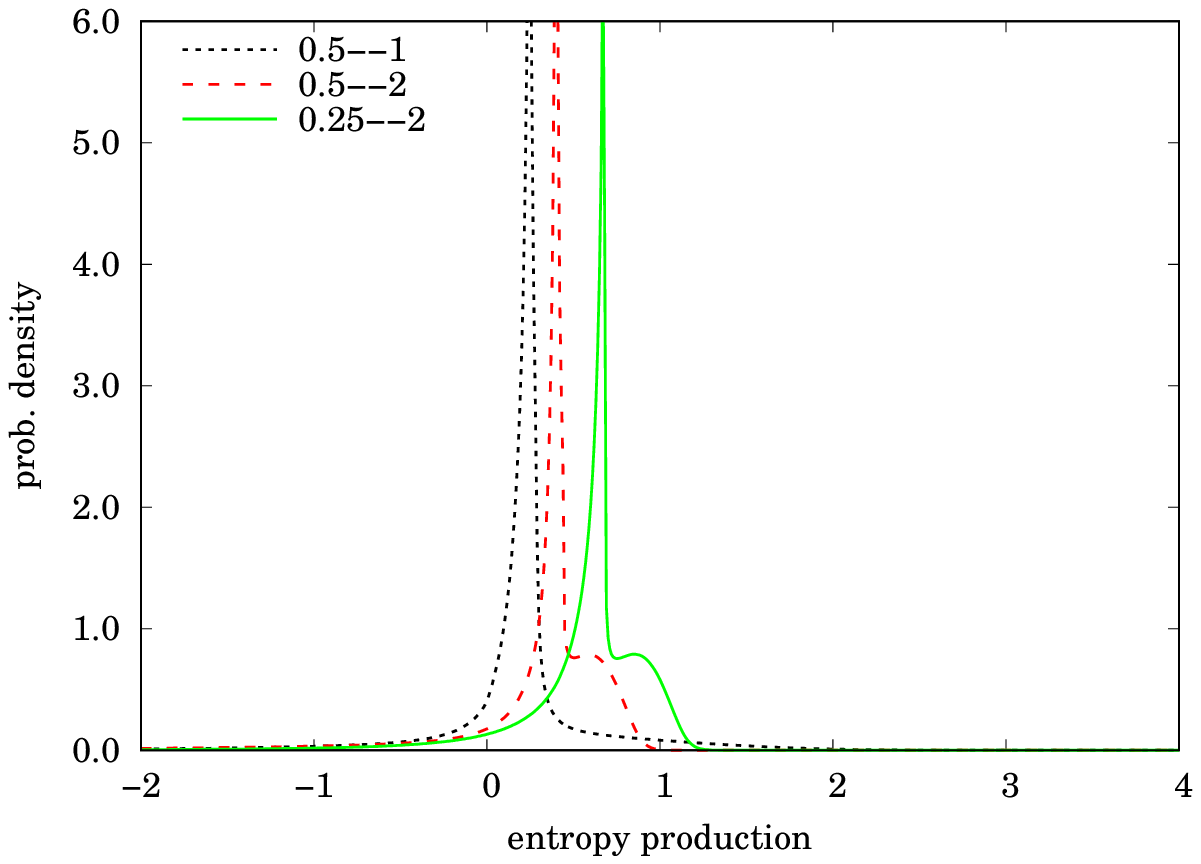}} &
(d)\scalebox{0.625}{\includegraphics*{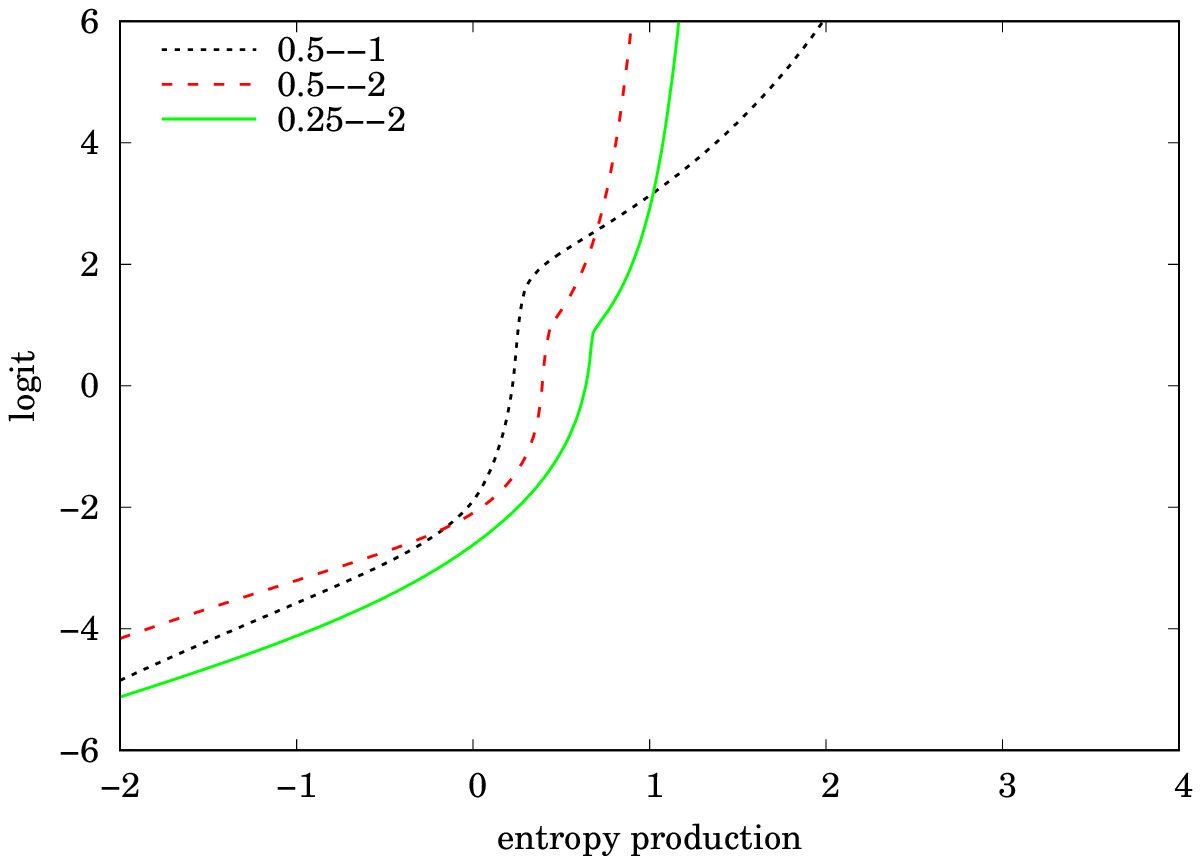}} \\
(e)\scalebox{0.625}{\includegraphics*{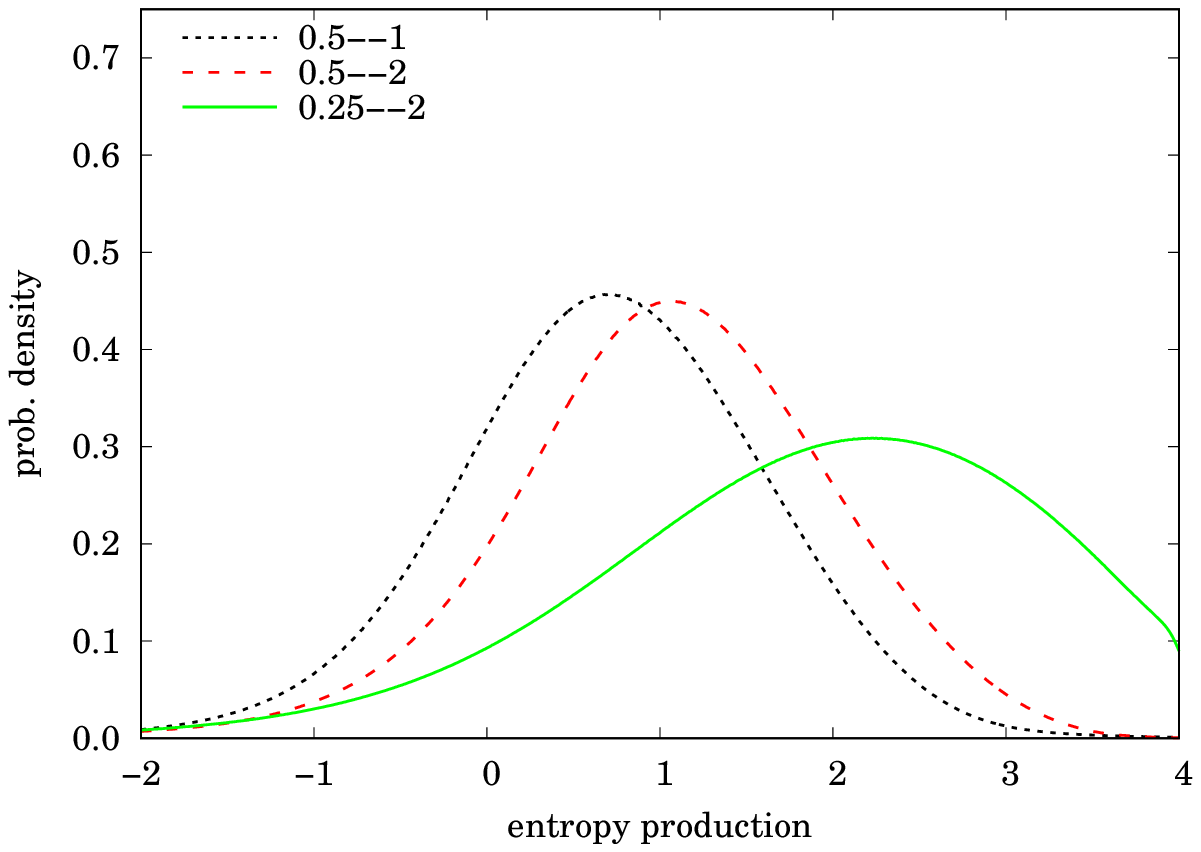}} &
(f)\scalebox{0.625}{\includegraphics*{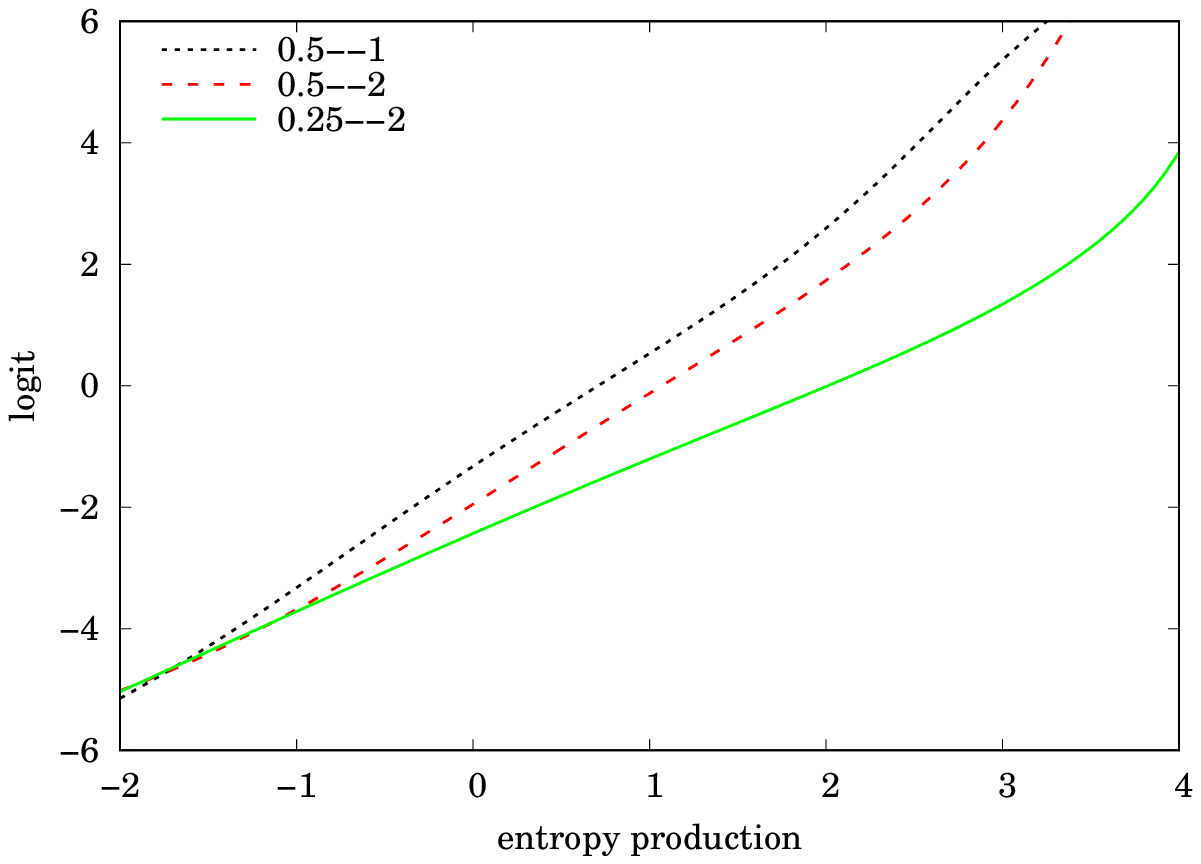}} 
\end{tabular}
\caption{\small Double-well example. (a,c,e) Approximate pdf of entropy production over various time periods, starting from $\yzero=0,1.5,3$. (b,d,f) As (a,c,e) but on logit scale.
}
\label{fig:dblwell}
\end{figure}

A useful general form for the stationary state for a double-well potential is
\[
\fy(\infty,y)=Ke^{-y^{2}/2}\frac{y^{2}+\gamma^{2}}{\big((y-\alpha_{1})^{2}+\beta_{1}^{2}\big)\big((y-\alpha_{2})^{2}+\beta_{2}^{2}\big)}
\]
from which the force field is 
\[
\ffield(y)=-y+\frac{2y}{y^{2}+\gamma^{2}}-\frac{2(y-\alpha_{1})}{(y-\alpha_{1})^{2}+\beta_{1}^{2}}-\frac{2(y-\alpha_{2})}{(y-\alpha_{2})^{2}+\beta_{2}^{2}}.
\]
The parameters act as follows: $\gamma\to0$ makes the
two wells disjoint; $\alpha_{1,2}$ control the location; letting $\beta_{1,2}\to0$ makes them deeper.
In principle the implied coefficient of normalisation $K$, and the quantity $\langle-\ffield'\rangle_{\infty}$, can be calculated directly using Dawson's integral, but the computational effort does not seem worthwhile, as a simple numerical calculation of the integrals is sufficient.
We consider, as in \cite{Martin18b}, the case  $\alpha_1=\alpha_2=2$, $\beta_1=\beta_2=1$, $\gamma=\frac{1}{\sqrt{2}}$, for which $\langle-\ffield'\rangle_{\infty}\approx1.557$.
Unsurprisingly, starting from $\yzero=0$, the point of \emph{unstable} equilibrium, generates \emph{more} entropy
than starting in either of the wells. Referring to Figure~\ref{fig:dblwell}, more entropy is generated in case (a) than in case (c), for any particular time period.

\subsection{Remarks}

We shall reserve our general conclusions for the end of the paper, but it is noticeable that the probability densities of entropy production, over various time intervals, have a number of features in common across different models. Most obvious is that there is a shift in probability mass to the right as time advances, which is expected from the integral fluctuation theorem. The cases where the particle motion involves relaxation towards a unimodal pdf over position have reasonably similar pdfs over entropy production, as might be expected.


\clearpage{}

\section{Theory in higher dimension}

\label{sec:md}

We start with the multidimensional OU model.

\subsection{Ornstein--Uhlenbeck}

\label{sec:oum}

This is similar in terms of tractability to the one-dimensional theory. Let the process be written as 
\begin{equation}
dY_t = -\kappa \gm Y_t \, dt + \sqrt{2\kappa} \, dW_t
\label{eq:mvou}
\end{equation}
where  italic bold letters are square matrices and $W_t$ is a $d$-dimensional standard Brownian motion, i.e.\ different coordinates are independent and so $\ex[dW_t \, dW_t\trs ]=\mI\, dt$.
The normal form (\ref{eq:mvou}) can be obtained from the more general form 
\[
dX_t = -\bm{a}_X X_t \, dt + \bm{b}_X \, dW_t
\]
by writing
\[
Y = \sqrt{2\kappa}\, \bm{b}_X\inv X, \qquad \kappa \gm = \bm{b}_X\inv  \bm{a}_X \bm{b}_X .
\]

Then $Y_t$ is multivariate Normal conditional on $\yzero$, with mean 
\[
\ex[Y_t \cdl \yzero]=e^{-\gm \tau}Y_0
\]
and covariance matrix
\[
\V[Y_t,Y_t\cdl \yzero] = \msigma(t) = \int_{0}^{\tau}  e^{-\gm s}  e^{-\gm\trs s}\,ds
\]
(recall that $\tau=\kappa t$).
Accordingly the pdf of $Y_t$ given $\yzero$ is 
\[
\fy(t,y\cdl \yzero)=\frac{1}{(2\pi)^{m/2}|\msigma(t)|^{1/2}}\exp\left(-\half\big(y-e^{-\gm \tau} \yzero \big)\trs \msigma(t)\inv\big(y-e^{-\gm \tau} \yzero \big)\right),
\]
and in the same notation as before 
\[
\gy (t,y\cdl \yzero)=\frac{|\msigma(\infty)|^{1/2}}{|\msigma(t)|^{1/2}}\exp\left(-\half\big(y-e^{-\gm \tau} \yzero\big)\trs  \msigma(t)\inv \big(y-e^{-\gm \tau}\yzero\big) + \half y\trs  \msigma(\infty)\inv y \right).
\]
This allows the mgf of the entropy production to be written as
\[
\left(\frac{|\msigma(t_2)|}{|\msigma(t_1)|}\right)^{\lambda/2}\frac{1}{(2\pi)^{m}|\msigma(t_1)|^{1/2}|\msigma(t_2-t_1)|^{1/2}}\int_{\R^d}\int_{\R^d} e^{-\mathcal{Q}/2} \,d[Y_{t_1}]\,d[Y_{t_2}]
\]
where $d[Y]$ denotes a volume element in $Y$-space and, as previously,  $\mathcal{Q}$ is a quadratic form.

We confine ourselves to the case in which $\gm$ is symmetric\footnote{We cannot simply write, in the expression for the covariance, $e^{-\gm\trs s}e^{-\gm s} = e^{-(\gm\trs +\gm)s}$, as $\gm,\gm\trs $ cannot be assumed to commute.}, corresponding to the gradient of a quadratic potential. In that case
\[
\msigma(t_1) = \frac{\mI - \mqone}{\gm}, \quad \msigma(t_2) = \frac{\mI - \mqtwo}{\gm}; \qquad 
\mqone = e^{-2\gm \tau_1}, \quad \mqtwo = e^{-2\gm \kappa  \tau_2}
\]
and $\mI$ the $d$-dimensional identity matrix\footnote{The reader might look askance at notation of the form $\gm/(\mI-\mqone)$, which looks like an attempt to divide matrices. However, as $\mI$, $\mqone$, $\mqtwo$ all lie in the commutative matrix ring $\R[\gm]$, it is legitimate to write a fraction in this form. In other words $\gm/(\mI-\mqone)$ is identical to the more `usual' expressions $\gm(\mI-\mqone)\inv$ or $(\mI-\mqone)\inv\gm$. The only proviso is that the denominator of such a fraction be nonsingular. There is also no ambiguity in writing $\sqrt{\mqone}$ or similar.}.
There are two ways of proceeding. The first is to redo the analysis of Section~\ref{sec:ou1} in a multidimensional setting, which we do next; the second idea will become apparent presently. 

Analogously to before,
\begin{equation}
\mathcal{Q} = 
\begin{bmatrix}\yzero\\y_1\\y_2\end{bmatrix}\trs 
\begin{bmatrix}
\frac{(\lambda+1)\gm\mqone}{\mI-\mqone} - \frac{\lambda\gm\mqtwo}{\mI-\mqtwo} 
&
- \frac{(\lambda+1)\gm\sqrt{\mqone}}{\mI-\mqone}
&
\frac{\lambda \gm \sqrt{\mqtwo}}{\mI-\mqtwo}
\\[\spc]
- \frac{(\lambda+1)\gm\sqrt{\mqone}}{\mI-\mqone}
&
\frac{\lambda \gm\mqone}{\mI-\mqone} + \frac{\gm \mqtwo}{\mqone-\mqtwo}
&
-\frac{\gm\sqrt{\mqone\mqtwo}}{\mqone-\mqtwo}
\\[\spc]
\frac{\lambda \gm \sqrt{\mqtwo}}{\mI-\mqtwo}
&
-\frac{\gm\sqrt{\mqone\mqtwo}}{\mqone-\mqtwo}
&
-\frac{\lambda\gm\mqtwo}{\mI-\mqtwo} + \frac{\gm\qone}{\mqone-\mqtwo}
\end{bmatrix}
\begin{bmatrix}\yzero\\y_1\\y_2\end{bmatrix}
\end{equation}
and if we write the matrix entries as $Q_{00}$ etc as before (only now these are $d\times d$ matrices rather than scalars) then this can also be written
\[
\mathcal{Q} = 
\begin{bmatrix} 1 \\ y_1 \\ y_2 \end{bmatrix}\trs 
\begin{bmatrix}
\yzero\trs  Q_{00} \yzero & \yzero\trs  Q_{01} & \yzero\trs  Q_{02} \\
Q_{10} \yzero & Q_{11} & Q_{12} \\
Q_{20} \yzero & Q_{21} & Q_{22} 
\end{bmatrix}
\begin{bmatrix} 1 \\ y_1 \\ y_2 \end{bmatrix}
\]
in which the top left-hand entry is a scalar but the other elements on the leading diagonal are $d \times d$ matrices, and so on.

Defining (much as before) the determinants
\[
\Delta_Q(\yzero) = \left|
\begin{matrix}
\yzero\trs  Q_{00} \yzero & \yzero\trs  Q_{01} & \yzero\trs  Q_{02} \\
Q_{10} \yzero & Q_{11} & Q_{12} \\
Q_{20} \yzero & Q_{21} & Q_{22} 
\end{matrix}
\right|, \qquad 
\delta_Q=\left|\begin{matrix}Q_{11} & Q_{12}\\
Q_{21} & Q_{22}
\end{matrix}\right|, 
\]
we have that the double-integral of $e^{-\mathcal{Q}/2}$ above evaluates to
\[
\frac{(2\pi)^d}{\delta_Q^{1/2}}\exp\biggr(-\frac{\Delta_Q(\yzero)}{2\delta_Q}\biggr)
\]
and so the mgf of the entropy production is 
\begin{equation}
M_{\Delta s}(\lambda) = \left(\frac{|\mI-\mqtwo|}{|\mI-\mqone|}\right)^{\lambda/2}
\frac{ \exp\big(-\Delta_Q(\yzero)/2\delta_Q\big)}{|\mI-\mqone|^{1/2} |\mI-\mqone/\mqtwo|^{1/2} \, \hat{\delta}_Q^{1/2} }
\label{eq:mgf4}
\end{equation}
with $\hat{\delta}_Q=\delta_Q/|\gm|^2$.

We have obtained, as desired, a multidimensional version of the work in \S\ref{sec:ou1}. At this juncture it is convenient to mention an alternative approach. Had we diagonalised $\gm$ at the outset, thereby rotating the coordinate axes so they they aligned with the eigenvectors of the covariance ellipsoid $\msigma(\infty)$, and effectively decoupling the dynamics into $d$ independent one-dimensional systems, we could have expressed the $d$-dimensional result as the convolution of $d$ (in general not identically distributed) one-dimensional results, i.e.\ pdfs of the form given in Prop.~\ref{prop:ou1_ne0}. This is because the transition density is simply a product of $d$ component densities in the principal directions. As the entropy production relates to the logarithm of the density, it is given by the sum of the entropies produced in each of the principal directions, which are independent, and so the pdf of the entropy production is the convolution of one-dimensional pdfs. Alternatively, the mgf of the entropy production will be a product of one-dimensional mgfs of the form (\ref{eq:mgf2}). With this in mind, it seems reasonable to complete the analysis by performing the necessary algebraic manipulations on the determinants in (\ref{eq:mgf4}) to make this factorisation apparent. We define the eigenvalues of $\gm$ to be $(\eval_r)_{r=1}^d$, and write $q_i^{(r)} = \exp(-2\eval_r \tau_i)$, $i=1,2$.

Analysing as before, the term on the front of (\ref{eq:mgf4}) represents a shift by an amount 
\[
\sstar = \half\ln\frac{|\mI-\mqtwo|}{|\mI-\mqone|}
= \half \sum_{r=1}^d \ln \frac{1-\qtwo^{(r)}}{1-\qone^{(r)}} ,
\label{eq:sstar2}
\]
which is clearly the sum of shifts in the principal directions.

Next consider $\yzero=0$. In that case the only thing to analyse is the determinant $\delta_Q$. Under an orthogonal change of basis $\gm$ is diagonalised and then all of the matrices $Q_{11}$, $Q_{12}$, $Q_{21}$, $Q_{22}$ are brought to diagonal form. It is convenient to define
\[
Q^\sharp = \begin{bmatrix} Q_{11} & Q_{12} \\ Q_{21} & Q_{22} \end{bmatrix}
\]
so that $|Q^\sharp|=\delta_Q$.
Now permute the rows and columns of $Q^\sharp$ by taking them in the order $1,m+1,2,m+2,\ldots$. Then the matrix consists of $2\times 2$ blocks along the leading diagonal and zeros elsewhere, and the determinant, which is unaffected by this operation, is given by
\[
\frac{\delta_Q}{|\gm|^2} = \prod_{r=1}^d \left| 
\begin{matrix}
\frac{\lambda \qone^{(r)}+1}{1-\qone^{(r)}} + \frac{\qtwo^{(r)}}{\qone^{(r)}-\qtwo^{(r)}} & -\frac{\sqrt{\qone^{(r)}\qtwo^{(r)}}}{\qone^{(r)}-\qtwo^{(r)}} \\[10pt]
-\frac{\sqrt{\qone^{(r)}\qtwo^{(r)}}}{\qone^{(r)}-\qtwo^{(r)}} & - \frac{\lambda \qtwo^{(r)}}{1-\qtwo^{(r)}} + \frac{\qone^{(r)}}{\qone^{(r)}-\qtwo^{(r)}}
\end{matrix}
\right|
\]
which is the product of $d$ results of the form (\ref{eq:lampm}).

When $\yzero\ne0$ we need to analyse $\Delta_Q$ and note that, by elementary row operations,
\[
\begin{bmatrix} 1 & \yzero\trs \sqrt{\mqone} & \yzero\trs \sqrt{\mqtwo} \\[\spc] 0 & \mI & 0 \\[\spc] 0 & 0 & \mI \end{bmatrix}
\begin{bmatrix}
\yzero\trs  Q_{00} \yzero & \yzero\trs  Q_{01} & \yzero\trs  Q_{02} \\[\spc]
Q_{10} \yzero & Q_{11} & Q_{12} \\[\spc]
Q_{20} \yzero & Q_{21} & Q_{22} 
\end{bmatrix}
= 
\begin{bmatrix} 0 & -\lambda\yzero\trs \sqrt{\mqone} & \lambda\yzero\trs \sqrt{\mqtwo} \\[\spc]
Q_{10}\yzero & Q_{11} & Q_{12} \\[\spc]
Q_{20}\yzero & Q_{21} & Q_{22} 
\end{bmatrix} 
 .
\]
By the block-determinant lemma \cite{Cohn84},
\[
\left|\begin{matrix}
A & B \\ C & D 
\end{matrix}\right|
=
|D| \, |A-BD\inv C| \qquad \mbox{($D$ square)}
\]
applied with $A=0$, $D=Q^\sharp$, we deduce
\[
-\frac{\Delta_Q(\yzero)}{2\delta_Q} =
\frac{\lambda \yzero\trs }{2\delta_Q} 
\gm \begin{bmatrix} -\sqrt{\mqone} & \sqrt{\mqtwo}  \end{bmatrix}
\mathrm{adj} (Q^\sharp)   
\begin{bmatrix} Q_{10} \\ Q_{20} \end{bmatrix} \yzero
\]
where adj denotes the adjugate (transpose of the matrix of cofactors; \cite{Cohn84}).
By applying the same permutation trick to write $Q^\sharp$ as an array of $2\times 2$ blocks, we can calculate the adjugate directly and then multiply the matrices out to obtain
\[
-\frac{\Delta_Q(\yzero)}{2\delta_Q} =
\frac{\lambda(\lambda+1)}{2} \sum_{r=1}^d
 \frac{\eval_r Y_{0,r}^2 \big(\qone^{(r)}-\qtwo^{(r)}\big)}{1+ \big(\qone^{(r)}-\qtwo^{(r)}\big) \lambda - \displaystyle\frac{\qtwo^{(r)}\big(\qone^{(r)}-\qtwo^{(r)}\big)}{1-\qtwo^{(r)}}\lambda^2}
\]
which is a sum of $d$ results of the form (\ref{eq:Dqdq}), as anticipated.

In summary:
\begin{prop}
\label{prop:oum}
In a $d$-dimensional OU model the pdf of the entropy generation is a $d$-fold convolution of one-dimensional models along the principal axes. 
If $\yzero=0$, then in the isotropic case $\gm=\theta \mI$, we have
\begin{equation}
p(\Delta s) =
\frac{(b^2-a^2)^{\nu+\half}}{\sqrt{\pi} \, \Gamma(\nu+\half)}
 e^{a(\Delta s-\sstar)} \left| \displaystyle \frac{\Delta s-\sstar}{2b} \right|^{2\nu} K_\nu(b|\Delta s-\sstar|)
\end{equation}
with $\nu=(d-1)/2$ and $a,b$ as in Prop.~\ref{prop:ou1_0} and
\[
\sstar = \frac{d}{2} \ln\frac{1-e^{-2\theta\kappa t_2}}{1-e^{-2\theta\kappa t_1}}.
\]
$\Box$ 
\end{prop}

\begin{figure}
\noindent
\begin{tabular}{rr}
(a) \scalebox{0.625}{\includegraphics*{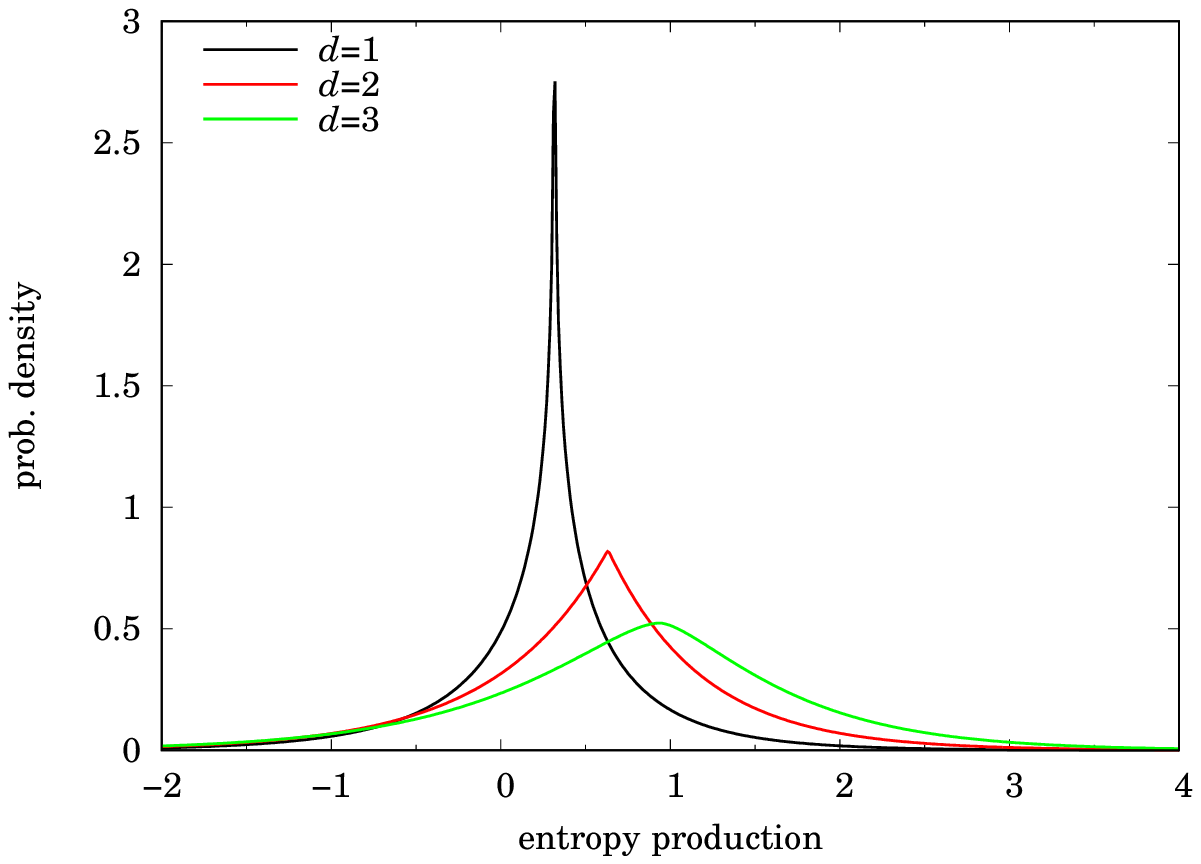}} &
(b) \scalebox{0.625}{\includegraphics*{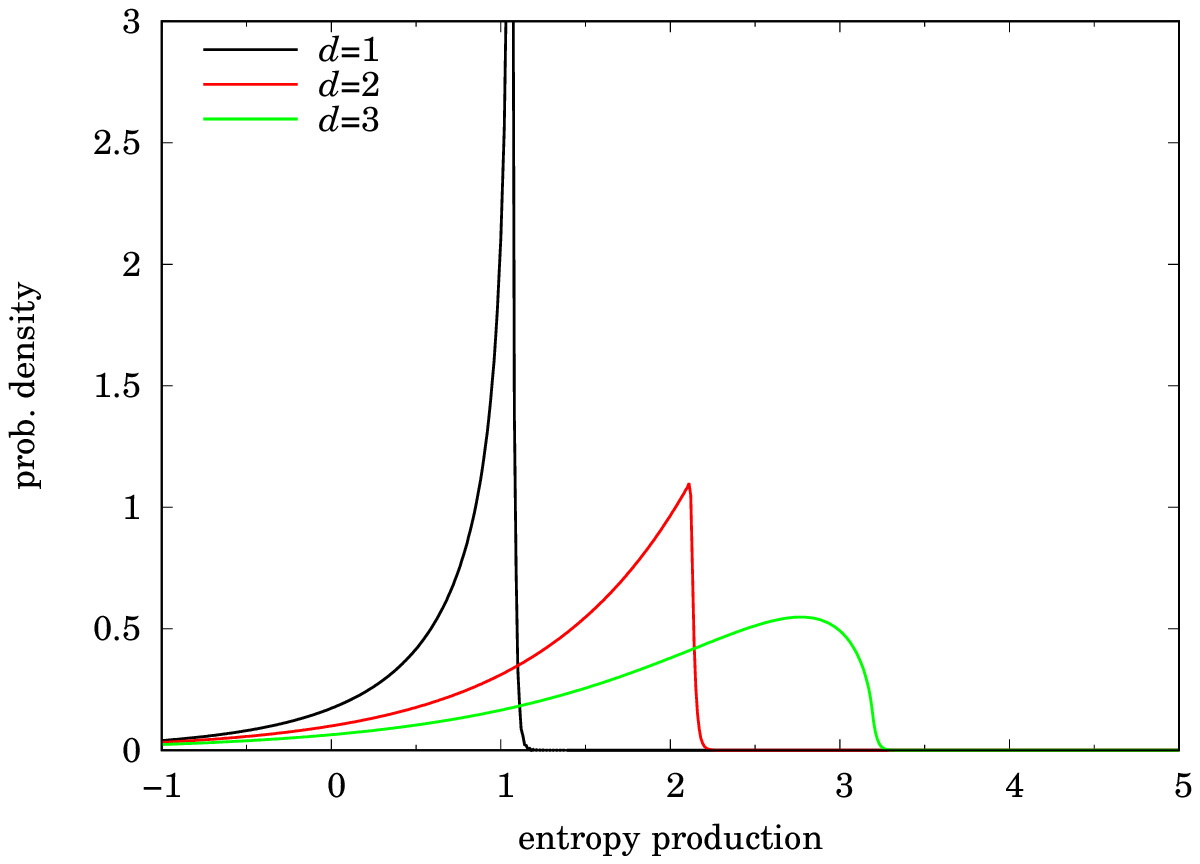}} \\
(c) \scalebox{0.625}{\includegraphics*{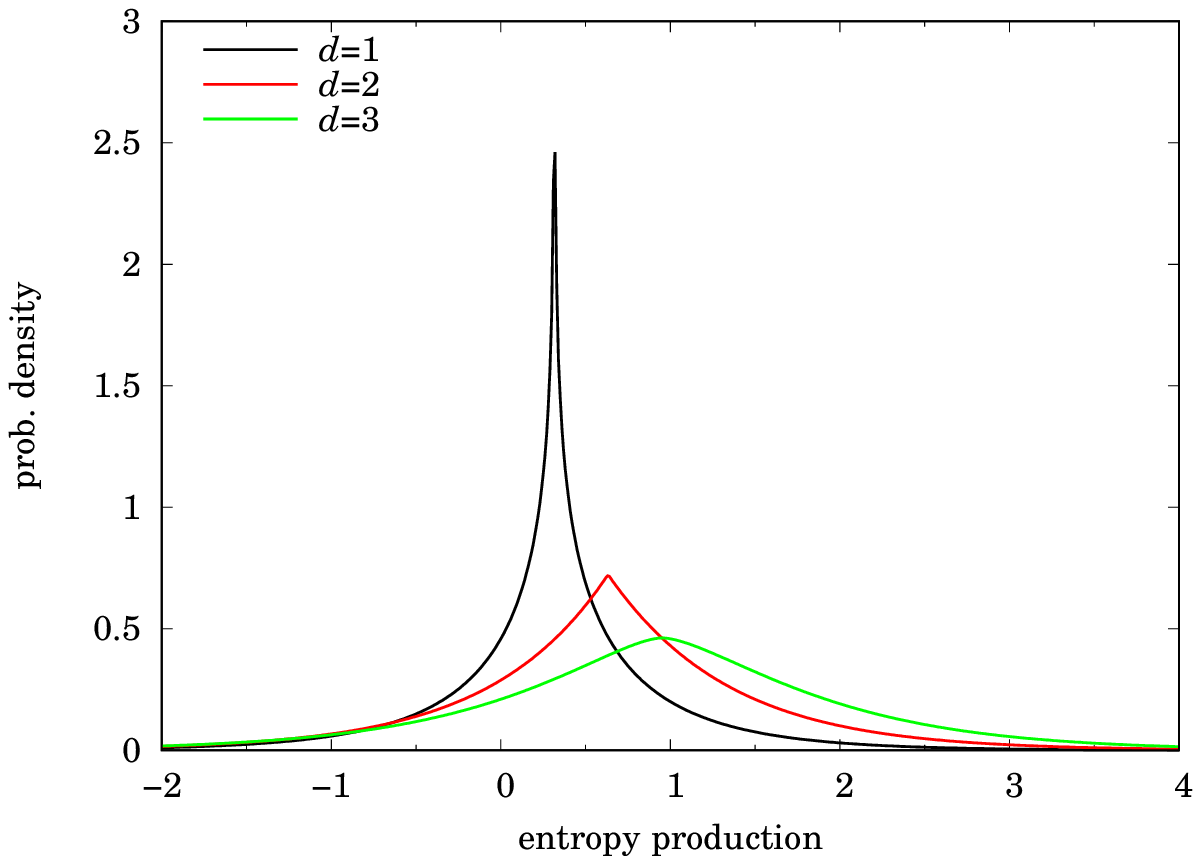}} &
(d) \scalebox{0.625}{\includegraphics*{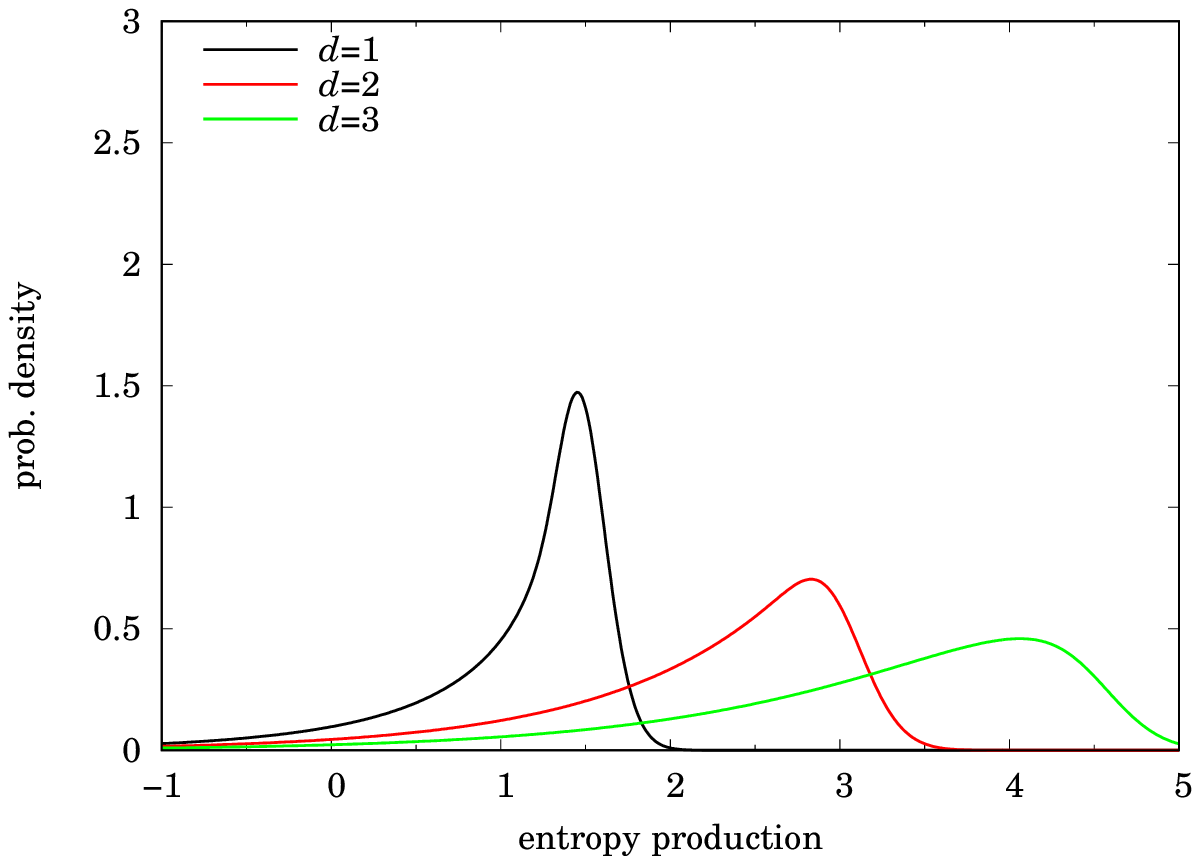}}
\end{tabular}
\caption{\small Probability density of entropy production for multivariate OU process ($\ffield(y)=-y$) starting from the origin, in dimension 1,2,3.
Time intervals: (a,c) $[0.0625,0.125]$; (b,d) $[0.0625,2]$.
Starting-points ($\yzero$): (a,b) $[0 \ldots 0]\trs$; (c,d) $[1 \ldots 1]\trs$.
}
\label{fig:oud}
\end{figure}

In general the effect of increasing the dimension is to make the pdf of the entropy production `less singular', i.e.\ smoother and closer to being Normally distributed. This is seen in Figure~\ref{fig:oud}, which shows some representative cases.

\subsubsection{Non-conservative OU model}

When $\gm$ is not symmetric, so that the force field is not  the gradient of a potential, solution of the Fokker--Planck equation is more difficult. Some discussion of this is given in \cite[\S4.3]{Martin18b}, which is summarised now.

First, although the steady-state density is still multivariate Gaussian with zero mean, its covariance matrix $\msigma_\infty$ is not $\gm\inv$: instead it is given by the Lyapunov equation,
\[
\gm \msigma_\infty + \msigma_\infty \gm\trs  = 2\mI
\]
which can be solved for $\msigma_\infty$ by writing it as a set of linear equations in its elements.
Secondly, recalling the general objective of \cite{Martin18b}, it is possible to find two OU models with the same short- and long-term behaviour, even if the medium-term behaviour is different. Indeed
\[
\gm = (\mI+\bm{u}) \msigma_\infty\inv, \qquad \bm{u}\in\mathfrak{A}
\]
where $\mathfrak{A}$ is the space of skew-symmetric matrices, give rise to the same behaviour in both limits.
Formally, define two generators $\gm$ to be equivalent (notation $\siminf$) if they give rise to the same long-time covariance matrix. Then any generator is equivalent under $\siminf$ to a unique symmetric one; we obtain $\msigma_\infty$ from the Lyapunov equation and then invert it to obtain this symmetric generator. After this, the equivalent symmetric generator can be used as a first approximation for entropy calculations.

As an example: when $\gm = \begin{bmatrix} 1 & a \\ 0 & 1 \end{bmatrix}$ we have
\[
\msigma(t) =  
\begin{bmatrix} 1-e^{-2\tau}+\shalf a^2\big(1-(1+2\tau+2\tau^2)e^{-2\tau}\big) & -\shalf a \big(1-(1+2\tau)e^{-2\tau}\big) \\ -\shalf a \big(1-(1+2\tau)e^{-2\tau}\big) & 1-e^{-2\tau} \end{bmatrix}
\]
and
\begin{equation}
\msigma_\infty = \begin{bmatrix} 1+a^2/2 & -a/2 \\ -a/2 & 1 \end{bmatrix}, \qquad 
\msigma_\infty\inv = \frac{1}{1+a^2/4} \begin{bmatrix} 1 & a/2 \\ a/2 & 1+a^2/2 \end{bmatrix}
\label{eq:example}
\end{equation}
and so the equivalence class of $\gm$ under $\siminf$ is
\[
[\gm]_{\siminf} = 
\frac{1}{1+a^2/4} \begin{bmatrix} 1-ab/2 & a/2-b-a^2b/2 \\ b+a/2 & 1+ab/2+a^2/2 \end{bmatrix}, \qquad b\in\R
\]
which contains the following elements in particular:
\[
\begin{bmatrix} 1 & a \\ 0 & 1 \end{bmatrix}; \quad 
\frac{1}{1+a^2/4} \begin{bmatrix} 1 & a/2 \\ a/2 & 1+a^2/2 \end{bmatrix}.
\]
It is easy to see that all elements of $[\gm]_{\siminf}$ have the same trace, which in this example is 2.


\subsection{General potential}

\label{sec:gpm}

The multivariate analogue of (\ref{eq:Yt}) is
\begin{equation}
dY_t = \kappa \ffield(Y_t) \, dt + \sqrt{2\kappa} \, dW_t
\label{eq:mvou_gen}
\end{equation}
where, in $d$ dimensions, $W_t, Y_t\in\R^d$ and $A:\R^d\to\R^d$ is the force field.
The corresponding Fokker--Planck equation is (with $\tau=\kappa t$ as before)
\begin{equation}
\pderiv{\fy}{\tau} = - \nabla \cdot (A \fy) + \nabla^2 \fy.
\label{eq:mfp_f}
\end{equation}
If $\ffield$ is conservative, i.e.\ the gradient of a potential, then
\begin{equation}
\ffield(y) = \nabla \ln \fy (\infty,y)
\end{equation}
and $\gy(t,y)=\fy(t,y)/\fy(\infty,y)$ obeys the backward equation
\begin{equation}
\pderiv{\gy}{\tau} = A\cdot \nabla \gy + \nabla^2 \gy ;
\label{eq:mfp_g}
\end{equation}
but neither of these last two equations is true if $\ffield$ is non-conservative (the statement of the backward equation is correct, it is just that $\gy$ does not satisfy it). We concentrate only on the conservative case from now on.

The analogue of $\theta$ in (\ref{eq:theta}) is now a symmetric matrix $\mtheta$ given by
\begin{equation}
\mtheta = \langle -\nabla \ffield \rangle_\infty = \langle \ffield \ffield \rangle_\infty
\end{equation}
which shows it to be positive-definite, and as before we write
\[
\mq = \exp(-2\mtheta \tau). 
\]
Again in the interest of stating our results upfront, we have analogously to (\ref{eq:fapprox},\ref{eq:gapprox}),
\begin{equation}
\fy(t,y) \sim 
\frac{1}{|\mI-\mq|^{1/2}}  
 \exp\left( - \half (y-\yzero)^\dagger \frac{\mtheta \!\sqrt{\mq}}{\mI-\mq} (y-\yzero) \right) 
\frac{\left( \frac{|\mtheta/2\pi|}{ f(\infty,\muinfty)^2 } \right)^{\rho(\tau)} \fy(\infty,y)}{\Omega(\tau,y) \Omega(\tau,\yzero)}.
\label{eq:newf2}
\end{equation}
and
\begin{equation}
\gy(t,y) \sim 
\frac{1}{|\mI-\mq|^{1/2}}  
 \exp\left( - \half (y-\yzero)^\dagger \frac{\mtheta \!\sqrt{\mq}}{\mI-\mq} (y-\yzero) \right) 
\frac{\left( \frac{|\mtheta/2\pi|}{ f(\infty,\muinfty)^2 } \right)^{\rho(\tau)}}{\Omega(\tau,y) \Omega(\tau,\yzero)} .
\label{eq:newg2}
\end{equation}
These are exact for the (conservative) multivariate OU model and generalise the one-dimensional work in a reasonably natural way.
The function $\rho$ is defined by
\begin{equation}
\rho(\tau) = \frac{1}{d}  \tr \, \frac{\sqrt{\mq}}{\mI+\sqrt{\mq}}.
\end{equation}
The function $\Omega$ is defined by
\begin{equation}
\Omega(\tau,y) = \exp \int_{\muinfty}^y dx \cdot \left(\frac{\sqrt{\mq}}{\mI+\!\sqrt{\mq}}  \ffield(x) \right),
\label{eq:Omega}
\end{equation}
where $\muinfty=\langle Y\rangle_\infty$ is the mean of the stationary distribution and the path of integration is a straight line; regardless of the dimension of the problem ($d$), this is still a one-dimensional integral, and in the examples considered it can be calculated in closed form.

The remainder of this section is devoted to details relating to the above results and the reader may omit it.

As before we define $H=-\gy\inv \nabla \gy$, which satisfies the vector equation
\begin{equation}
\pderiv{H}{\tau} = \nabla (A\cdot H + \nabla \cdot H - H \cdot H).
\label{eq:mfp_H}
\end{equation}
Analogously to the univariate case, and also following from the multivariate OU model, for which the following is exact, we adopt the ansatz 
\begin{equation}
H(t,y) = \frac{\mtheta \!\sqrt{\mq}}{\mI-\mq} (y-\yzero) +  \frac{\sqrt{\mq}}{\mI+\!\sqrt{\mq}}  \ffield(y)  + \sqrt{\mq} \, o(1) .
\label{eq:hvec1}
\end{equation}
The first term integrates to give a Gaussian, which is immediately visible in (\ref{eq:newf2},\ref{eq:newg2}) and to be expected.
The second term is more difficult and the function $\Omega$ arises from integrating it. The main analytical features of it are (i) it tends to $\ffield(y)/2$ as $t\to0$, which is seen from dominant balance in (\ref{eq:mfp_H}), and (ii) it vanishes as $t\to\infty$.
But despite its links to the OU model and to the one-dimensional theory given earlier, this second term is not in general a conservative field. To see why this is, consider its gradient: defining the symmetric matrix $\bm{\phi}=\sqrt{\mq}/(\mI+\!\sqrt{\mq})$, the gradient is $\bm{\gamma}$ given by
\[
\gamma_{ij} = \phi_{jk} \partial_i \ffield_k
\]
(summing over the repeated suffix $k$ in the usual way).
Now $\ffield$ is conservative so $\partial_i \ffield_k = \partial_k \ffield_i$, and so the above expression is the product of two symmetric matrices. Such a product is a symmetric matrix iff the two matrices commute, which in turn holds iff their principal axes (eigenvectors) are in alignment.

We point out that in certain cases this condition will already be met as the necessary commutation already holds. One is the OU model, for which $\bm{\phi}$ and $\nabla \ffield$ are both in $\R[\gm]$ (though we know that there cannot be a problem as the expression for $H$ is exact). Another is any spherical model, i.e.\ one in which $\fy(\infty,y)$ is a function of $(y-\mu)^\dagger(y-\mu)$ for some constant vector $\mu$: in that case $\mtheta$, and hence $\bm{\phi}$, are scalar multiples of the identity matrix.

In general, though, some alteration of the second term in (\ref{eq:hvec1}) is in principle necessary. By rotating $\bm{\phi}$ by a `small' amount---more specifically replacing $\bm{\phi}$ with $\bm{w}\trs\bm{\phi}\bm{w}$, where $\bm{w}$ is orthogonal---will align its principal axes with those of $\nabla \ffield$ to make $\bm{\gamma}$ symmetric, and hence make (\ref{eq:hvec1}) a conservative field.
While this is a solution, there are potential difficulties with it: we must find $\bm{w}$, which depends on $y$, and is also not as yet well-defined; and then we must integrate it, which would probably have to be done numerically.

Importantly, the antisymmetric part (or curl) of $\bm{\gamma}$, i.e.\ $\gamma_{ij}-\gamma_{ji}$, which measures how non-conservative is the second term in (\ref{eq:hvec1}), is $O(\tau)$ as $\tau\to0$. This error is commensurate with, or possibly smaller than, the error incurred by ignoring the term marked $\sqrt{\mq} \, o(1)$ in (\ref{eq:hvec1}). Put differently, fixing the `non-conservativeness' of the second term in (\ref{eq:hvec1}) may well not give a significantly more accurate result.
Also the curl vanishes as $\tau\to\infty$, and it vanishes on average, i.e.\ if $y$ is integrated over the stationary distribution $\fy(\infty,\cdot)$.
Besides, the main objective of this work is to provide an approximation that is reasonably simple to calculate. This is why we persist with (\ref{eq:hvec1}) and its consequences, even though the expression is not theoretically ideal.

\subsection{Examples}

\begin{figure}
\noindent
\begin{tabular}{rr}
(a)\scalebox{0.625}{\includegraphics*{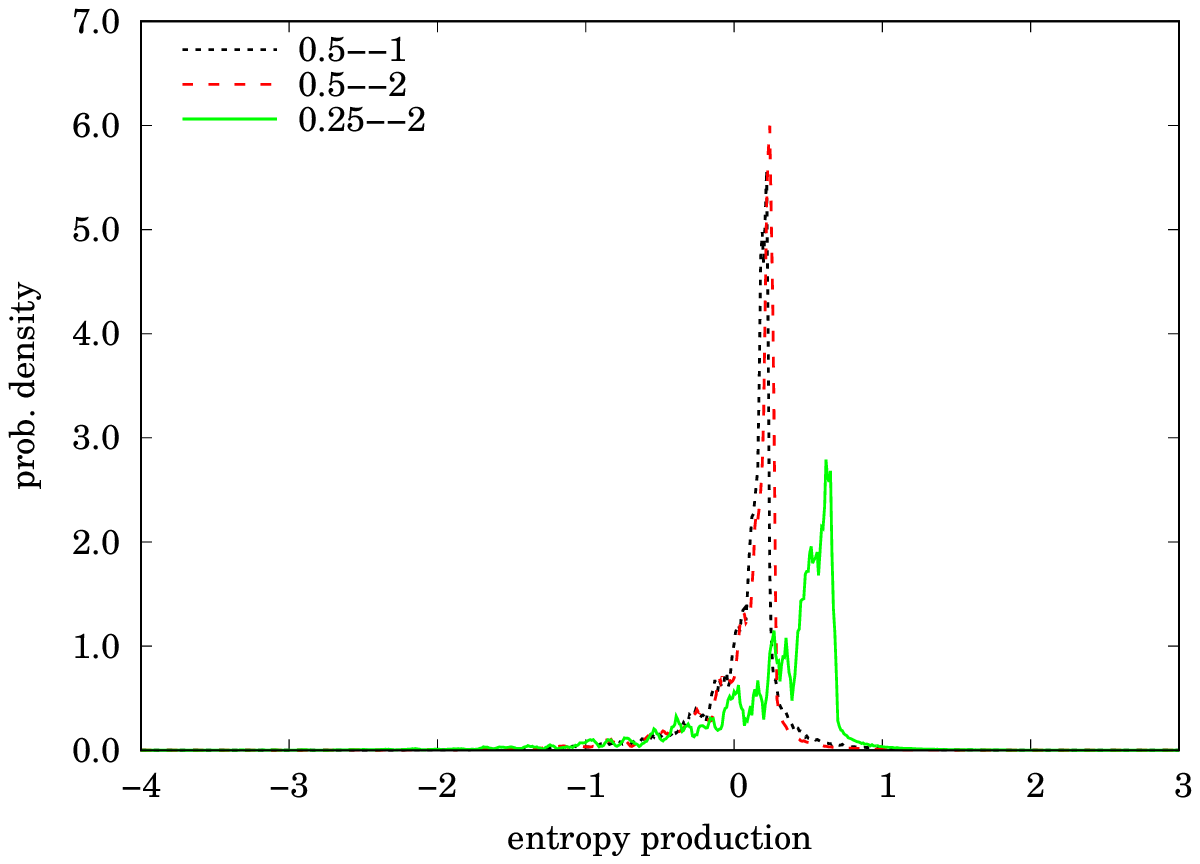}} &
(b)\scalebox{0.625}{\includegraphics*{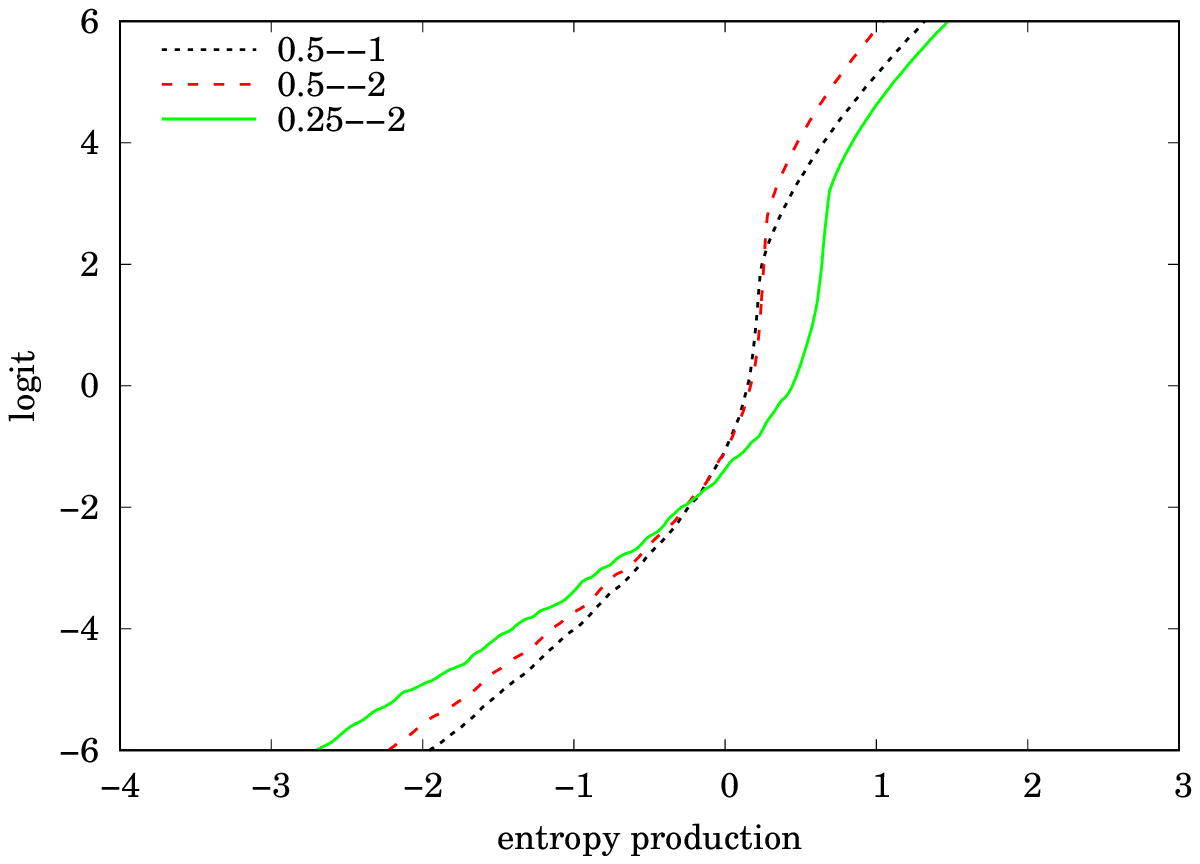}} \\
(c)\scalebox{0.625}{\includegraphics*{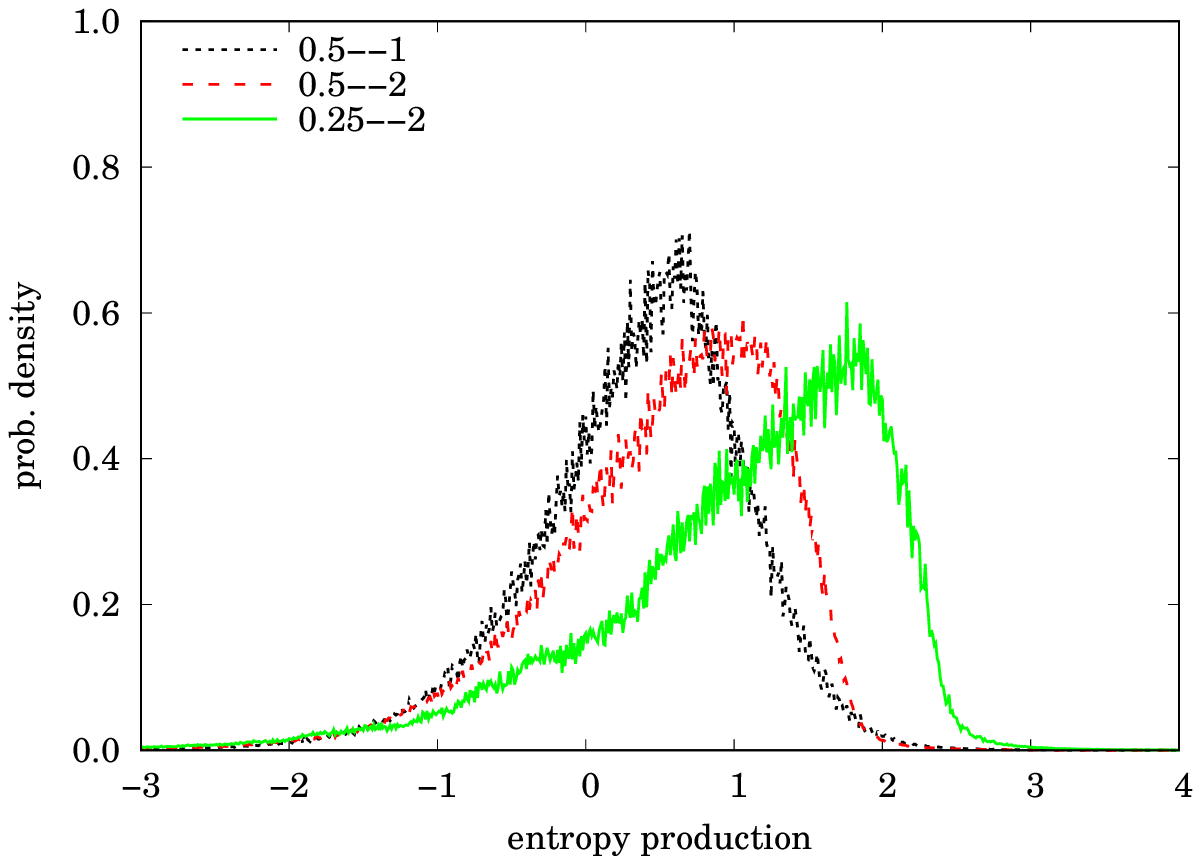}} &
(d)\scalebox{0.625}{\includegraphics*{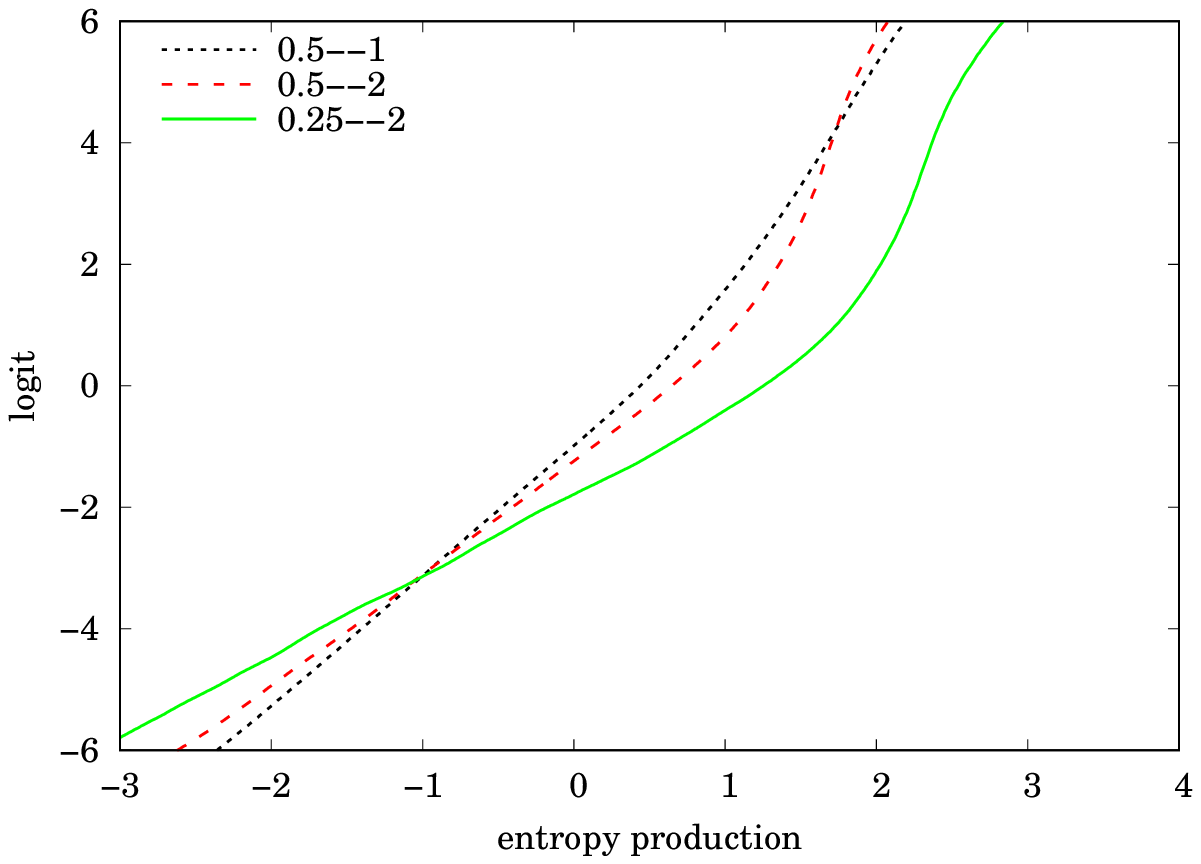}} \\
(e)\scalebox{0.625}{\includegraphics*{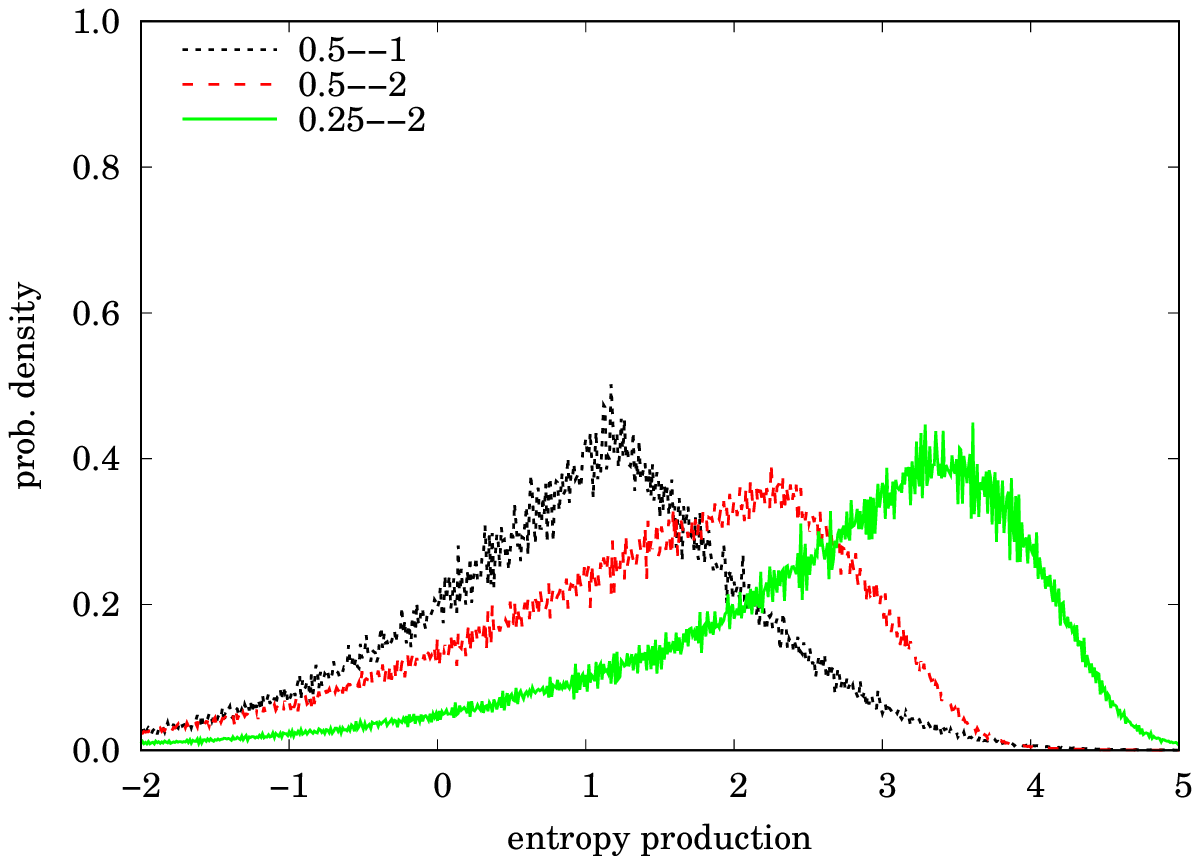}} &
(f)\scalebox{0.625}{\includegraphics*{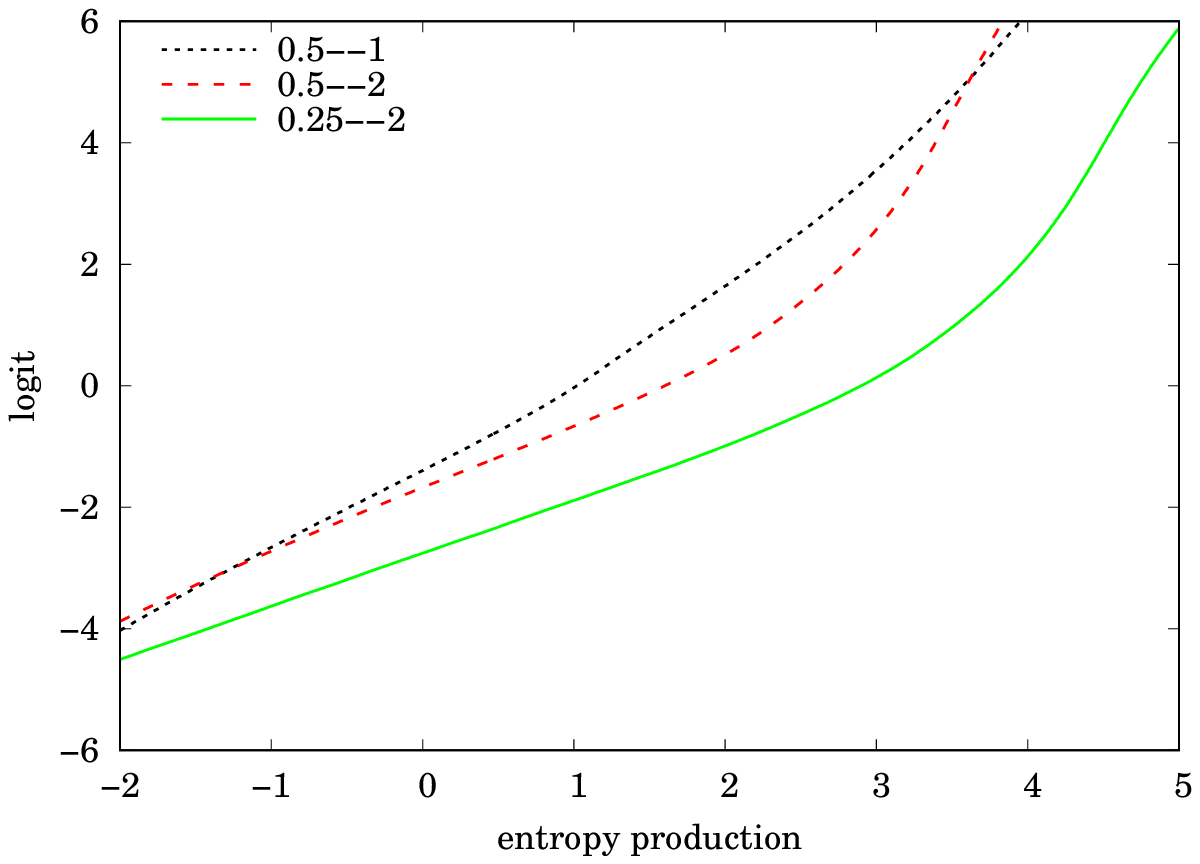}} 
\end{tabular}
\caption{\small Bivariate Student~$\mathrm{t}_3$ example. (a,c,e) Approximate pdf of entropy production over various time periods, starting from $\yzero=[0\;0]\trs$, $[1\;1]\trs$, $[2\;2]\trs$. (b,d,f) As (a,c,e) but on logit scale.
}
\label{fig:student2d}
\end{figure}

\begin{figure}
\noindent
\begin{tabular}{rr}
(a)\scalebox{0.625}{\includegraphics*{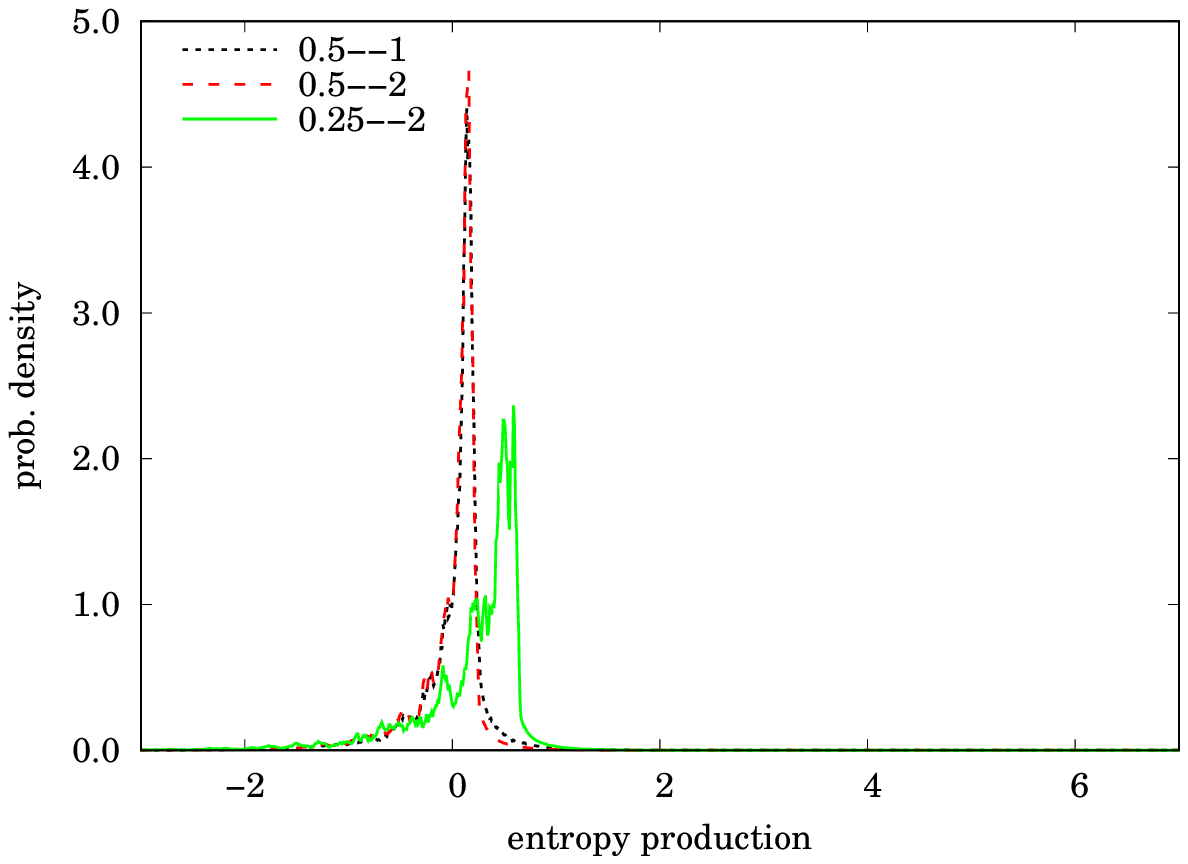}} &
(b)\scalebox{0.625}{\includegraphics*{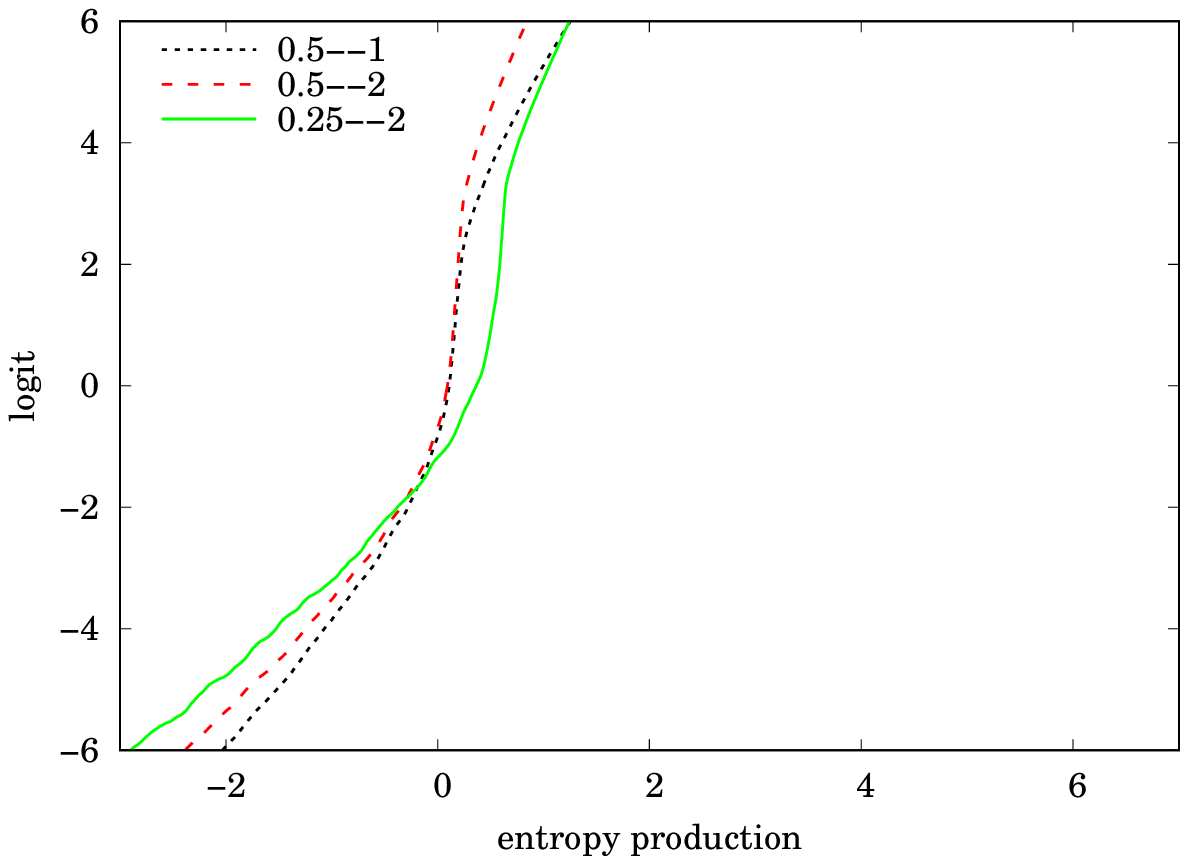}} \\
(c)\scalebox{0.625}{\includegraphics*{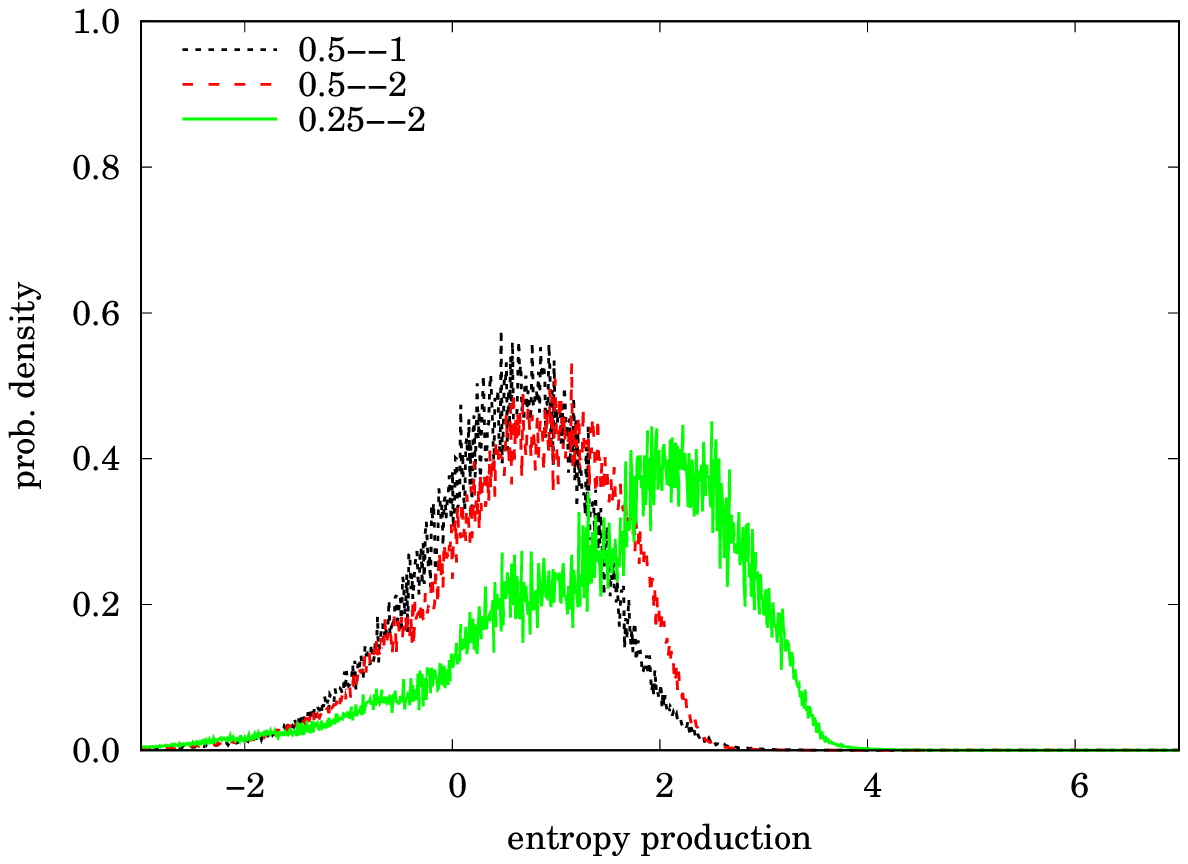}} &
(d)\scalebox{0.625}{\includegraphics*{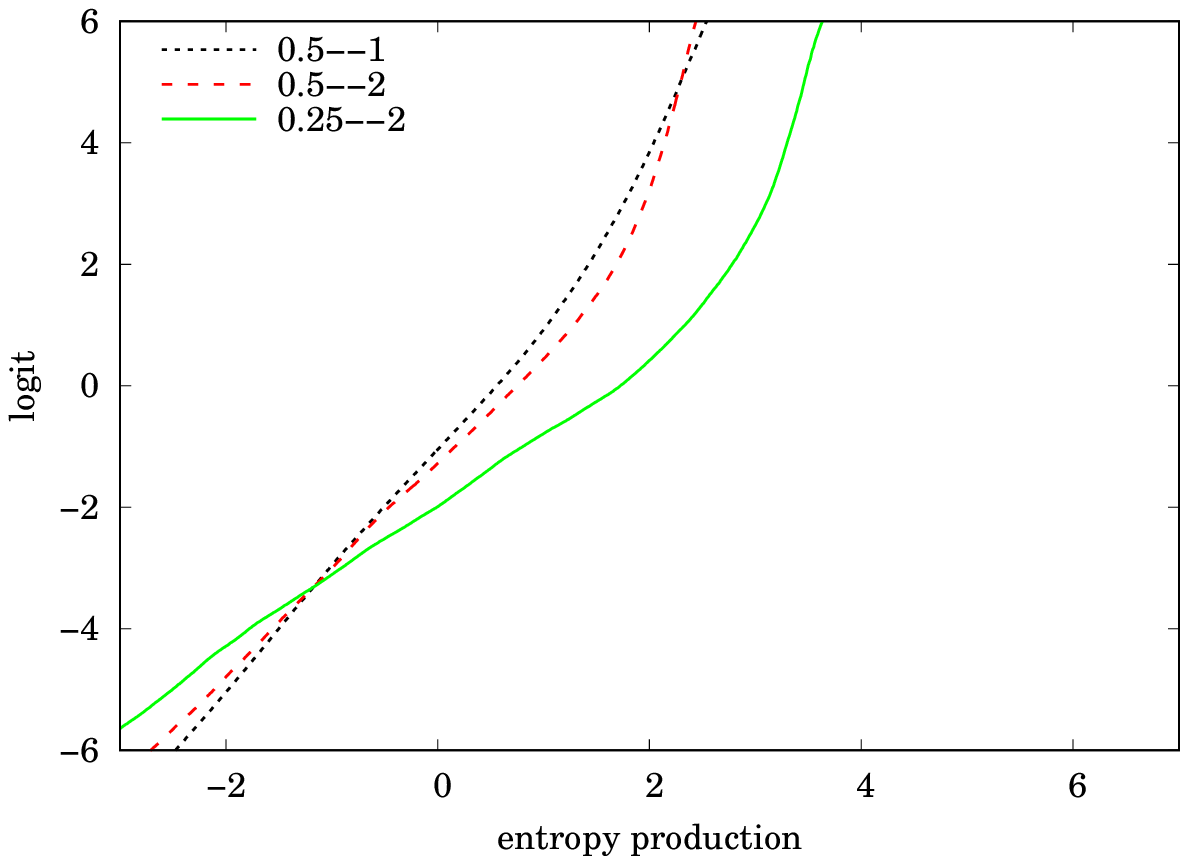}} \\
(e)\scalebox{0.625}{\includegraphics*{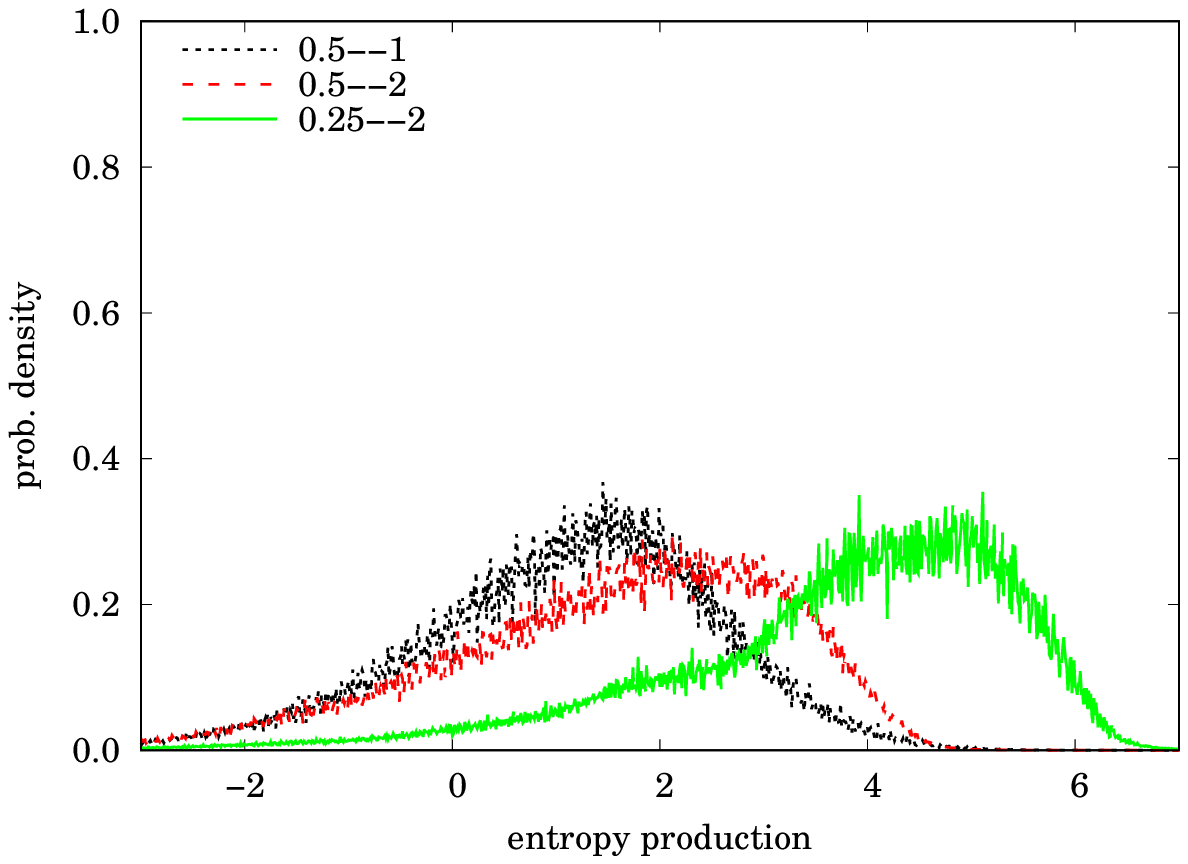}} &
(f)\scalebox{0.625}{\includegraphics*{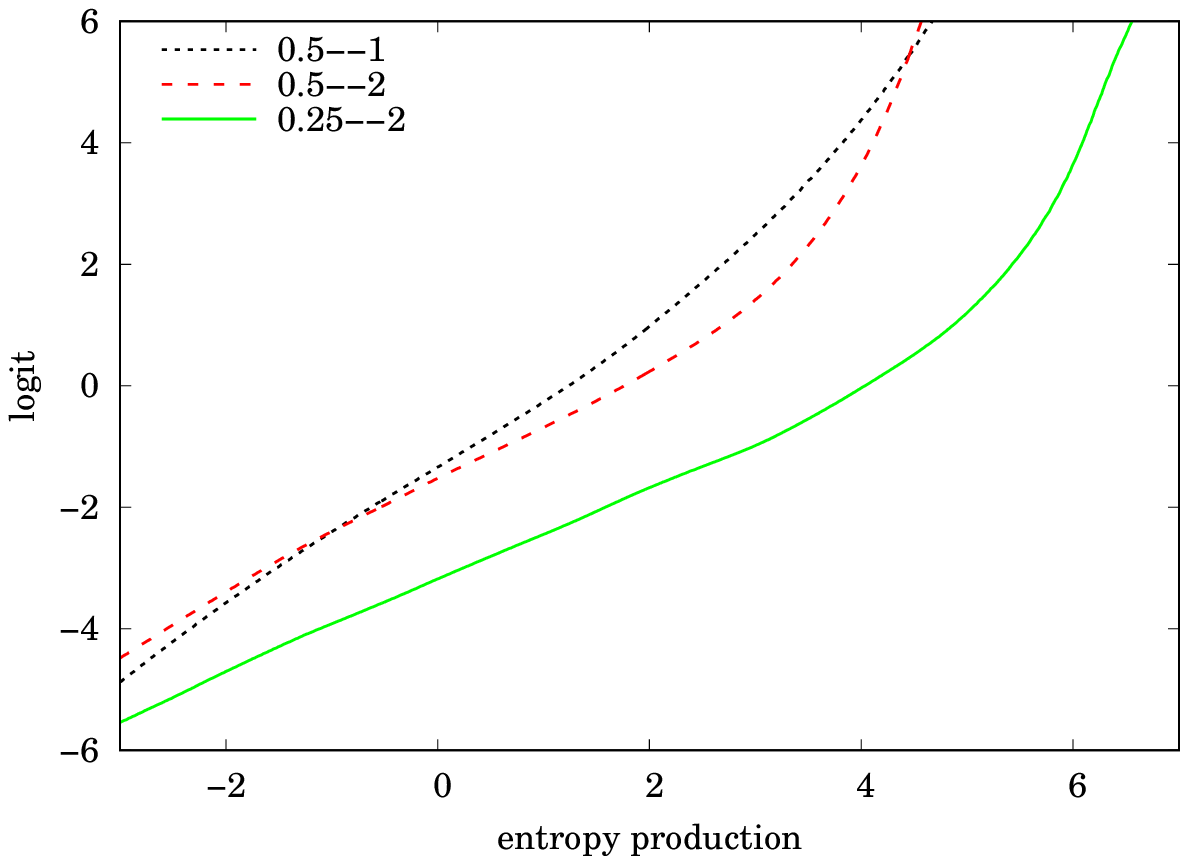}} 
\end{tabular}
\caption{\small Trivariate Student~$\mathrm{t}_3$ example. (a,c,e) Approximate pdf of entropy production over various time periods, starting from $\yzero=[0\;0\;0]\trs$, $[1\;1\;1]\trs$, $[2\;2\;2]\trs$. (b,d,f) As (a,c,e) but on logit scale.
}
\label{fig:student3d}
\end{figure}

\subsubsection{Student~t}

The multivariate form of the one-dimensional model we considered earlier is
\[
\fy(\infty,y) = \frac{\Gamma(\frac{\nu+d}{2})|\gm|^{1/2}}{\Gamma(\frac{\nu}{2})(\nu\pi)^{d/2}} \left(1 + \frac{y\trs \gm y}{\nu}\right)^{-\frac{\scriptstyle \nu+d}{\scriptstyle 2}} 
\]
for which
\[
\ffield(y) = \frac{-\frac{\nu+d}{\nu} \, \gm y}{1+y\trs \gm y/\nu}; \qquad
\langle -\nabla \ffield \rangle_\infty =  {\textstyle\frac{\nu+d}{\nu+d+2}} \, \gm;
\]
$\gm$ is related to the steady-state covariance matrix by $\msigma_\infty = \frac{\nu}{\nu-2} \gm\inv$, provided $\nu>2$.
Also
\[ 
\Omega(\tau,y) = \left( \frac{f_{\infty}(y)}{f_{\infty}(0)} \right) ^{\eta(\tau,y)}, \qquad
\eta(\tau,y) = y\trs \frac{\gm \sqrt{\mq}}{\mI+\sqrt{\mq}} \, y \biggr/ y\trs \gm y
\]
so (\ref{eq:newf2}) is explicit. 
The density starts off circularly-symmetric, and ends up ellipsoidal (for a numerical demonstration see the examples in \cite{Martin18b}). Like the univariate Student~t it has power-law tails.

Keeping $\nu=3$ as before we take two cases:
\begin{itemize}
\item
Bivariate ($d=2$) with generator and starting points
\[
\gm=\begin{bmatrix}1&0 \\ 0&2\end{bmatrix}; \qquad 
\yzero= \begin{bmatrix} 0 \\ 0 \end{bmatrix}, \begin{bmatrix} 1 \\ 1 \end{bmatrix}, \begin{bmatrix} 2 \\ 2 \end{bmatrix}
\]
The results are shown in Figure~\ref{fig:student2d}.
\item
Trivariate ($d=3$) with generator and starting points
\[
\gm=\begin{bmatrix}1&0&0 \\ 0&2&0 \\ 0&0&3 \end{bmatrix}; \qquad 
\yzero= \begin{bmatrix} 0 \\ 0 \\ 0 \end{bmatrix}, \begin{bmatrix} 1 \\ 1 \\ 1 \end{bmatrix}, \begin{bmatrix} 2 \\ 2 \\ 2\end{bmatrix}.
\]
The results are shown in Figure~\ref{fig:student3d}.
\end{itemize}

The results are qualitatively similar to those in Figure~\ref{fig:oud}, which is unsurprising in view of the qualitative similarity of the invariant density to the multivariate Gaussian.


\clearpage

\section{Conclusions and Final Remarks}

\label{sec:conc}

We have considered the problem of entropy production in diffusive systems evolving away from a given starting-point at time zero.
Entropy production is understood to be a measure of the extent to which time reversal symmetry is broken within a context of stochastic dynamics: it expresses the sense that certain patterns of evolution are more likely than the exact reverse behaviour. More precisely, it assesses the likelihood of observing reverse behaviour over a period subsequent to that in which the forward behaviour took place. In finding the distribution of entropy production over a certain time interval, we characterise the reversibility of the evolution of a system exposed to ill-determined environmental forces. 
Whereas macroscopic systems admit no reversibility of behaviour in these circumstances, and obey a firm requirement that the entropy production be non-negative, small systems can undergo fluctuations that allow them to retrace their steps, and such events give rise to negative entropy production. The modern formulation of the second law of thermodynamics can accommodate such behaviour.

In the OU model, if the starting-point coincides with the equilibrium level (i.e.\ where the force field vanishes)  then the distribution of  entropy production is the $K$-distribution, expressible in terms of the $K_\nu$ Bessel function. Otherwise the best route to obtaining the pdf is to use the inverse Fourier integral, as the moment-generating function is known in closed form. This is true regardless of dimension. The convenience of these results as well as the centrality of the OU model justifies the effort devoted to understanding it here. The result for a drifted Brownian motion (limit of zero mean reversion) is not materially simpler and is most easily found by the algebra of the OU derivation, suggesting that the OU process is the right way to approach this special case.

For nonlinear force fields (non-quadratic potentials), numerical simulation methods are required, together with a new and powerful analytical approximation to the transition density; the necessity for this machinery is particularly clear in problems of dimension $>1$. While potentials that are qualitatively similar to the OU model---in effect, unimodal potentials---produce qualitatively similar distributions of entropy production, investigation of the finer details requires numerical techniques. Further, when the divergence from the OU case is substantial, as for example in the double-well potential considered earlier, one has no choice but to go down the numerical route. This justifies the effort devoted in the paper to numerico-analytical work.

An obvious conclusion from any of our graphical results is that some realisations of the dynamics violate the classical thermodynamic behaviour, in the sense that there is always positive probability of negative entropy production. 
Another general result is that as time advances there is a shift to the right in probability mass of the entropy production, which is to be expected from the integral fluctuation theorem.
We also note that the pdf of entropy production often contains singularities, and that these are sometimes softened as time progresses; they are also less pronounced in higher-dimensional systems. The explicit pattern of entropy production can be decidedly complex, and certainly not symmetrical about the origin. Such are the statistics of the second law at the level of a small system evolving under the influence of its complex environment.

Further work in this field might consider more complex dynamics: an obvious idea is to incorporate jumps in both directions, thereby considering so-called  L\'evy processes. Typically these are considerably more difficult to analyse than simple diffusions, as the forward equation is no longer a parabolic PDE but instead an integro-differential equation. 
A general introduction to such processes is provided by  \cite{Sato02}, and \cite{Martin18a} shows how to use calculate certain functionals of L\'evy processes, as a way of generalising the Brownian motion.
There is, therefore, a broad scope for further work in this field, and we hope that the ideas demonstrated herein will provide fresh insight into stochastic thermodynamics, and allow concrete results to be obtained on difficult and analytically intractable models of the world.

\section*{Acknowledgements}

We thank Michael Kearney and Frank Kelly for helpful discussions.

\section*{Contacts and ORCIDs}

\begin{tabular}{ll}
{\tt richard.martin1@imperial.ac.uk} &   0000-0002-9272-1360 \\
{\tt i.ford@ucl.ac.uk}  & 0000-0003-2922-7332
\end{tabular}

\notthis{

\appendix
\section{Appendix}

\subsection{Notes on the OU process in one dimension}

The calculations pertaining to $M_{\Delta s}$ for the OU process
are facilitated by the following identity: 
\[
\int_{0}^{\infty}e^{-pt}K_{0}(t)\,dt=\left\{ \begin{array}{rl}
{\displaystyle \frac{\arccosh\,{p}}{\sqrt{p^{2}-1}},} & p>1\\
1, & p=1\\
{\displaystyle \frac{\arccos{p}}{\sqrt{1-p^{2}}},} & -1<p<1
\end{array}\right.
\]
using the standard branch of arccos, i.e.\ $[-1,1]\to[0,\pi]$. 

} 

\bibliographystyle{plain}
\bibliography{}

\begin{thebibliography}{10}

\bibitem{Abate95}
J.~Abate and W.~Whitt.
\newblock Numerical inversion of {L}aplace transforms of probability
  distributions.
\newblock {\em ORSA J. Comp.}, 7(1):36--43, 1995.

\bibitem{Abramowitz64}
M.~Abramowitz and I.~A. Stegun.
\newblock {\em Handbook of Mathematical Functions}.
\newblock Dover, New York, 1964.

\bibitem{Auconi19}
A.~Auconi.
\newblock {\em Fluctuations, irreversibility and causal influence in time
  series}.
\newblock PhD thesis, Humboldt U. of Berlin, 2019.

\bibitem{Bender78}
C.~M. Bender and S.~A. Orszag.
\newblock {\em Advanced mathematical methods for scientists and engineers}.
\newblock McGraw-Hill, New York, 1978.

\bibitem{Brown09}
H.~R. Brown, W.~Myrvold, and J.~Uffink.
\newblock Boltzmann's ${H}$-theorem, its discontents, and the birth of
  statistical mechanics.
\newblock {\em Stud. Hist. Phil. Mod. Phys.}, 40:174--191, 2009.

\bibitem{Carberry04}
D.~M. Carberry, J.~C. Reid, G.~M. Wang, E.M. Sevick, D.~J. Searles, and D.~J.
  Evans.
\newblock Fluctuations and irreversibility: {A}n experimental demonstration of
  a second-law-like theorem using a colloidal particle held in an optical trap.
\newblock {\em Phys. Rev. Lett.}, 92:140601, 2004.

\bibitem{Chatterjee10}
D.~Chatterjee and B.~J. Cherayil.
\newblock Exact path-integral evaluation of the heat distribution function of a
  trapped {B}rownian oscillator.
\newblock {\em Phys. Rev. E}, 82:051104, 2010.

\bibitem{Cohn84}
P.~M. Cohn.
\newblock {\em Algebra I}.
\newblock Wiley, 1984.

\bibitem{Crisanti17}
A.~Crisanti, A.~Sarracino, and M.~Zannetti.
\newblock Heat fluctuations of {B}rownian oscillators in nonstationary
  processes: {F}luctuation theorem and condensation transition.
\newblock {\em Phys. Rev. E}, 95:052138, 2017.

\bibitem{Daniels87}
H.~E. Daniels.
\newblock Tail probability approximations.
\newblock {\em International Statistical Review}, 55(1):37--48, 1987.

\bibitem{Deffner10}
S.~Deffner, O.~Abah, and E.~Lutz.
\newblock Quantum work statistics of linear and nonlinear parametric
  oscillators.
\newblock {\em Chem. Phys.}, 375:200--208, 2010.

\bibitem{Deffner08}
S.~Deffner and E.~Lutz.
\newblock Nonequilibrium work distribution of a quantum harmonic oscillator.
\newblock {\em Phys. Rev. E}, 77:021128, 2008.

\bibitem{Feuerverger00}
A.~Feuerverger and A.~C.~M. Wong.
\newblock Computation of value-at-risk for nonlinear portfolios.
\newblock {\em J. of Risk}, 3(1):37--55, 2000.

\bibitem{Ford13}
I.~J. Ford.
\newblock {\em Statistical Physics: An Entropic Approach}.
\newblock Wiley, 2013.

\bibitem{Ford15}
I.~J. Ford.
\newblock Measures of thermodynamic irreversibility in deterministic and
  stochastic dynamics.
\newblock {\em New J. Phys.}, 17:075017, 2015.

\bibitem{Ford12}
I.~J. Ford, D.~S. Minor, and S.~J. Binnie.
\newblock Symmetries of cyclic work distributions for an isolated harmonic
  oscillator.
\newblock {\em Eur. J. Phys.}, 33:1789--1801, 2012.

\bibitem{Gradshteyn94}
I.~S. Gradshteyn and I.~M. Ryzhik.
\newblock {\em {T}able of {I}ntegrals, {S}eries and {P}roducts}.
\newblock Academic, fifth edition, 1994.

\bibitem{Harris07}
R.~J. Harris and G.~M. Sch{\"u}tz.
\newblock Fluctuation theorems for stochastic dynamics.
\newblock {\em J. Stat. Mech. \textup{P07020}}, 2007.

\bibitem{Imparato05}
A.~Imparato and L.~Peliti.
\newblock Work-probability distribution in systems driven out of equilibrium.
\newblock {\em Phys. Rev. E}, 72:046114, 2005.

\bibitem{Imparato07}
A.~Imparato, L.~Peliti, G.~Pesce, G.~Rusciano, and A.~Sasso.
\newblock Work and heat probability distribution of an optically driven
  {B}rownian particle: {T}heory and experiments.
\newblock {\em Phys. Rev. E}, 76:050101, 2007.

\bibitem{Jackel02}
P.~J\"ackel.
\newblock {\em Monte Carlo Methods in Finance}.
\newblock Wiley, 2002.

\bibitem{Kelly79}
F.~P. Kelly.
\newblock {\em Reversibility and Stochastic Networks}.
\newblock Wiley, Chichester, 1979.

\bibitem{Lebowitz99}
J.~L. Lebowitz.
\newblock A century of statistical mechanics: a selective review of two central
  issues.
\newblock {\em Rev. Mod. Phys.}, 71:346, 1999.

\bibitem{Manikandan17}
S.~K. Manikandan and S.~Krishnamurthy.
\newblock Asymptotics of work distributions in a stochastically driven system.
\newblock {\em {\tt arXiv:1706.06489}}, 2017.

\bibitem{Martin11b}
R.~J. Martin.
\newblock Saddlepoint methods in portfolio theory.
\newblock In A.~Lipton and A.~Rennie, editors, {\em The Oxford Handbook of
  Credit Derivatives}, chapter~15. Oxford University Press, 2011.
\newblock {\tt arXiv:1201.0106}.

\bibitem{Martin15b}
R.~J. Martin, R.~V. Craster, and M.~J. Kearney.
\newblock Infinite product expansion of the {F}okker--{P}lanck equation with
  steady-state solution.
\newblock {\em Proc. R. Soc. A}, 471(2179):20150084, 2015.

\bibitem{Martin18b}
R.~J. Martin, R.~V. Craster, A.~Pannier, and M.~J. Kearney.
\newblock Asymptotic approximation to the multidimensional {F}okker--{P}lanck
  equation with steady state.
\newblock {\em J. Phys. A: Math. Theor.}, 52:085002, 2019.
\newblock {\tt arXiv:1810.08401}.

\bibitem{Martin19a}
R.~J. Martin, R.~V. Craster, A.~Pannier, and M.~J. Kearney.
\newblock Long- and short-time asymptotics of the {O}rnstein--{U}hlenbeck and
  other mean-reverting processes.
\newblock {\em J. Phys. A: Math. Theor.}, 52(13):134001, 2019.
\newblock {\tt arXiv:1810.13010}.

\bibitem{Martin18a}
R.~J. Martin and M.~J. Kearney.
\newblock Time since maximum of {B}rownian motion and asymmetric {L}\'evy
  processes.
\newblock {\em J. Phys. A: Math. Theor.}, 51:275001, 2018.

\bibitem{Nicolis17}
G.~Nicolis and Y.~{De Decker}.
\newblock Stochastic thermodynamics of {B}rownian motion.
\newblock {\em Entropy}, 19:434, 2017.

\bibitem{Niederreiter92}
H.~Niederreiter.
\newblock {\em Random Number Generation and Quasi-Monte Carlo Methods}.
\newblock SIAM, 1992.

\bibitem{Pal13}
A.~Pal and S.~Sabhapandit.
\newblock Work fluctuations for a {B}rownian particle in a harmonic trap with
  fluctuating locations.
\newblock {\em Phys. Rev. E}, 87:022138, 2013.

\bibitem{Pal14}
A.~Pal and S.~Sabhapandit.
\newblock Work fluctuations for a {B}rownian particle driven by a correlated
  external random force.
\newblock {\em Phys. Rev. E}, 90:052116, 2014.

\bibitem{NRC}
W.~H. Press, B.~P. Flannery, S.~A. Teukolsky, and W.~T. Vetterling.
\newblock {\em Numerical Recipes in C++}.
\newblock Cambridge University Press, 2002.

\bibitem{Qian13}
H.~Qian.
\newblock A decomposition of irreversible diffusion processes without detailed
  balance.
\newblock {\em {\tt arXiv:1204.6496}}, 2013.

\bibitem{Saha09}
A.~Saha, S.~Lahiri, and A.~M. Jayannavar.
\newblock Entropy production theorems and some consequences.
\newblock {\em Phys. Rev. E}, 80:011117, 2009.

\bibitem{Sato02}
K.-I. Sato.
\newblock {\em L\'evy Processes and Infinitely Divisible Distributions}.
\newblock Cambridge University Press, 2002.

\bibitem{Seifert05}
U.~Seifert.
\newblock Entropy production along a stochastic trajectory and an integral
  fluctuation theorem.
\newblock {\em Phys. Rev. Lett.}, 95:040602, 2005.

\bibitem{Seifert08}
U.~Seifert.
\newblock Stochastic thermodynamics: principles and perspectives.
\newblock {\em Eur. Phys. J. B}, 64(3-4):423--431, 2008.

\bibitem{SpinneyFord12a}
R.~E. Spinney and I.~J. Ford.
\newblock Fluctuation relations: a pedagogical overview.
\newblock In R.~J. Klages, W.~Just, and C.~Jarzynski, editors, {\em
  Nonequilibrium Statistical Physics of Small Systems: Fluctuation Relations
  and Beyond}. Wiley-VCH, 2012.

\bibitem{Talkner08}
P~Talkner, P.~{Sekhar Burada}, and P.~H\"anggi.
\newblock Statistics of work performed on a forced quantum oscillator.
\newblock {\em Phys. Rev. E}, 78:011115, 2008.

\bibitem{Touchette10a}
H.~Touchette, E.~Van der Straeten, and W.~Just.
\newblock Brownian motion with dry friction: {F}okker--{P}lanck approach.
\newblock {\em J. Phys A: Math. Theor.}, 43:445002, 2010.
\newblock {\tt arXiv:1008.3331v2}.

\bibitem{Uhlenbeck30}
G.~E. Uhlenbeck and L.~S. Ornstein.
\newblock On the theory of {B}rownian motion.
\newblock {\em Phys. Rev.}, 36:823--841, 1930.

\bibitem{Weeks66}
W.~T. Weeks.
\newblock Numerical inversion of {L}aplace transforms using {L}aguerre
  functions.
\newblock {\em J. ACM}, 13:419--426, 1966.

\bibitem{Wibisono19}
A.~Wibisono and V.~Jog.
\newblock Convexity of mutual information along the {O}rnstein--{U}hlenbeck
  flow.
\newblock {\em {\tt arXiv:1805.01401}}, 2019.

\end{thebibliography}

\end{document}